\documentclass[twocolumn]{aastex63}
\usepackage{breqn}
\usepackage{amsmath}
\submitjournal{ApJ. Accepted on November 9, 2020}

\shorttitle{SAMI: bulge and disk stellar populations}
\shortauthors{Barsanti et al.}
\graphicspath{{./}{figures/}}

\begin{document}

\title{The SAMI Galaxy Survey: bulge and disk stellar population properties in cluster galaxies}

\correspondingauthor{Stefania Barsanti}
\email{stefania.jess@gmail.com}

\author{S. Barsanti}
\affiliation{Department of Physics and Astronomy, Macquarie University, NSW 2109, Australia}
\affiliation{Astronomy, Astrophysics and Astrophotonics Research Centre, Macquarie University, Sydney, NSW 2109, Australia}

\author{M. S. Owers}
\affiliation{Department of Physics and Astronomy, Macquarie University, NSW 2109, Australia}
\affiliation{Astronomy, Astrophysics and Astrophotonics Research Centre, Macquarie University, Sydney, NSW 2109, Australia}

\author{R. M. McDermid}
\affiliation{Department of Physics and Astronomy, Macquarie University, NSW 2109, Australia}
\affiliation{Astronomy, Astrophysics and Astrophotonics Research Centre, Macquarie University, Sydney, NSW 2109, Australia}

\author{K. Bekki}
\affiliation{ICRAR, The University of Western Australia, 35 Stirling Highway, Crawley, WA 6009, Australia}
\author{J. Bland-Hawthorn}
\affiliation{Sydney Institute for Astronomy (SIfA), School of Physics, University of Sydney, NSW 2006, Australia}

\author{S. Brough}
\affiliation{School of Physics, University of New South Wales, NSW 2052, Australia}
\affiliation{ARC Centre of Excellence for All Sky Astrophysics in 3 Dimensions (ASTRO 3D), Australia}

\author{J. J. Bryant}
\affiliation{Sydney Institute for Astronomy (SIfA), School of Physics, University of Sydney, NSW 2006, Australia}
\affiliation{ARC Centre of Excellence for All Sky Astrophysics in 3 Dimensions (ASTRO 3D), Australia}
\affiliation{Australian Astronomical Optics, AAO-USydney, School of Physics, University of Sydney, NSW 2006, Australia}

\author{L. Cortese}
\affiliation{ICRAR, The University of Western Australia, 35 Stirling Highway, Crawley, WA 6009, Australia}
\affiliation{ARC Centre of Excellence for All Sky Astrophysics in 3 Dimensions (ASTRO 3D), Australia}

\author{S. M. Croom}
\affiliation{Sydney Institute for Astronomy (SIfA), School of Physics, University of Sydney, NSW 2006, Australia}
\affiliation{ARC Centre of Excellence for All Sky Astrophysics in 3 Dimensions (ASTRO 3D), Australia}

\author{C. Foster}
\affiliation{Sydney Institute for Astronomy (SIfA), School of Physics, University of Sydney, NSW 2006, Australia}
\affiliation{ARC Centre of Excellence for All Sky Astrophysics in 3 Dimensions (ASTRO 3D), Australia}
\author{J. S. Lawrence}
\affiliation{Australian Astronomical Optics - Macquarie, Macquarie University, NSW 2109, Australia}

\author{\'A. R. L\'opez-S\'anchez}
\affiliation{Department of Physics and Astronomy, Macquarie University, NSW 2109, Australia}
\affiliation{ARC Centre of Excellence for All Sky Astrophysics in 3 Dimensions (ASTRO 3D), Australia}
\affiliation{Australian Astronomical Optics - Macquarie, Macquarie University, NSW 2109, Australia}

\author{S. Oh}
\affiliation{ARC Centre of Excellence for All Sky Astrophysics in 3 Dimensions (ASTRO 3D), Australia}
\affiliation{Research School of Astronomy and Astrophysics, Australian National University, Canberra, ACT 2611, Australia}

\author{A. S. G. Robotham}
\affiliation{ICRAR, The University of Western Australia, 35 Stirling Highway, Crawley, WA 6009, Australia}
\affiliation{ARC Centre of Excellence for All Sky Astrophysics in 3 Dimensions (ASTRO 3D), Australia}

\author{N. Scott}
\affiliation{Sydney Institute for Astronomy (SIfA), School of Physics, University of Sydney, NSW 2006, Australia}
\affiliation{ARC Centre of Excellence for All Sky Astrophysics in 3 Dimensions (ASTRO 3D), Australia}

\author{S. M. Sweet}
\affiliation{ARC Centre of Excellence for All Sky Astrophysics in 3 Dimensions (ASTRO 3D), Australia}
\affiliation{School of Mathematics and Physics, University of Queensland, Brisbane, QLD 4072, Australia}
\author{J. van de Sande}
\affiliation{Sydney Institute for Astronomy (SIfA), School of Physics, University of Sydney, NSW 2006, Australia}
\affiliation{ARC Centre of Excellence for All Sky Astrophysics in 3 Dimensions (ASTRO 3D), Australia}




\begin{abstract}

We explore stellar population properties separately in the bulge and the disk of double-component cluster galaxies to shed light on the formation of lenticular galaxies in dense environments. We study eight low-redshift clusters from the Sydney-AAO Multi-object Integral field (SAMI) Galaxy Survey, using 2D photometric bulge-disk decomposition in the $g$, $r$ and $i$-bands to characterize galaxies. For 192 double-component galaxies with $M_{*}>10^{10~}M_{\odot}$ we estimate the color, age and metallicity of the bulge and the disk. The analysis of the $g-i$ colors reveals that bulges are redder than their surrounding disks with a median offset of 0.12$\pm$0.02 mag, consistent with previous results. To measure mass-weighted age and metallicity we investigate three methods: (i) one based on galaxy stellar mass weights for the two components, (ii) one based on flux weights and (iii) one based on radial separation. The three methods agree in finding 62\% of galaxies having bulges that are 2-3 times more metal-rich than the disks. Of the remaining galaxies, 7\% have bulges that are more metal-poor than the disks, while for 31\% the bulge and disk metallicities are not significantly different. We observe 23\% of galaxies being characterized by bulges older and 34\% by bulges younger with respect to the disks. The remaining 43\% of galaxies have bulges and disks with statistically indistinguishable ages. Redder bulges tend to be more metal-rich than the disks, suggesting that the redder color in bulges is due to their enhanced metallicity relative to the disks instead of differences in stellar population age.
\end{abstract}



\section{Introduction} 
Lenticular (S0) galaxies were introduced as a separate morphology for the first time by \citet{Hubble1936}. They are characterized by a central bulge, similar to elliptical galaxies, and a disk, similar to spiral galaxies. However, unlike spiral galaxies, the disks of S0 do not contain spiral arms and ongoing star formation. Due to these features S0s are located between ellipticals and spirals in the Hubble Sequence and are considered a transitional phase. S0s show heterogeneous physical properties (e.g. \citealp{Laurikainen2010, Cappellari2011, Barway2013}), causing their origin still to be highly debated. Historically, S0s were speculated to be formed from faded spirals due to the similarities between their structural and kinematical properties \citep{vandenBergh1976,Bedregal2006,Moran2007,Prochaska2011}. However, some works show that some properties of S0s differ from those of
spiral galaxies \citep{Christlein2004,Williams2010,Falcon2015}, indicating that major mergers can be responsible for their formation \citep{Spitzer1951,Bekki1998,Tapia2017}.  

S0s are mainly found in groups and clusters of galaxies while spirals populate lower density regions, implying that the environment plays a fundamental role in their formation \citep{Dressler1980}. This result suggests a scenario where spiral galaxies are transformed into S0s when they enter the cluster environment, where their star formation is quenched (e.g. \citealt{Dressler1997,Fasano2000,Kormendy2012}). The fraction of S0 galaxies compared to spirals in clusters is observed to increase with time, providing evidence for the transformation of spirals into S0s \citep{Poggianti2009}. Groups and pre-processing mechanisms can also play an important role for this transformation, since the changing fraction of S0s relative to spirals as a function of time is found to increase for lower halo masses \citep{Poggianti2009}. This suggests that mechanisms that are not simply due to cluster processes might also be involved. 

To explore the possible formation of S0 galaxies, it is fundamental to understand the individual stellar populations of the bulges and the disks of S0 cluster galaxies. The study of the age and metallicity of the separate stellar populations help us to distinguish between scenarios exclusively caused by cluster processes and those due to mechanisms that may also occur in lower density environments. Environment processes in clusters include interactions with the intra-cluster medium when the galaxy travels through the cluster. In this scenario the cold gas in the disk is stripped by ram pressure stripping \citep{Gunn1972}, or the hot halo reservoir is removed by strangulation \citep{Larson1980}. These processes act on the disks and leave the old stellar populations in the bulge unperturbed \citep{Quilis2000}. On the other hand, a strongly concentrated star formation activity in the galaxy central region is observed when neighbouring galaxies interact, causing gas stripping by galaxy harassment in clusters \citep{Moore1996} or minor mergers in lower density environments \citep{Mihos1994}. These scenarios imprint different fossil records in the stellar populations suggesting that they may be distinguished via analyses of spectra.

In recent years, much effort has been devoted to the study of S0 galaxies and the properties of the stellar populations of their components, such as colors, ages and metallicities. This allows the development of new methodologies in order to identify S0 galaxies, decompose them in their components and analyze their properties both in galaxy clusters \citep{Hudson2010,Johnston2014,Head2014,Johnston2020} and in the field environment \citep{FraserMcKelvie2018,Tabor2019,Mendez2019,Mendez2019b,Johnston2020}. Differences between the bulge and disk properties of cluster and field galaxies may indicate different formation scenarios for S0s. In this paper, we focus on the stellar populations of S0s in clusters in order to understand the role that environment plays in their formation.  

While previous studies of S0s in clusters have provided important clues that help to understand the impact of the cluster environment on S0s, they have been limited by their focus on photometric investigations \citep{Head2014}, or by the relatively small number of objects with resolved spectroscopy \citep{Johnston2014}. \citet{Head2014} studied 200 S0 galaxies in the Coma cluster and performed 2D photometric bulge-disk decomposition in the $u$, $g$ and $i$-bands. They analyzed the colors of the bulges and the disks, observing that the bulges are redder than the disk components. This color difference is expressed as linear combinations of stellar population ages and metallicities. They concluded that bulges are either 2-3 times older or $\sim$2 times more metal-rich than the disks, suggesting a scenario of disk fading caused by cluster processes. However, the study of the colors alone is not sufficient to break the age-metallicity degeneracy, hence \citet{Head2014} were not able to distinguish between the two scenarios.

\citet{Johnston2014} studied 21 S0s in the Virgo cluster, decomposing long-slit spectroscopic data to build separated one-dimensional spectra of the bulge and the disk for each galaxy. 13 S0s could be reliably decomposed using this method. They analyzed the Lick indices of these spectra and the $\alpha$-element abundances, observing younger and more metal-rich stellar populations in bulges than those in the disks. They suggested that the most recent episode of star formation in S0s occurs in the bulge, caused by gas previously enriched in the disk which has been dumped in the bulge during the processes of stripping and quenching of the disk. This population of more metal-rich bulges is consistent with the results of \citet{Head2014}, however only a handful of galaxies in one cluster were analyzed by \cite{Johnston2014}. In order to overcome these limitations, the large statistical samples available from previous photometric studies need to be combined with the spectroscopic precision achieved on small samples. 

In this context, the aim of this paper is to study the ages and metallicities of the individual stellar populations of the S0 galaxy components for a large sample in the cluster environment in order to understand the formation of these galaxies. We make use of SAMI Galaxy Survey \citep{Bryant2015} since it is the only integral field spectroscopic survey that specifically targets cluster regions, collecting resolved spectroscopy for 906 galaxies in eight low-redshift high-mass clusters \citep{Owers2017}. The statistically significant number of galaxies in the cluster sample of the SAMI Galaxy Survey combined with its spatially-resolved spectroscopic information allow us to disentangle the age-metallicity degeneracy and to better understand the separate bulge-disk stellar populations. We use 2D photometric bulge-disk decomposition to identify the double-component galaxy sample, i.e. galaxies characterized by both a bulge and a disk. Barsanti et al. in preparation perform the decomposition for the SAMI cluster sample in the $g$, $r$ and $i$-bands using the image analysis package {\sc ProFound} and the photometric galaxy profile fitting package {\sc ProFit} \citep{Robotham2017,Robotham2018}. We test three methods to estimate the age and metallicity of the bulge and the disk, combining the 2D photometric bulge-disk decomposition with the stellar population information from the SAMI spectra:
\begin{enumerate}
\item We develop a new method based on galaxy stellar mass weights for the two components. We use the 2D photometric bulge-disk decomposition results to estimate for each spatial bin the contributions of the bulge and the disk to the total galaxy stellar mass. These bulge-disk stellar mass fractions are used as weights for the galaxy age and metallicity derived from the SAMI spectra.   
\item We adapt a method similar to that of \citet{Mendez2019} based on flux weights for the two components. We perform 2D photometric bulge-disk decomposition on the SAMI datacubes at each SAMI wavelength to obtain separate datacubes and 1D spectra for the bulge and the disk.
\item We adapt a method similar to that of \citet{FraserMcKelvie2018} based on radial separation. We consider as the 1D bulge spectrum the galaxy spectrum from the most central bin, while the representative 1D disk spectrum is the galaxy spectrum from the outermost spatial bin.  
\end{enumerate}

This paper is organized as follows. We present our SAMI cluster sample and the selection of double-component galaxies in Section~\ref{sec:Data}. Section~\ref{Stellar population properties} describes the three different methods to estimate mass-weighted single-age and single-metallicity values of the bulge and the disk. In Section~\ref{sec:Results} we present our results, analyzing the comparison between the stellar population properties of the bulges and the disks from the three methods. In Section~\ref{sec:Discussion} we compare our results with previous works and we discuss the physical implications. Finally, we summarize and conclude in Section~\ref{Summary and conclusions}. Throughout this work, the uncertainties on the percentages are estimated using the method of \citet{Cameron2011} to measure confidence intervals on binomial population proportions. The bin sizes of the histograms are estimated according to the Scott's rule of thumb \citep{doi:10.1002/wics.103}. We assume $\Omega_{m}=0.3$, $\Omega_{\Lambda}=0.7$ and $H_{0}=70$ km $\rm s^{-1} Mpc^{-1}$ as cosmological parameters.

\section{Data and Sample Selection}
\label{sec:Data}
In this Section we describe the details of the SAMI Galaxy Survey and the considered sample of SAMI cluster galaxies. Finally, we select the sample of SAMI cluster galaxies modelled by a photometric double-component profile. 

\subsection{SAMI Galaxy Survey}
\label{SAMI galaxy survey}
SAMI was mounted on the 3.9m Anglo-Australian Telescope \citep{Croom2012}. The instrument is characterized by 13 fused optical fibre bundles (hexabundles), each one containing 61 fibres of 1.6$^{\prime \prime}$ diameter resulting in each Integral Field Unit (IFU) having a 15$^{\prime \prime}$ diameter \citep{Bland2011,Bryant2014}. The SAMI fibres are fed into the two arms of the AAOmega spectrograph \citep{Sharp2006}. The SAMI Galaxy Survey uses the 580V grating in the blue arm resulting in a resolution R=1812 and wavelength coverage of 3700-5700~\AA, and the 1000R grating in the red arm resulting in the higher resolution R=4263 over the range 6300-7400~\AA. The median full-width-at-half-maximum values for each arm are FWHM$_{blue}$=2.65~\AA\hspace{0.5mm} and FWHM$_{red}$=1.61~\AA\hspace{0.5mm} \citep{vandeSande2017}.

The SAMI Galaxy Survey is a spatially-resolved spectroscopic survey of more than 3000 galaxies collected during 2013-2018, with stellar mass range $\log{(M_{*}/M_{\odot})}=8-12$ and redshift range $0.004<z\leq 0.115$ \citep{Bryant2015}. The data are reduced using the SAMI {\sc python} package (\citealp{Allen2014}) which includes the 2d{\sc fdr} package \citep{2015ascl.soft05015A}. We make use of the internal data release v0.11 (for a complete description of the data reduction from raw frames to datacubes see \citealp{Sharp2015,Allen2015,Green2018,Scott2018}). The final datacubes are characterized by a grid of 0.5$^{\prime \prime}\times0.5^{\prime \prime}$ spaxels, where the blue and red spectra have pixel scales of 1.03~\AA\hspace{1mm}and 0.56~\AA, respectively. 

\newpage
\subsection{SAMI cluster sample}
\label{sec:cluster sample}
The cluster component of the SAMI Galaxy Survey contains 906 galaxies selected from eight low-redshift massive clusters with virial masses $\log{(M_{200}/M_{\odot})}=14.25-15.19$ at $0.029<z<0.058$ (for a detailed description see \citealp{Owers2017}). In summary, the galaxies are selected using: redshift-dependent stellar mass cuts with a lowest limit of $\log{(M_{*}/M_{\odot})}=9.5$, projected cluster-centric distance $R<R_{200}$ and peculiar velocity $\mid v_{pec}\mid <3.5\sigma_{200}$ where $\sigma_{200}$ is the cluster velocity dispersion within $R_{200}$.

The SAMI cluster sample contains the four clusters APMCC0917, EDCC0442, A3880 and A4038 selected from the 2dFGRS catalogue \citep{DePropris2002} with the photometry provided by the VLT Survey Telescope's ATLAS (VST/ATLAS) survey \citep{Shanks2013,Shanks2015}, and the four clusters A85, A168, A119 and A2399 in the SDSS regions with the photometric data from SDSS DR9 \citep{Ahn2012}. The imaging data are used in Section~\ref{Galaxy characterization} for galaxy characterization with the purpose of selecting only double-component cluster galaxies. 

\subsection{Galaxy characterization}
\label{Galaxy characterization}
Our primary aim is to understand the bulge and disk stellar populations of S0 galaxies. To do so, we must first identify those SAMI galaxies that can be decomposed into double-component systems. We use a purely photometric approach to separate the bulge and the disk components. This can cause semantic confusion compared to the classical ``bulge'' and ``disk'' definitions. In this study the ``disk'' is defined as the exponential component, while the ``bulge'' corresponds to the S\'ersic component representing the light excess over the exponential component. We decide to nominate the two components as the ``bulge'' and the ``disk'' in line with the previous literature and in order to offer easier comparisons with the previous works of \citet{Hudson2010,Head2014,Johnston2014,FraserMcKelvie2018,Mendez2019}.

We make use of the two-dimensional (2D) photometric bulge-disk decomposition in the $g$-, $r$- and $i$-bands presented by Barsanti et al. in preparation for the SAMI cluster sample. They use the VST/ATLAS imaging for the APMCC0917, EDCC0442, A3880 and A4038 clusters and the SDSS imaging for the A85, A168, A119 and A2399 clusters, as introduced in Section~\ref{sec:cluster sample}. More details about the imaging data for the clusters are reported in \citet{Owers2017}. The 2D bulge-disk decomposition is performed for 1204 SDSS and 591 VST/ATLAS cluster galaxies, for a total of 1795. This sample includes galaxies characterized by stellar mass $M_{*}\geq10^{9.5}\,M_{\odot}$, projected cluster-centric distance within $2.5\,R_{200}$ and both with and without SAMI observations. For source finding and image analysis they use the {\sc ProFound} package \citep{Robotham2018}. {\sc ProFound} offers the key input requirements to {\sc ProFit}, a package for Bayesian 2D photometric galaxy profile modelling \citep{Robotham2017}. Barsanti et al. in prep. fit single-component S\'ersic profile and multiple double-component bulge+disk profiles for each galaxy in the $r$-band. The disk profile is fixed to exponential, while the S\'ersic bulge profile varies with increasing complexity. The results for the structure parameters from the $r$-band are used as inputs for the single/double-component fits in the $g$- and $i$-bands. The pipelines {\sc ProFound}+{\sc ProFit} work in the $r$-, $g$- and $i$-bands for 1730/1795 cluster galaxies. For these galaxies the model selection is performed in the $r$-band using the Bayes Factor (BF). The double-component sample has $\ln(\rm{BF})>60$. Double-component galaxies for which the separation of the bulge and disk measurements is unreliable, i.e. they are dominated by one component having bulge-to-total flux ratio B/T$<0.2$ and B/T$>0.8$, or for which the fits are unphysical, i.e. the fitted parameters are pegged at the fitting boundaries, are excluded. The final number of double-component cluster galaxies is 469, including galaxies with $M_{*}\geq10^{9.5}\,M_{\odot}$, within $2.5\,R_{200}$ and both with and without SAMI observations. 

From the 469 double-component galaxies, we select only the SAMI primary targets, i.e. galaxies with SAMI observations within $R_{200}$. We consider only galaxies with stellar mass $M_{*}\geq10^{10}\,M_{\odot}$; for galaxies with $M_{*}$ lower than this limit the SAMI data contain only a small number of spaxels with signal-to-noise ratio $>$ 3 \citep{Owers2019}. We include galaxies with maximum velocity dispersion $\sigma<400$ km $\rm{s^{-1}}$ as measured by \citet{vandeSande2017}. High velocity dispersions are typical of the brightest cluster galaxies. Massive and bright ellipticals are excluded since their complex light profiles can contaminate the double-component galaxy selection. Applying these selection criteria we find 192 SAMI double-component galaxies. 

Since we want to study S0 galaxies, we consider the spectroscopic, kinematic and morphological properties of the 192 double-component galaxies. Using the spectroscopic classification of \citet{Owers2019}, we find that 170/190 double-component galaxies are in the passive class, while for 2 galaxies the measurements are not available. \citet{vandeSande2017b} measured the kinematic properties for 175/192 double-component galaxies, finding that 160/175 galaxies are fast rotators according to the definition of \citet{Cappellari2016}. Using the morphology classification of \citet{Cortese2016} we find that 146/191 are S0s/possible S0s. Thus, our double-component cluster galaxy sample is mainly characterized by passive, fast-rotating S0s. 

\begin{figure*}
\centering
\includegraphics[width=\columnwidth]{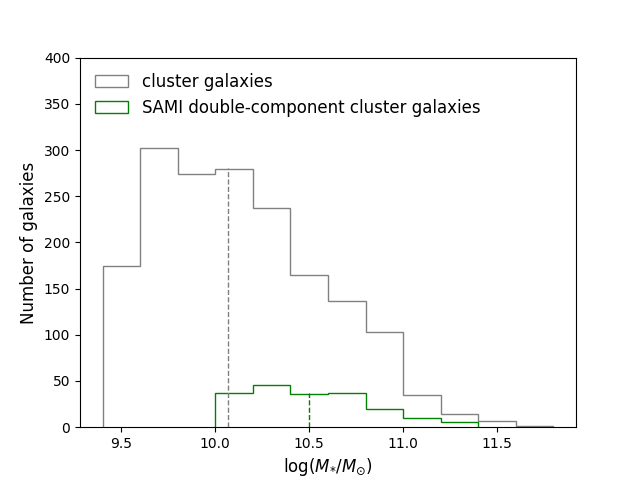}
\includegraphics[width=\columnwidth]{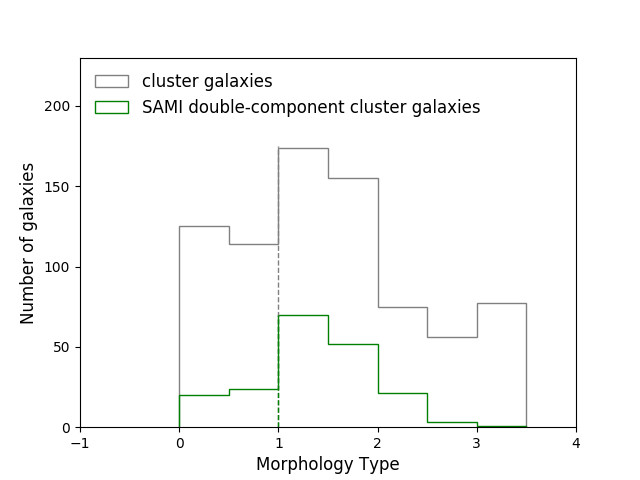}
\includegraphics[width=\columnwidth]{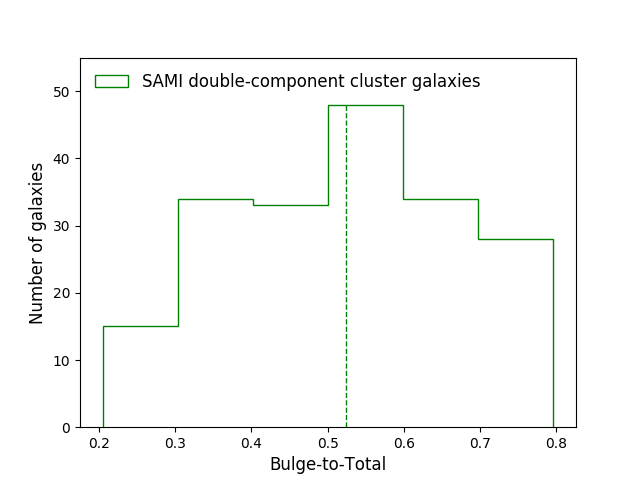}
\includegraphics[width=\columnwidth]{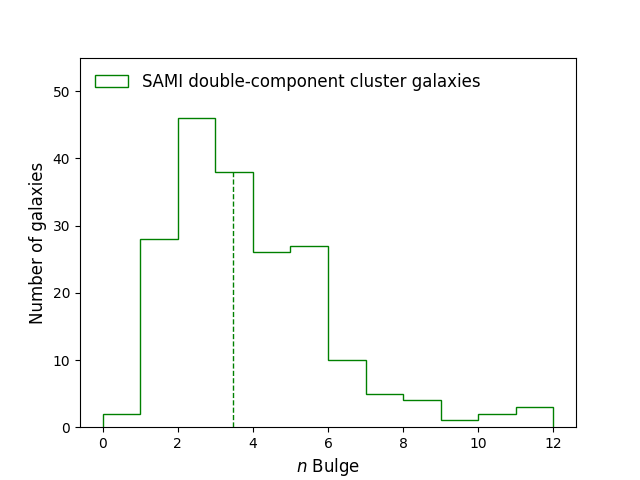}
\caption{Distributions in $M_{*}$, morphology, bulge-to-total flux ratio and bulge S\'ersic index for the 192 SAMI double-component galaxies. Stellar mass and morphology histograms are also plotted for the cluster galaxy sample of Barsanti et al. in preparation. The dashed lines represent the median values. The medians at morphology=1=S0, B/T=0.52 and $n_{bulge}=3.48$ show the reliability of the model selection.}
\label{Histograms192}
\end{figure*}

Figure~\ref{Histograms192} shows the distributions in stellar mass, morphology, B/T and bulge S\'ersic index for the 192 double-component galaxies. The histograms for stellar mass and morphology are compared to those for the 1730 cluster galaxies analyzed in Barsanti el al. in preparation. The only systematic bias is due to our choice to select massive galaxies with $M_{*}\geq10^{10}\,M_{\odot}$. According to the morphological classification of \citet{Cortese2016}, the galaxy morphology is represented by: elliptical=0, elliptical/S0=0.5, S0=1, S0/early-spiral=1.5, early-spiral=2, early/late-spiral=2.5 and late-spiral=3. The morphological distribution shows a peak for S0 galaxies, which is what expected for double-component galaxies and for dense environments \citep{Silchenko2012}. The median values B/T=0.52 and $n_{bulge}$=3.48 highlight the reliability of the model selection, since we expect B/T not to assume extreme high or low values as implemented in the filtering process and $n_{bulge}$=4 for a typical galaxy with two components.

\section{Methods for separating bulge and disk stellar population properties}
\label{Stellar population properties}
The aim of this paper is to analyze the individual stellar population properties of the bulge and the disk. This is a challenging problem, with projection effects, beam smearing, and imperfect modelling all impeding the separation of bulge and disk light. There have been recent attempts to overcome these problems using Integral Field spectroscopy, including those that decompose IFU spectra wavelength-by-wavelength \citep{Johnston2017,Mendez2019}, and those that combine spectra from regions dominated by bulge and disk light \citep{FraserMcKelvie2018}. Here, we develop a new method based on the combination of stellar mass maps and IFU spectroscopy. We also apply the two previously used methods of \citet{Mendez2019} and \citet{FraserMcKelvie2018} to compare our results.

\subsection{Method based on $M_{*}$ weights}
\label{A new method}
We implement a new mass-weighted method to separate the bulge and the disk stellar population properties. At each observed position in a SAMI cube, we assume that the observed spectrum is a result of the combination of at most two simple stellar populations coming from either or both a bulge and disk component. We assume that the mass-weighted contributions of the two populations to each observed galaxy spectrum add approximately linearly. Under these assumptions, we are able to disentangle the relative contributions from the bulge and disk stellar populations in a spectrum if we have knowledge of the fractional contribution of mass each population makes to that spectrum. The resolved spectroscopy offers the advantage that at each position the fractional contribution of the bulge and the disk varies in a known way that can be determined from our 2D bulge-disk decompositions. In this context, the combination of the 2D bulge-disk decomposition and the 2D IFU spectroscopic information allow us to recover the bulge and disk stellar population properties. 

For each galaxy we estimate the mass-weighted age and metallicity using each observed spectrum in the galaxy SAMI datacube. Taking advantage of the 2D bulge-disk decomposition, we estimate for each position the separate contributions to the galaxy stellar mass from the bulge and the disk. To measure the bulge-disk ages and metallicities, we consider a linear model: 
\begin{equation}
\small
Age[x,y]=f^{bulge}_{M_{*}}[x,y] \times Age_{bulge} +f^{disk}_{M_{*}}[x,y] \times Age_{disk}
\end{equation}
\begin{equation}
\small
[M/H][x,y]=f^{bulge}_{M_{*}}[x,y] \times [M/H]_{bulge} + f^{disk}_{M_{*}}[x,y] \times [M/H]_{disk}
\end{equation}
where $x$ and $y$ are the spatial positions of the measured spectrum, $Age[x,y]/[M/H][x,y]$ is the galaxy mass-weighted age/metallicity at this position, $f^{bulge}_{M_{*}}[x,y]$ and $f^{disk}_{M_{*}}[x,y]$ are the fractions of stellar mass due to the bulge and disk components at the same position respectively, and $Age_{bulge}/[M/H]_{bulge}$ and $Age_{disk}/[M/H]_{disk}$ are the fitted parameters representing the single-age/metallicity values of the galaxy components. We aim to find the combination of bulge and disk age/metallicity that best reproduce the whole galaxy age/metallicity spatial maps. We fit for the single-age/metallicity values of both galaxy components using the least-squares method. We minimize the sums of the squared residuals in the parameter space: 
\begin{equation}
S(Age[x,y])=\sum(Age[x,y]_{e}-Age[x,y]_{m})^{2}
\end{equation}
\begin{equation}
S([M/H][x,y])\sum([M/H][x,y]_{e}-[M/H][x,y]_{m})^{2}
\end{equation}
where $Age[x,y]_{e}/[M/H][x,y]_{e}$ are the estimated galaxy age/metallicity from the fitting and $Age[x,y]_{m}/[M/H][x,y]_{m}$ are the measured values.

This model is based on several assumptions that may limit the outcomes. We assume that galaxies are characterized by only two components: a central bulge and a surrounding disk. We consider single values for the mass-weighted age and metallicity of the bulge and the disk, without considering radial gradients. Finally, the 2D projected mass-weighted quantities are separated into linear combinations, as described by Equations (1) and (2). Despite these limitations, this method offers insights to disentangle the bulge-disk mass-weighted stellar population properties without being time consuming. It is also the starting point to develop more complicated models, e.g. adding radial gradients. A similar method has been explored by \citet{Hudson2010} (see their Appendix), where they combined 2D bulge-disk decomposition with single-fibre spectra. They measure the bulge stellar population properties using spectral index measurements, making some assumptions on the disk properties. Since our data are spatially-resolved and cover a larger fraction of the galaxy light, we can directly fit for the disk stellar properties.

The following Sections describe the building blocks of the method: the spatial binning, the kinematic corrections applied to the SAMI spectra, determining $f^{bulge}_{M*, bin}$ and $f^{disk}_{M*, bin}$, determining $Age_{bin}$ and $[M/H]_{bin}$ and finally estimating $Age_{bulge/disk}$ and $[M/H]_{bulge/disk}$.

\subsubsection{Spatial binning of the SAMI spectra}
\label{binning}
Determining $Age[x,y]$ and $[M/H][x,y]$ for our selected galaxies represents a crucial step toward our goal of measuring separately the stellar population parameters for the bulge and disks. They are obtained from full-spectrum fitting using the Penalized Pixel-Fitting software (pPXF; \citealt{Cappellari2004}, \citealt{Cappellari2017}).

The SAMI galaxy spectra are spatially binned to improve the signal-to-noise ratio (S/N) for the spectral fitting analysis and make more reliable measurements. The S/N is measured taking the median value of the flux divided by the noise in the rest-frame wavelength range 4600-4800~\AA\hspace{0.5mm}. The noise is defined as the square root of the variance at each wavelength from the SAMI datacube, including contributions from the covariance. This range is selected because it is clear of skylines and is fully contained within the SAMI blue arm \citep{Scott2018}. 

\begin{figure*}
\centering
\includegraphics[width=19cm]{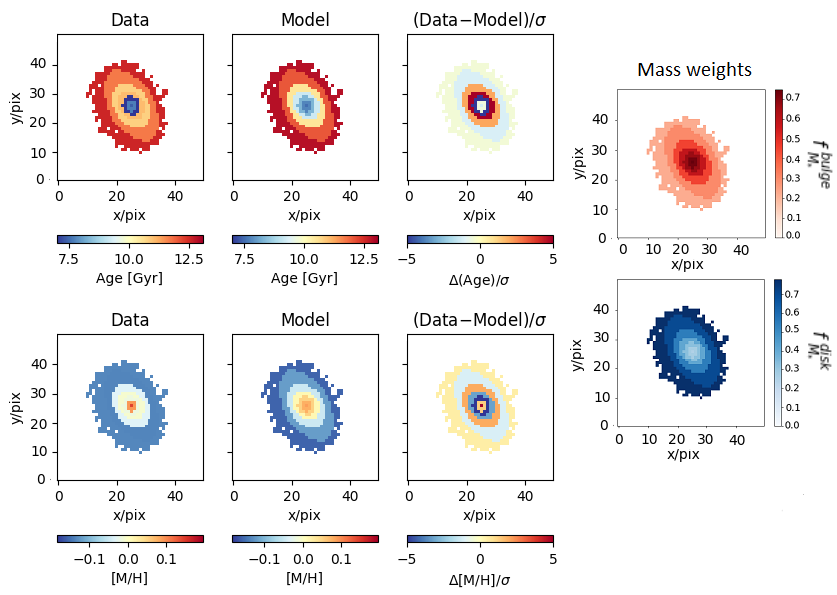}
\caption{Mass-weighted age (\textit{left upper panels}) and metallicity (\textit{left lower panels}) maps for the 9091700038 galaxy, representing the data (\textit{left}), model (\textit{middle}) and residuals (\textit{right}). For one spaxel of each spatial bin $Age_{bin}$ and $[M/H]_{bin}$ are linearly fitted for $Age_{bulge/disk}$ and $[M/H]_{bulge/disk}$ weighted for $f^{bulge}_{M*, bin}$ and $f^{disk}_{M*, bin}$ (\textit{right panels}). On average the residuals agree within 2$\sigma$, but for some bins the agreement is only at 5$\sigma$. This is consistent with radial gradients within the galaxy that cannot be fitted from the liner model represented by Equations (1) and (2).}
\label{DataModelResidual}
\end{figure*}

The spatial binning is based on the 2D bulge-disk decomposition results obtained in Barsanti et al. in preparation. The magnitudes of the double-component fits are corrected for the Milky Way extinction. For each galaxy the 2D model of the total flux in the $g$-band is rebinned onto 50 $\times$ 50 pixels matching the SAMI grid in order to generate a SAMI-like image. The model image is convolved with the SAMI PSF, which is modelled as a Moffat profile \citep{Allen2015}.  We choose the $g$-band since it is contained within the wavelength range covered by the SAMI blue arm spectra. The flux distribution of this model image is used to generate annular bins for stacking the SAMI spectra. We follow the isophotes of the model and adaptively grow annular bins until the S/N=20, which is the minimum S/N required to reliably measure age and metallicity (see Appendix~\ref{sec:Required Signal-to-Noise}). This methodology for spatial binning allows for better separation of bulge and disk spectra in the stacking, with respect to using a single ellipticity and position angle to define annular bins. Figure~\ref{DataModelResidual} shows an example of the binning for the galaxy 9091700038. In this case, a fixed elliptical annulus would not follow the relatively round central bulge component. 

The outer regions of the galaxies have low S/N and it is possible that the outer bin never meets the requirement of S/N$\geq$20. If this is the case, we still consider the outer bin if its S/N$>$10 otherwise we exclude it. For the stacking of the SAMI spectra we use the binning code from the SAMI software package, which sums the fluxes of the spaxels for each bin and weights them accounting for covariances \citep{Sharp2015,Allen2015}. 

\subsubsection{Kinematic corrections}
\label{sec:Corrections for velocity and velocity dispersion}
Large velocity and velocity dispersion fields can produce complicated line profiles after stacking. Thus, for each spaxel the spectrum is corrected for velocity and velocity dispersion prior to stacking. Correcting each spectrum for the velocity field has the advantage of aligning emission and absorption features prior to stacking. This increases the S/N post-stacking compared with non-corrected spectra. 
Finally, the improved S/N of absorption lines helps with determining stellar population properties.

We use the 2D SAMI velocity $V_{pPXF}$ and velocity dispersion $\sigma_{pPXF}$ maps measured by \citet{vandeSande2017}. Following \citet{Johnston2014}, the velocity dispersion is corrected bringing each spectrum to the same maximum value $\sigma_{max}$ measured within the galaxy. Each spectrum is convolved with the appropriate Gaussian having a corrected $\sigma_{cor}$ due to:
\begin{equation}
\sigma_{cor}=\frac{\Delta\sigma\lambda}{\delta_{\lambda}}\times\frac{(1+z)}{c} 
\end{equation}
where $\Delta\sigma=\sqrt{\sigma_{max}^2-\sigma_{pPXF}^{2}}$, $\lambda$ is the SAMI blue or red wavelength range, $\delta_{\lambda}$ is the spectral pixel scale of 1.03~\AA\hspace{1mm} or 0.56~\AA\hspace{1mm} for the SAMI blue or red arm, $z$ is the galaxy redshift and $c$ is the speed of light. The velocity correction is applied by considering the spectrum of each spaxel at the corresponding wavelengths $\lambda_{cor}=\lambda/(1+V_{pPXF}/c)$. Then, we interpolate to a common SAMI wavelength range for each spectrum. To be consistent we also apply these kinematic corrections to the noise. In this case the corrected sigma for the Gaussian used in the convolution is $\sigma_{cor}/\sqrt{2}$. The final corrected noise is then obtained dividing the original noise by 2$\sqrt{\pi}\sigma_{cor}$. We do not account for the covariance introduced by the smoothing. 

For each galaxy the corrected spectra and associated noise in each spaxel are used to build the final corrected blue and red SAMI datacubes. Finally, the limits and the accuracy of these corrections are tested in Appendix~\ref{Test for kinematics corrections}. If the kinematic corrections are applied correctly, then bulges and disks are expected to have matched zero velocity and maximum velocity dispersion measured within the galaxy. We perform the fitting for the kinematic components of the 1D spectra of the bulge and the disk obtained with the flux weights method in the following Section~\ref{A method based on light weights}. We find that velocities and velocity dispersions for bulges and disks follow one-to-one relationships, but there are some offsets due to the low signal-to-noise ratio of the 1D disk spectra.

\subsubsection{Determining $f^{bulge}_{M*, bin}$ and $f^{disk}_{M*, bin}$}
\label{sec:Bulge-disk stellar mass maps}
From the 2D photometric bulge-disk decomposition results described in Barsanti et al. in preparation, we generate projected stellar mass maps for the bulge and disk components. These stellar mass maps are required to determine the $f^{bulge}_{M*, bin}$ and $f^{disk}_{M*, bin}$ fractions in Equations (1) and (2). We use {\sc ProFit} to reconstruct separate 2D bulge and disk flux maps in the $g$- and $i$-bands within the SAMI-like image. The following formula for the SAMI galaxies is used to find the projected bulge-disk stellar mass maps from the bulge-disk $g$- and $i$-band flux maps \citep{Bryant2015,Owers2017}:   
\begin{equation}
\begin{split}
\log_{10}(M_{*}/M_{\odot})=-0.4i+2\log_{10}(D_{L}/10)-\log_{10}(1+z) \\ +(1.2117- 0.5893z)+(0.7106-0.1467z)\times(g-i)
\end{split}
\end{equation}
where $z$ is the galaxy redshift and $D_{L}$ is the luminosity distance measured in parsec using the cluster redshift.

\begin{figure*}
\includegraphics[scale=0.55]{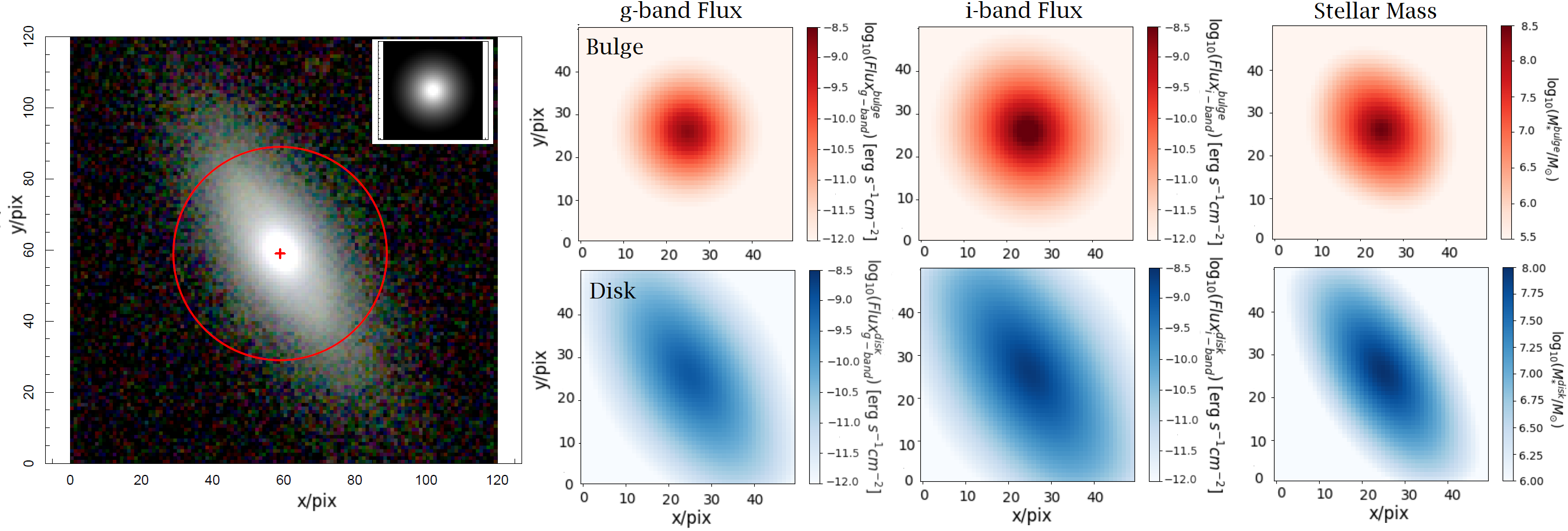}
\caption{The \textit{left panel} shows the VST/ATLAS $gri$ color image for the 9091700038 galaxy. The red cross marks the galaxy centre, the red circle
    shows the 15$^{\prime \prime}$ diameter of a SAMI hexabundle and the top right image is the SAMI PSF. The \textit{right panels} show the $g$-band flux, $i$-band flux and stellar mass maps for the bulge (\textit{top}) and the disk (\textit{bottom}). The projected maps are built within 50 $\times$ 50 pixels to match the SAMI grid. }
\label{griSDSSandSAMI}
\end{figure*}

The \textit{left panel} of Figure~\ref{griSDSSandSAMI} shows the VST/ATLAS $gri$ color image for the 9091700038 galaxy, with the red circle showing the diameter of a SAMI hexabundle. The \textit{right panels} show for the bulge and the disk the $g$-band and $i$-band flux maps from the 2D bulge-disk decomposition. These flux maps are used in Equation (6) to build the projected stellar mass maps (\textit{right-most panels}) for the two components.

The stellar mass maps are spatially binned as outlined in Section~\ref{binning} in order to match the binning performed on the SAMI cubes. Finally, they are used to estimate the bulge-disk contributions to the galaxy stellar mass. For each bin we estimate $f^{bulge}_{M_{*}, bin}$ as the sum of the bulge stellar masses of the spaxels divided by the summed total bulge+disk stellar masses for the same spaxels:
\begin{equation}
f^{bulge}_{M_{*}, bin}=\frac{\sum_{i} M^{bulge}_{*,i}}{\sum_{i}(M^{bulge}_{*,i}+M^{disk}_{*,i})}
\end{equation}
while the disk stellar mass fraction is $f^{disk}_{M_{*}, bin}=1-f^{bulge}_{M_{*}, bin}$. The most \textit{right panels} of Figure~\ref{DataModelResidual} show the projected bulge and disk stellar mass fraction maps for the 9091700038 galaxy. 

\subsubsection{Determining $Age_{bin}$ and $[M/H]_{bin}$}
\label{sec:Galaxy age and metallicity maps}
The SAMI spectra corrected for velocity and velocity dispersion are spatially binned as described in Section~\ref{binning}. These spectra are used to determine the mass-weighted age and metallicity associated with each bin, $Age_{bin}$ and $[M/H]_{bin}$ in Equations (1) and (2). We fit the corrected SAMI spectra with pPXF, making use of the MILES single stellar population (SSP) templates \citep{Vazdekis2010}. We consider the isochrones developed by the Padova group \citep{Girardi2000}, and the unimodal initial mass function having logarithmic slope 1.3. To fit for galaxy age and metallicity, we follow the first three steps of \citet{Kacharov2018}. Since we are not interested in star-formation history, we do not use regularization which finds the smoothest solution in agreement with the data \citep{McDermid2015,Cappellari2017}. Regularization might introduce some bias, smoothing the differences between the bulge-disk properties. \citet{Woo2019} found that the accuracy of the mass-weighted age and metallicity for the regularized solution is similar to the unregularized one. Thus, we decide to estimate unregularized solutions, which are also much faster to compute \citep{Woo2019}. 

The procedure we apply to build the mass-weighted age and metallicity maps for each galaxy is the following:
\begin{enumerate}
\item We combine the blue and the red spectra of each spatial bin following the approach of \citet{vandeSande2017}. The associated blue and red noises are combined in the same way.
\item We fit for the stellar kinematic components. We de-redshift the spectrum to a rest-frame wavelength using the galaxy redshift estimated by \citet{vandeSande2017}. We fit for the kinematic components even if the spectrum is corrected for velocity and velocity dispersion because of some limits in the applied kinematic corrections (see Figure~\ref{velsigma} in Appendix~\ref{Test for kinematics corrections}). We assume a Gauss-Hermite line-of-sight-velocity-distribution, fitting for galaxy velocity, velocity dispersion, skewness and kurtosis. We use a 12th order additive polynomial to minimise template mismatch and to correct for sky subtraction imperfections and scattered light. For the inputs of the kinematic components we use zero for the velocity and the maximum value for the velocity dispersion. 
\item We derive the extinction correction curve. We keep the stellar kinematic parameters fixed to those obtained at step (2). We re-fit using pPXF and we adopt the reddening law of \citet{Calzetti2000}. For the remaining steps, the observed galaxy spectrum is corrected for the best-fitting extinction curve. 
\item We find the best-fit linear weighted combination of SSP templates for the spectrum. We use a 12th order multiplicative polynomial to correct for any mismatch between the templates and spectrum due to, e.g., data reduction anomalies. In the fit we also include gas emission lines which can be well constrained since we include the red spectrum. A more reliable estimate of the noise is performed considering the residuals between the spectrum and the best-fit model at different wavelength ranges. We estimate the dispersion of the residuals around the mean wavelength value of each range and then we interpolate the dispersions through the entire wavelength interval as the new noise. 
\item We perform a second fit following the previous steps (2)-(4) using the new estimated noise. We determine $Age_{bin}$ and $[M/H]_{bin}$ using the weights determined by pPXF and Equations (1) and (2) of \citet{McDermid2015}. 
\item Uncertainties on $Age_{bin}$ and $[M/H]_{bin}$ are determined using a Monte Carlo approach. We add noise to the best-fit model from step (5), using a random Gaussian with sigma determined from the rescaled noise. We run 1000 simulations. We correct the SSP templates for the kinematic components and for the multiplicative polynomial determined in step (2) and (5), respectively. Thus, in each simulation we fit only for the weights of each SSP template that is linearly combined to form the best-fit model.  Finally, 1001 $Age_{bin}$ and $[M/H]_{bin}$ are obtained for the same bin and the standard deviations ($\sigma$) of these distributions are estimated for each bin.
\end{enumerate}

\begin{figure*}
\includegraphics[width=18cm,height = 21cm]{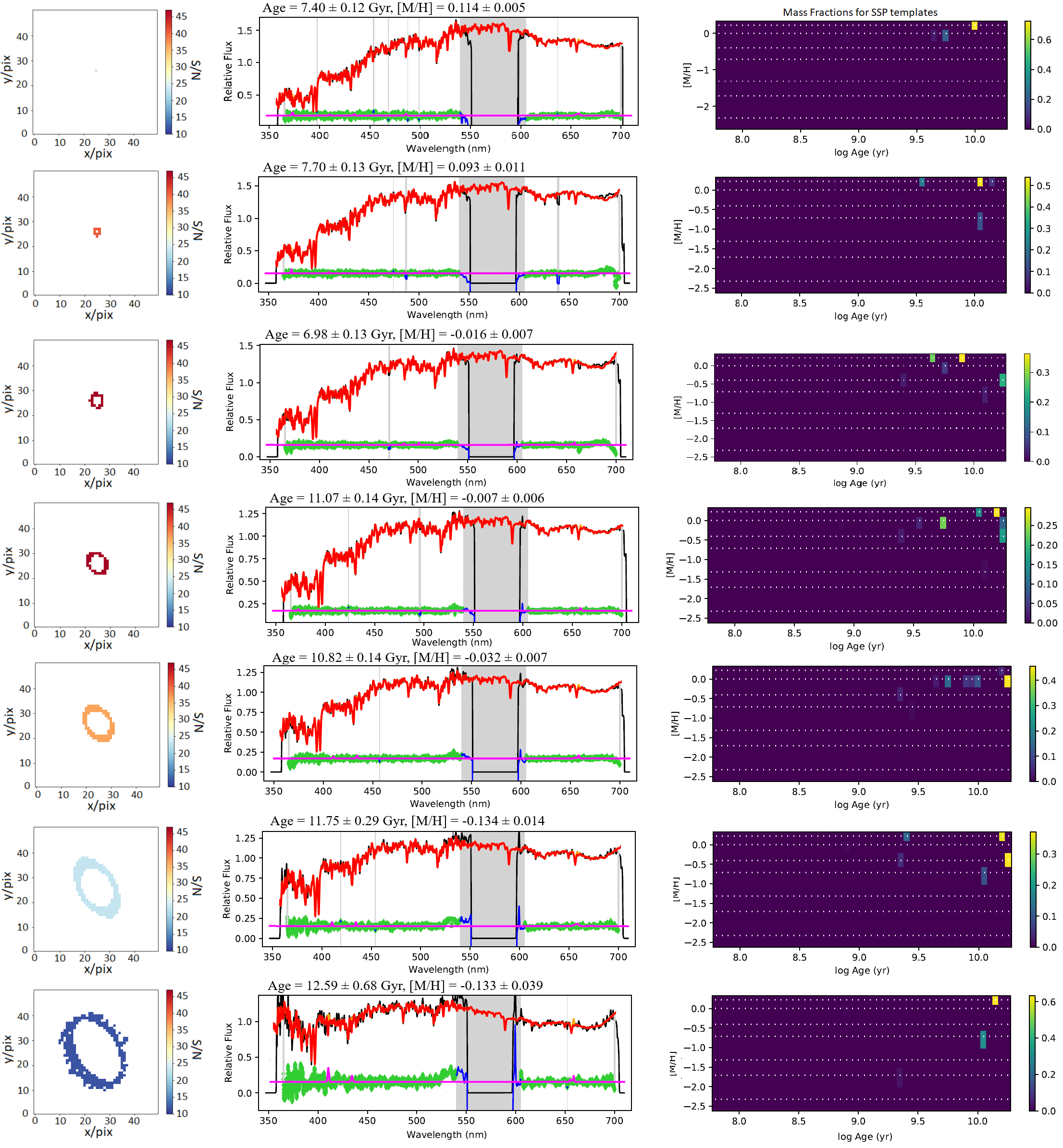}
\caption{Output plots from pPXF representing the spectra of the seven spatial bins for the 9091700038 galaxy. From \textit{left top} to \textit{bottom} the spectra are from the most central to the outermost bin. The \textit{middle panels} show the input spectrum (black), the best-fit model (red), the best-fitted gas emission lines (orange), the noise (green) and the masked pixels (grey). The \textit{right panels} displays the age-metallicity grid with the corresponding weights of the best linear combination of SSP templates which result in the best-fit model (see Equations (1) and (2) of \citealt{McDermid2015}).}
\label{9091700038_ppxf}
\end{figure*}

Figure~\ref{9091700038_ppxf} shows an example of the output plots from pPXF representing the spectra of each spatial bin for galaxy 9091700038. We display the spatial bins (\textit{left}), the best-fit spectra (\textit{middle}) and the obtained estimates in the age-metallicity grid (\textit{right panels}). The spectrum associated to the most external bin is characterized by the highest systematic noise since it is obtained from the combination of low S/N spaxels. This can be seen in the unphysical upturn at the beginning of the blue spectrum. However, the final results do not change if we exclude the initial blue upturns in the pPXF fitting process.

\subsubsection{Putting everything together: determining $Age_{bulge/disk}$ and $[M/H]_{bulge/disk}$}
\label{Age and metallicity of bulges and disks}
With the estimates for $f^{bulge}_{M*, bin}$, $f^{disk}_{M*, bin}$, $Age_{bin}$ and $[M/H]_{bin}$ in hand, we can now use Equations (1) and (2) to determine which combination of bulge and disk stellar population parameters best reproduce the observations, i.e. the whole galaxy age/metallicity maps. We attempt to reproduce the observed spatial variation in galaxy age and metallicity using the bulge and disk fraction mass maps by assigning a single spatially constant age and metallicity to each component. For each bin, we fit for the best $(Age,[M/H])_{bulge}$ and $(Age,[M/H])_{disk}$ that, when combined, most closely reproduced the observed galaxy $Age_{bin}$ and $[M/H]_{bin}$, determined in the previous Section~\ref{sec:Galaxy age and metallicity maps}.

We perform the fitting using the \textit{R optim} function with the ``L-BFGS-B'' algorithm, minimizing the squared sum of the difference between the estimated and the known galaxy age/metallicity maps. Ages and metallicities of both components are constrained to the lower and upper limits of the SSP templates used in Section~\ref{sec:Galaxy age and metallicity maps}: [0.0631, 17.7828] Gyr for ages and [-2.32, 0.22] for metallicities. The upper age limit leads to the possibility of SSP ages that are higher than the age of the Universe $\sim 13.8$ Gyr. We explore this caveat in Section~\ref{Caveats}. 

For the 9091700038 galaxy Figure~\ref{DataModelResidual} shows the fitting results: the data representing the whole galaxy age/metallicity maps, the galaxy model obtained from the fitting of Equations (1) and (2), and the residuals measured with respect to the $\sigma$ map estimated from the Monte Carlo simulations at step (6) of Section~\ref{sec:Galaxy age and metallicity maps}. Without any complicated modelling we are able to reproduce in this galaxy the gradients from outer regions being older and more metal-poor to inner regions being younger and more metal-rich. On average the residuals agree within 2$\sigma$, however for few spatial bins the agreement is only at 5$\sigma$. Further discussion about this agreement is reported in Section~\ref{Caveats}.

\begin{figure*}
\centering
\includegraphics[width=0.9\columnwidth]{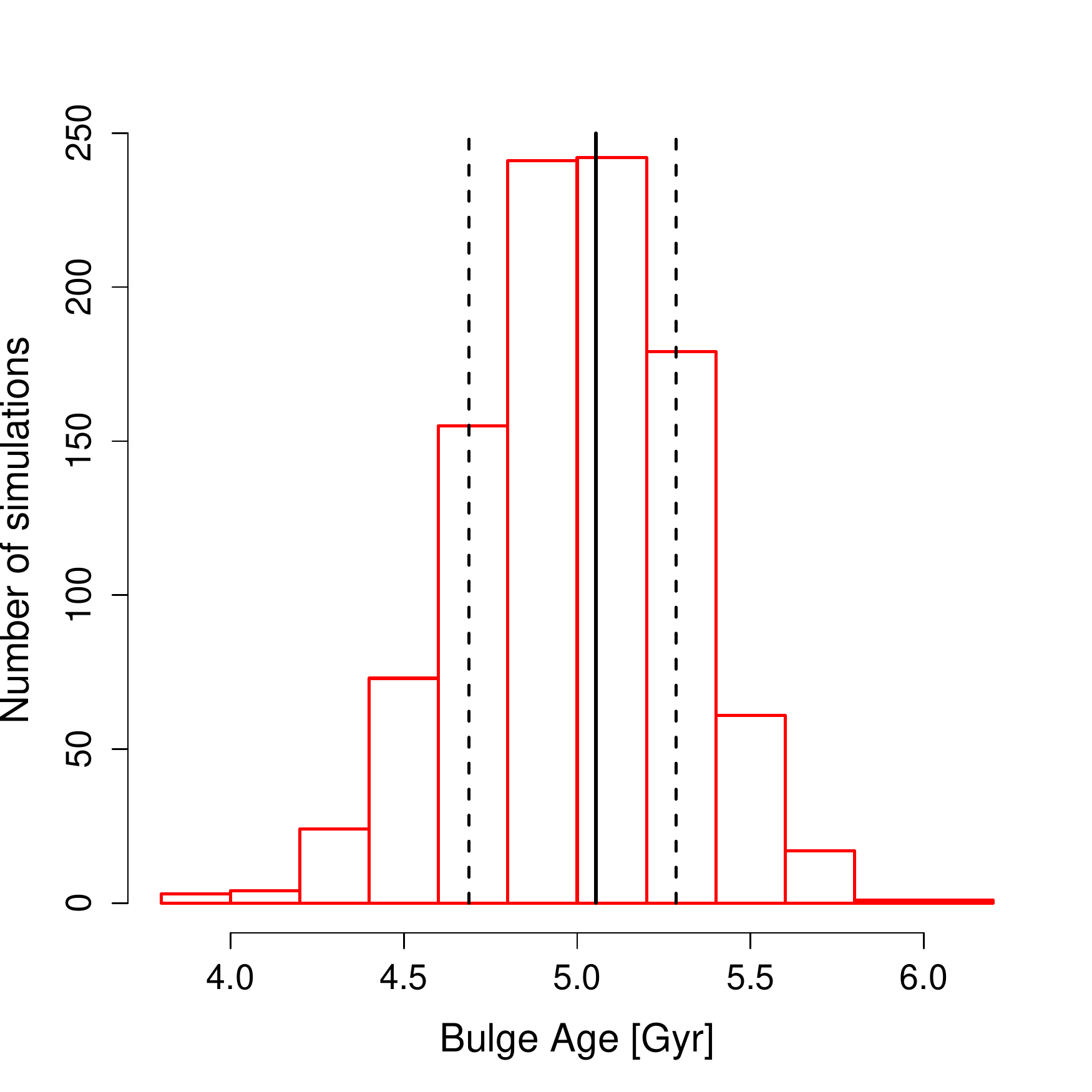}
\includegraphics[width=0.9\columnwidth]{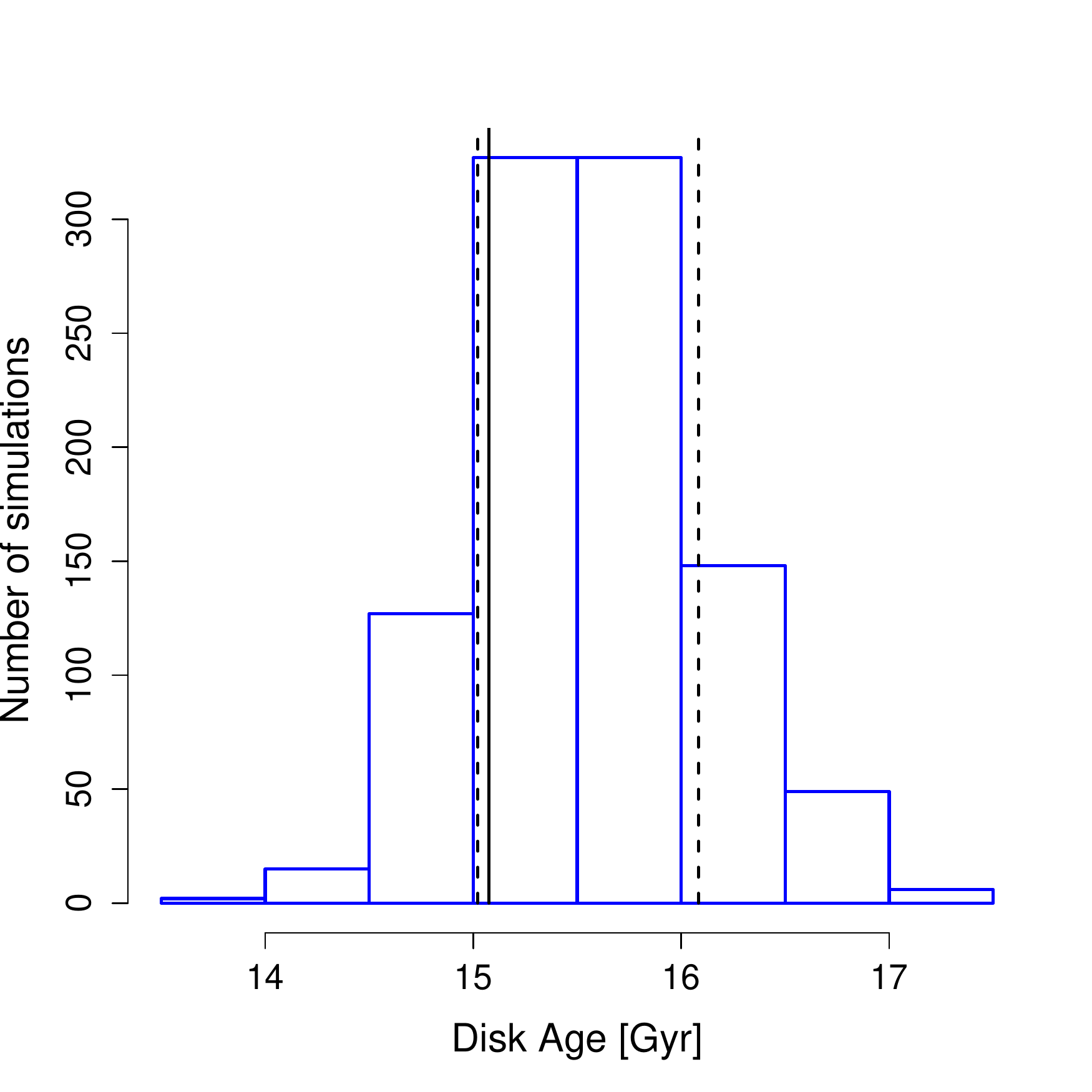}
\includegraphics[width=0.9\columnwidth]{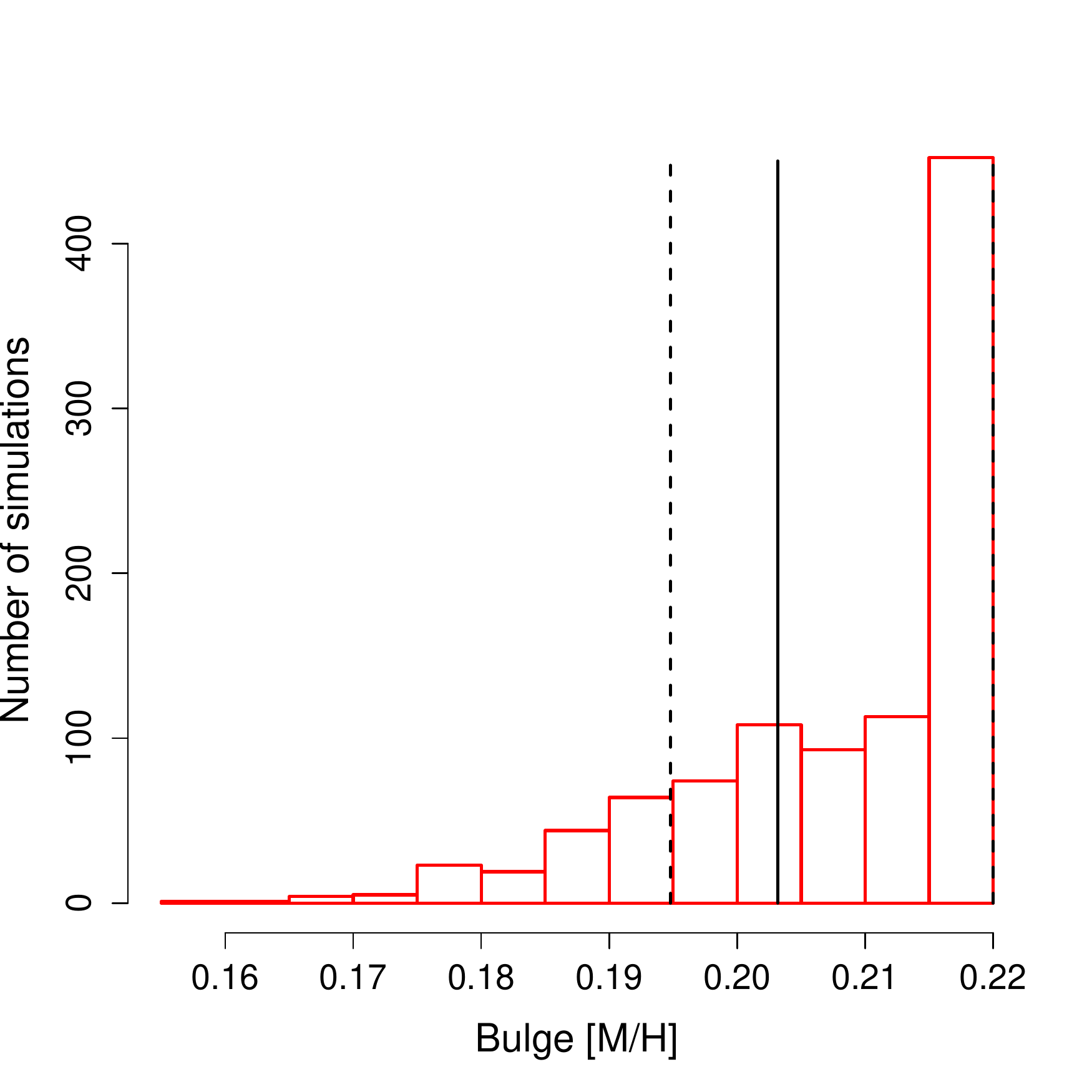}
\includegraphics[width=0.9\columnwidth]{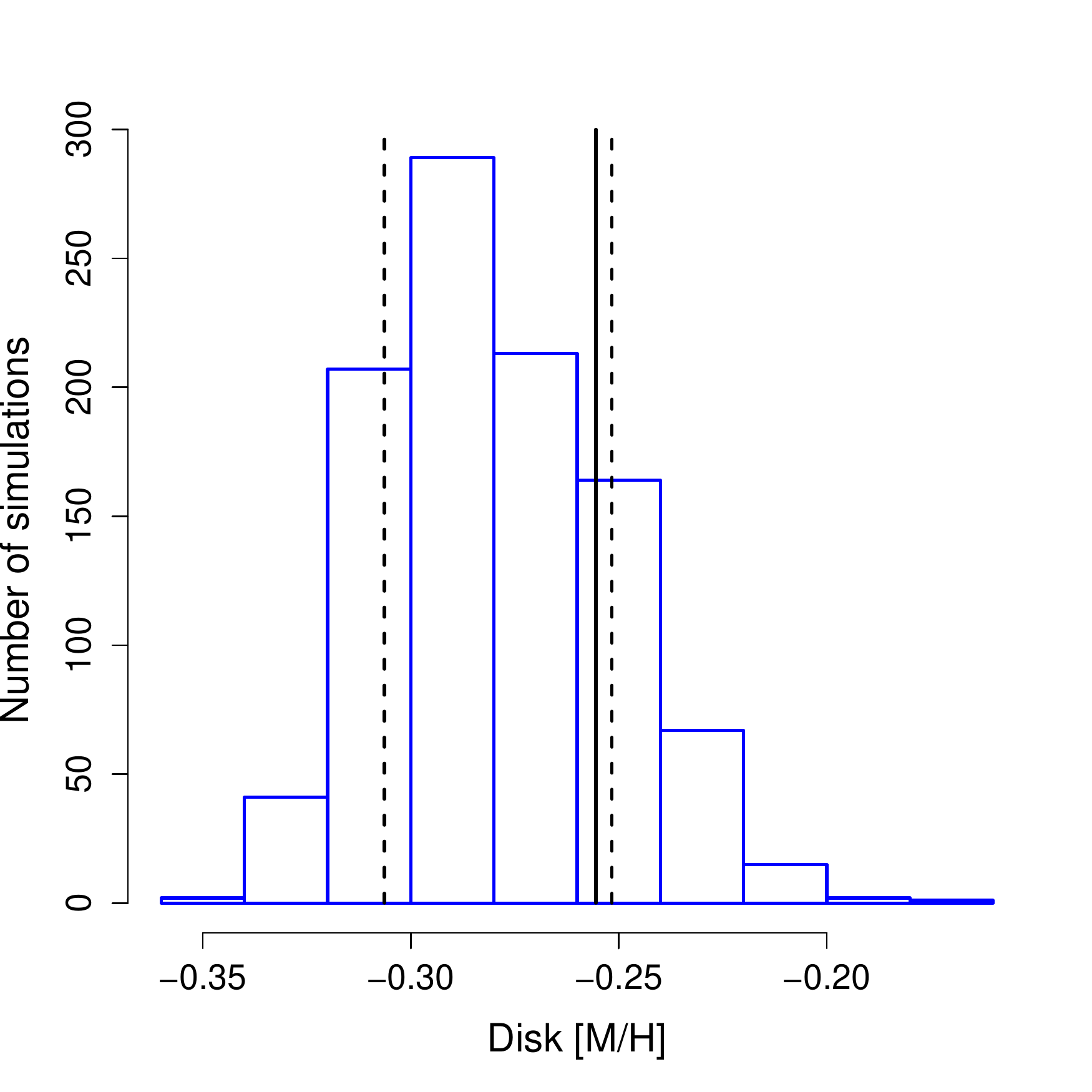}
\caption{Age (\textit{top panels}) and metallicity (\textit{bottom panels}) distributions of bulge (red) and disk (blue) of the 9091700038 galaxy. The filled black line represents $Age_{bulge/disk}$ and $[M/H]_{bulge/disk}$, the dashed black lines are the associated uncertainties from the 16th and 84th percentiles.}
\label{2DAgeMetalMaps}
\end{figure*}

For the uncertainties on $Age_{bulge/disk}$ and $[M/H]_{bulge/disk}$, we repeat this procedure fitting the remaining 1000 galaxy age/metallicity maps obtained from the Monte Carlo simulations in Section~\ref{sec:Galaxy age and metallicity maps}. The 16th and 84th percentiles of the $Age_{bulge/disk}$ and $[M/H]_{bulge/disk}$ distributions are taken as the lower and upper uncertainties. Figure~\ref{2DAgeMetalMaps} shows an example of these distributions for the 9091700038 galaxy. The results $(Age,[M/H])_{bulge}$=(5.05 Gyr, 0.20) and $(Age,[M/H])_{disk}$=(15.06 Gyr, $-$0.26) from fitting the data are shown as the black filled lines, while the dashed black lines display the associated 16th and 84th percentiles.

\begin{figure*}
\centering
\includegraphics[width=19cm]{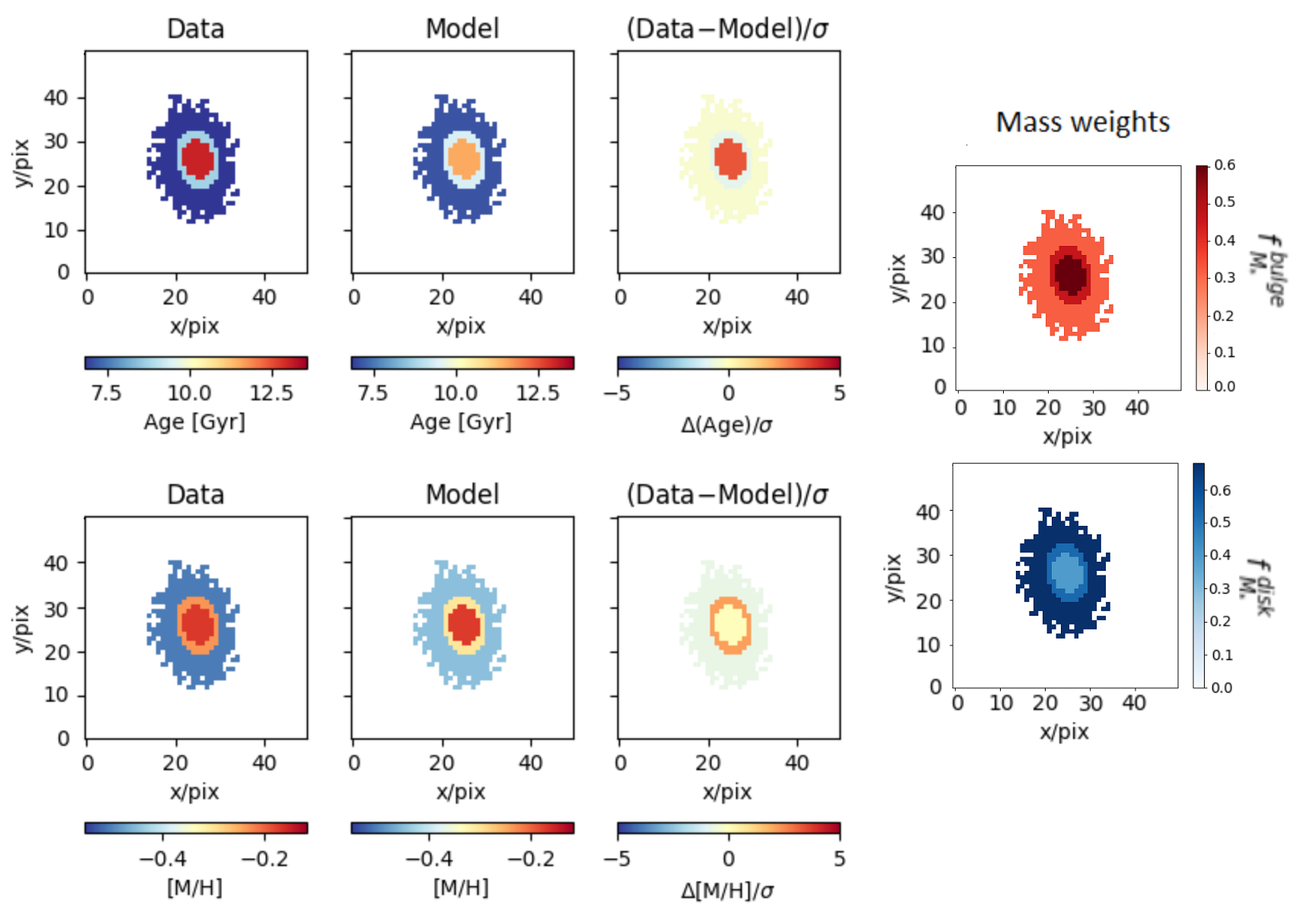}
\caption{Mass-weighted age (\textit{left upper panels}) and metallicity (\textit{left lower panels}) maps for the 9091700076 galaxy, representing the data (\textit{left}), model (\textit{middle}) and residuals (\textit{right}). For one spaxel of each spatial bin $Age_{bin}$ and $[M/H]_{bin}$ are linearly fitted for $Age_{bulge/disk}$ and $[M/H]_{bulge/disk}$ weighted for $f^{bulge}_{M*, bin}$ and $f^{disk}_{M*, bin}$ (\textit{right panels}). This galaxy has a bulge older and more metal-rich than the disk, oppositely to the 909170038 galaxy that shows a bulge younger than the disk in Figure~\ref{DataModelResidual}. The residuals agree within 3$\sigma$ for age and within 2$\sigma$ for metallicity.}
\label{DataModelResidual2}
\end{figure*}

In Figure~\ref{DataModelResidual2} we show another example of fitting results for the 9091700076 galaxy. As in Figure~\ref{DataModelResidual}, we show the data representing the whole galaxy age/metallicity maps, the galaxy model obtained from the fitting of Equations (1) and (2), and the residuals. Oppositely to the 9091700038 galaxy, the 9091700076 galaxy has an older and more metal-rich bulge when compared to its disk. Due to the low S/N per pixel, only three spatial bins are obtained with S/N=20. The residuals agree within 3$\sigma$ for age and within 2$\sigma$ for metallicity.

All 192 double-component galaxies have been decomposed with this method. However, 109/192 have at least one pegged solution to the lower or upper limits for $Age_{bulge/disk}$ and $[M/H]_{bulge/disk}$. The pegged solutions occur for the bulge metallicity in 48 galaxies and the disk age in 64 galaxies. Bulges can have $[M/H]_{bulge}>0.2$ \citep{Johnston2014}, thus these pegged metallicities at the upper limit are allowed. For the disks, the majority of the pegged ages also occur at the upper age limit. This is mainly for galaxies with low $f^{disk}_{M_{*},bin}$, suggesting that low $M_{*}$ weights affect the fits. The pegged solutions can occur because this model does not account for radial gradients in the age/metallicity pushing the fitting results to the limits. Moreover, low $M_{*}$ weights make the results harder to constrain. In Appendix~\ref{Simulations} we show that the reliability of the results depends on the bulge-to-total flux ratio B/T of the galaxy and it is affected by the $M_{*}$ weights. Low $f^{bulge/disk}_{M_{*}, bin}$ values imply that the results are more uncertain and might lead to pegged results at the lower or upper fit limits. The age/metallicity of the bulge are better constrained for galaxies with high B/T, while the results for the disk are better constrained for galaxies with low B/T. The exclusion of the pegged solutions does not change the results, thus we decide to consider the full sample of 192 galaxies.

\subsection{Method based on flux weights}
\label{A method based on light weights}
We perform a method similar to the methodology presented by \citet{Mendez2019,Mendez2019b}. \citet{Mendez2019} studied three early-type galaxies from the CALIFA survey \citep{Sanchez2012}, performing spectro-photometric multi-component decomposition of IFU datacubes. At each wavelength, the galaxy surface brightness distribution is fitted with a 2D bulge-disk photometric decomposition. This method provides both a 1D spectrum and a spatially-resolved spectroscopic datacube for each galaxy component. \citet{Johnston2017,Johnston2020} introduced a similar approach for 2D bulge-disk decomposition of IFU datacubes to extract the spectral properties of bulges and disks. 

\begin{figure*}
\includegraphics[width=\columnwidth]{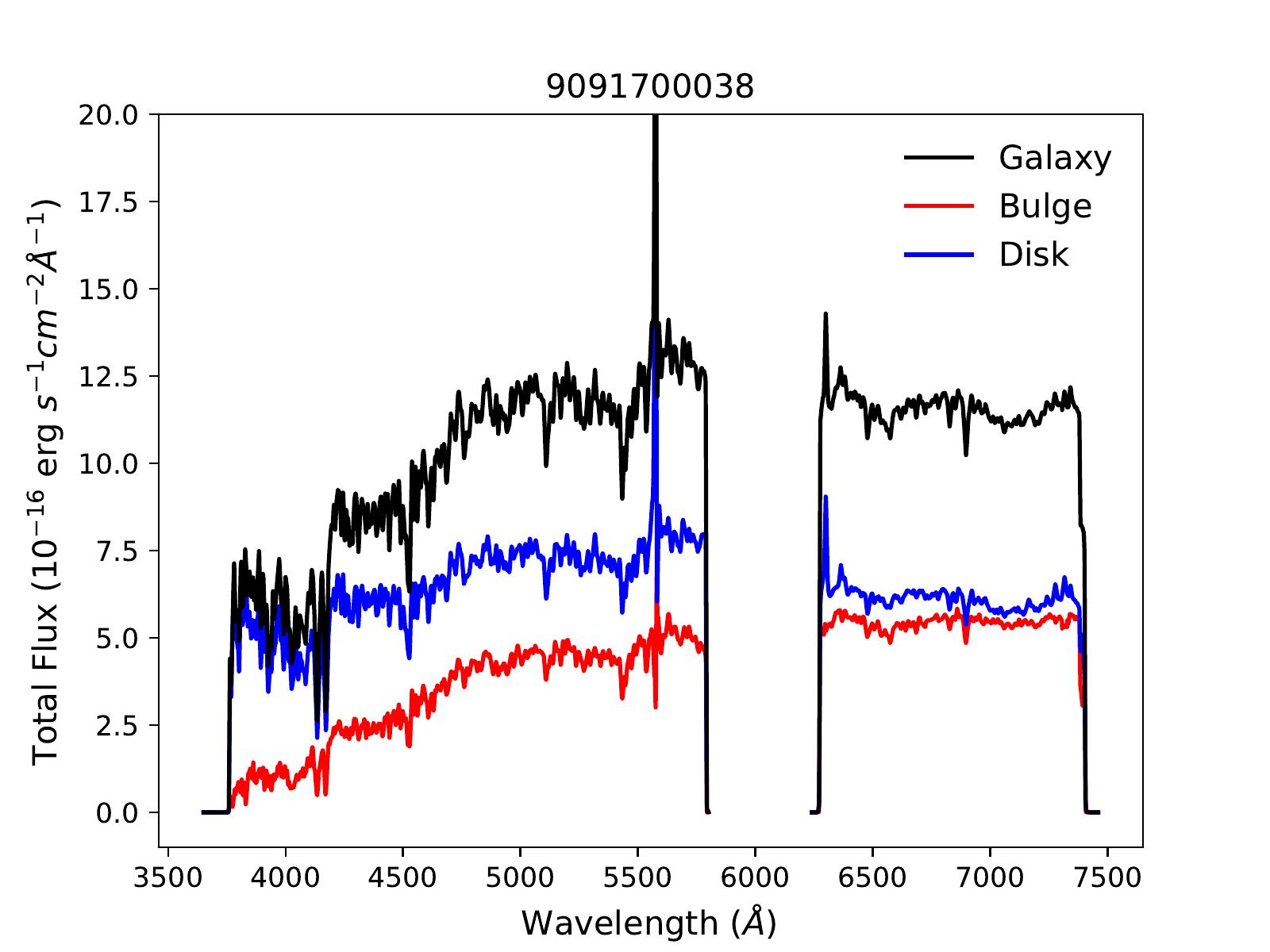}
\includegraphics[scale=0.525]{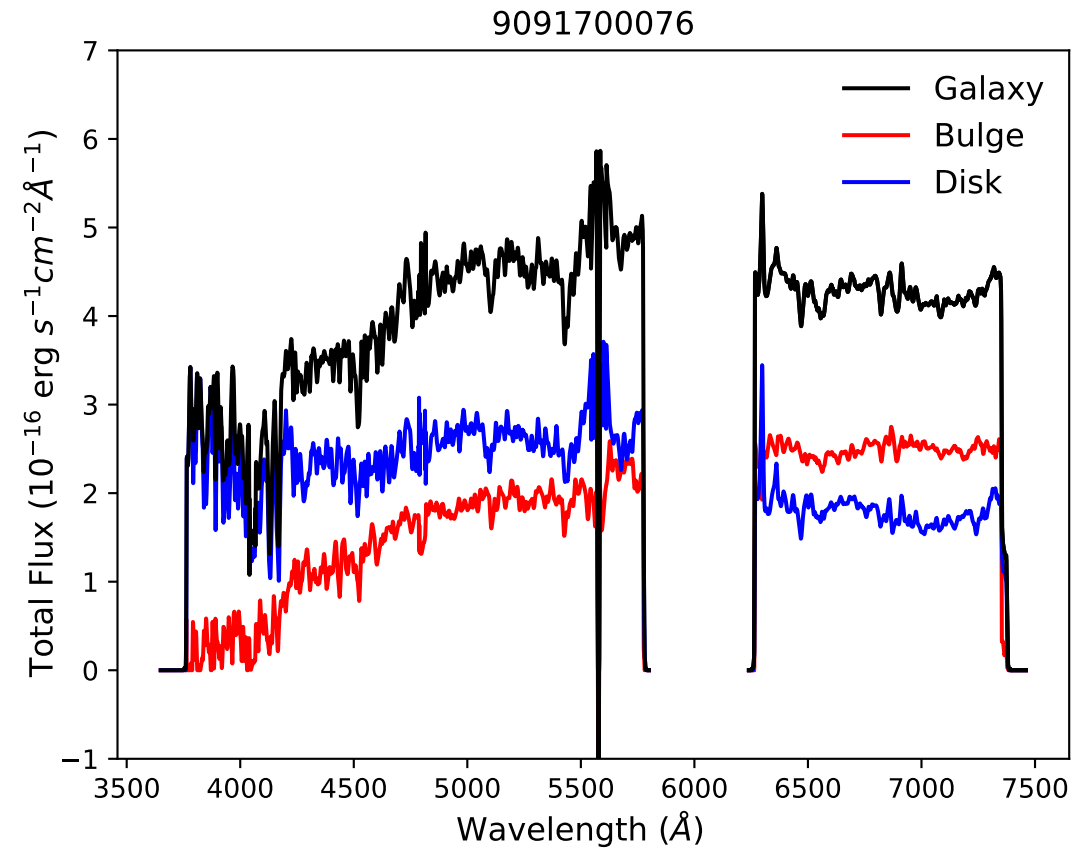}
\caption{1D spectra of the whole galaxy (black), the bulge (red) and the disk (blue) for the 9091700038 galaxy (\textit{left panel}) and the 9091700076 galaxy (\textit{right panel}), respectively.}
\label{1Dspectra}
\end{figure*}

We adapt the \citet{Mendez2019} method in the following way. 
\begin{enumerate}
\item We slice the blue/red SAMI galaxy datacubes, which have been corrected for velocity and velocity dispersion as described in Section~\ref{sec:Corrections for velocity and velocity dispersion}, into 2D narrow-band images. For the blue SAMI wavelength range 3650-6000~\AA\hspace{0.5mm}we build 7 narrow-band images of $\sim$300~\AA\hspace{0.5mm}each, while for the red SAMI wavelength range 6240-7450~\AA\hspace{0.5mm}we build 5 narrow-band images of $\sim$200~\AA\hspace{0.5mm}each. The images at each wavelength bin are made from the datacubes using the median flux per pixel. Using the same procedure we build the 2D noise map associated with each narrow-band image. We consider only pixels with S/N$>$1 to limit systematic effects. 
\item We use the 2D bulge-disk decomposition results in the $g$, $r$ and $i$-bands described in Barsanti et al. in preparation. The \textit{left panel} of Figure~\ref{griSDSSandSAMI} shows for the 9091700038 galaxy the VST/ATLAS $gri$ color image, where the red circle marks the size of a SAMI hexabundle. In Figure~\ref{griSDSSandSAMI} we also show the $g$-band and $i$-band flux projected maps for the two components from the 2D bulge-disk decomposition for this galaxy. For the $g$, $r$ and $i$-bands SDSS and VST/ATLAS show similar filter + telescope + atmospheric transmissions (see Figure 2 of \citealt{Shanks2015}). Thus, for both SDSS and VST/ATLAS data each $g$, $r$ and $i$-band image is assumed to be centred at 4770~\AA, 6231~\AA\hspace{0.5mm}and 7625~\AA, respectively. The galaxy photometric and shape parameters are found to be dependent on the wavelength \citep{Vika2013}. We use the results from the $g$, $r$, and $i$-band fits to linearly interpolate estimates for the bulge-to-total flux ratio, the effective radii of both components and the S\'ersic index of the bulge at the wavelengths of the narrow-band images. The S\'ersic index of the disk is fixed to 1 and the angle and axial ratio of both components are fixed to the $r-$band results. 
\item We use {\sc ProFit} to make a 2D double-component galaxy model of each narrow band image based on its assigned photometric and shape parameters from the interpolations at step (2). Each model is convolved with the single SAMI PSF image for that galaxy \citep{Allen2015}. Then, it is fitted using the 2D noise map as sigma map. We fix all the parameters to the interpolated values, except for the bulge-to-total flux ratio which is used to estimate initial values for the magnitudes of the bulge and the disk. 
\item  We build monochromatic images using the blue/red SAMI step in wavelength of 1.05/0.59~\AA\hspace{0.5mm}and obtaining a total of 4046 images. Each monochromatic image has assigned values of the photometric and shape parameters, obtained fitting linearly the results of the 2D bulge-disk decomposition for the narrow band images. We also construct the 2D noise map associated with each monochromatic image. We perform 2D bulge-disk decomposition on each monochromatic image using {\sc ProFit} and fitting only for the magnitude of both components, as in step (3).
\item Using the results of the 2D bulge-disk decomposition for each monochromatic image, we build the separate 2D models of the bulge and the disk convolved with the SAMI PSF at each wavelength. We use the 2D total, bulge and disk flux maps to estimate the 2D bulge-to-total and disk-to-total maps at each wavelength which act as flux weights for the two components. We multiply each monochromatic image with the corresponding 2D bulge-to-total and disk-to-total maps to obtain two separate 2D bulge and disk monochromatic images. Finally, stacking all the 2D bulge and disk monochromatic images we obtain two separate 3D blue/red datacubes for the two components. 
\item We build the separate 1D bulge/disk spectra for each galaxy. Using the 3D bulge/disk datacubes we sum the flux from all the spaxels to obtain the flux as a function of the wavelength. Figure~\ref{1Dspectra} shows the 1D spectra of the galaxy, bulge and disk for the 9091700038 and 9091700076 galaxies. 
\item We use pPXF to fit the 1D bulge/disk spectra separately in order to estimate the mass-weighted single-age/metallicity of the two components. We follow the same procedure described in Section~\ref{sec:Galaxy age and metallicity maps}, using the 1000 Monte Carlo simulations to estimate the 16th and 84th percentiles as uncertainties on the ages and metallicities. We approximate the bulge/disk noise performing a first fit of the bulge/disk spectrum using the galaxy noise. We consider the residuals binned in wavelength and we estimate the dispersion around the mean of each wavelength bin. Then, we interpolate the bulge/disk noise for the whole wavelength range.
\end{enumerate}

\begin{figure*}
\centering
\includegraphics[width=18cm]{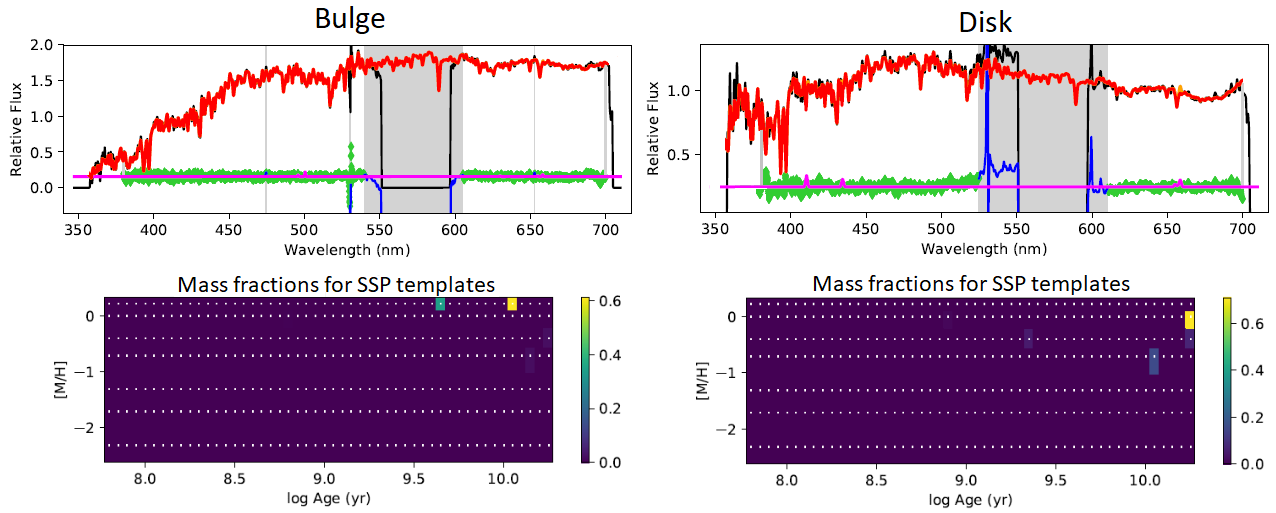}
\caption{1D bulge/disk spectra for the 9091700038 galaxy built with the method based on flux weights and fitted with pPXF.}
\label{1DspectraPPXFmendez}
\end{figure*}

181/192 double-component galaxies have been reliably decomposed with this method. The results are based on the collapsed information from the 1D SAMI aperture spectra. This method is based on fewer assumptions than the one based on $M_{*}$ weights, however it requires high signal-to-noise ratio in each wavelength slice and high physical spatial resolution which are strictly limited for the SAMI Galaxy Survey \citep{Sanchez2017,Mendez2019}. Since the SAMI spectra are not characterized by high signal-to-noise ratio in each wavelength slice, some systematic noise is generated and it can be seen in Figure~\ref{1Dspectra} as the upturn at the blue ends of the disk spectrum. We limit the systematic effects by excluding the initial blue upturns in the pPXF fits at step (7). Figure~\ref{1DspectraPPXFmendez} shows the 1D bulge/disk spectra fitted with pPXF for the 9091700038 galaxy. The obtained results are $(Age,[M/H])_{bulge}$=(8.71 Gyr, 0.20) and $(Age,[M/H])_{disk}$=(14.44 Gyr, $-$0.17).

\begin{figure*}
\centering
\includegraphics[width=0.7\columnwidth]{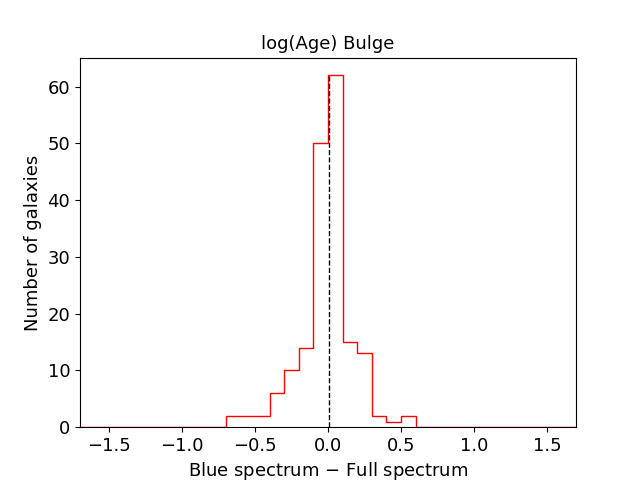}
\includegraphics[width=0.7\columnwidth]{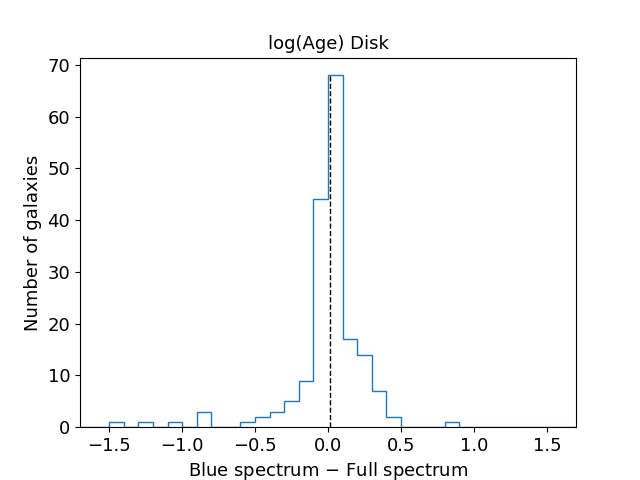}\\
\includegraphics[width=0.7\columnwidth]{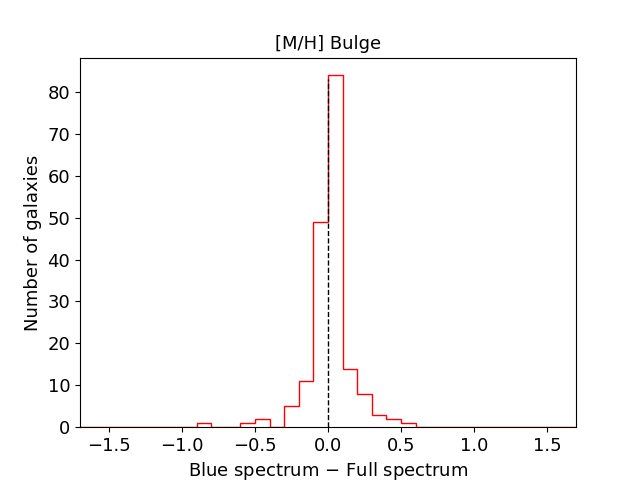}
\includegraphics[width=0.7\columnwidth]{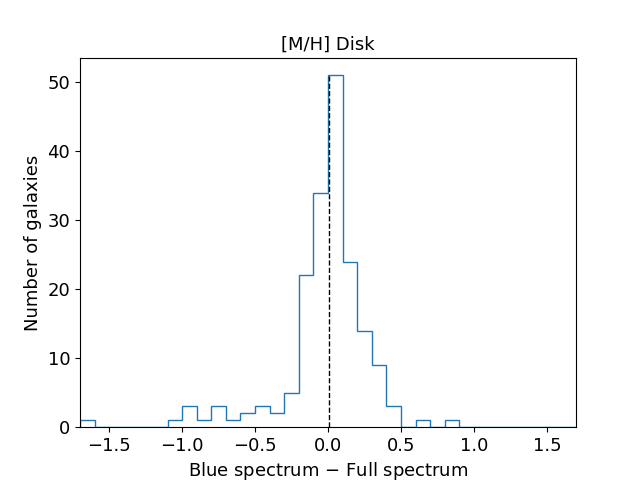}
\caption{Differences between the bulge/disk ages and metallicities obtained fitting the 1D blue spectra and the 1D full spectra with the flux weights method. The distributions are Gaussian with medians $\sim$0 dex and standard deviations consistent with random errors, meaning that the fitted wavelength range does not affect the results.}
\label{BluevsBlueRed}
\end{figure*}

\subsubsection{Effect of flux calibration}
The visual inspection of the 1D bulge and disk spectra in Figure~\ref{1Dspectra} for the 9091700038 galaxy suggests that the age of the disk is younger compared to the age of the bulge. The disk spectrum appears very blue with the red flux being lower than the blue flux. However, despite the shape of the spectra, we measure a disk older than the bulge in Figure~\ref{1DspectraPPXFmendez}. This might be due to the difference in normalisation of the blue and red portions of the spectrum introduced in the flux calibration during the SAMI data reduction process. 

We investigate that the results are not affected by the difference in normalisation of the blue and red spectral portions. To this end, we perform the pPXF fitting procedure for the flux weights method by using only the blue portion of the SAMI galaxy spectra and excluding the red portion. Then, we compare these results for the 181 galaxies with those previously obtained by fitting the full blue+red SAMI spectra. This comparison gave us an idea of the systematics due to the wavelength range fitted by pPXF.

Figure~\ref{BluevsBlueRed} shows the distributions of the differences between the results obtained by fitting only the blue spectra and by fitting the full spectra. The distributions are Gaussian with medians centred at 0 dex, showing that the fitted wavelength range does not affect the results. The few largest offsets are measured for galaxies for which the result from the blue spectrum fitting is close to the upper/lower fitting boundaries compared to the one from the full spectrum which is not. The average standard deviations on the age and metallicity are 0.08 dex for the bulges and 0.13 dex for the disks. Since the fittings of the blue and full spectra contain substantial overlap in the wavelength range, these standard deviations are not purely systematic. They contain contributions from both the random errors on the results obtained by the blue spectra and the full spectra. The random errors are estimated using the 1000 Monte Carlo simulations in the pPXF fitting process. The average random errors on the bulge and disk results from the full spectra are 0.03 dex and 0.06 dex, respectively. The average random errors from the blue spectra are 0.06 dex for bulges and 0.09 dex for disks. Larger errors are measured for the disks with respect to the bulges since the the disk results are harder to constrain. The random errors from the blue spectra are larger compared to those from the full spectra since a narrow wavelength range is fitted. The standard deviations of the difference distributions are consistent with the random errors. This implies that the systematic effect due to the different fitted wavelength ranges is negligible.

The blue spectrum contains the typical absorption lines used to estimate the stellar population properties. The inclusion of the red portion is used to obtain better fits of the stellar kinematic components, the gas emission lines and the extinction correction curve. In particular, the fitting for the extinction correction curve is less constrained without the red arm data \citep{Scott2017}. Overall, the results obtained with the flux weights method by fitting the full spectra are not affected by the chosen wavelength range and by the difference in normalisation of the blue and red portions of the spectrum introduced in the flux calibration.

\subsection{Method based on radial separation}
\label{A method based on radial separation}
We perform a method similar to that of \citet{FraserMcKelvie2018} to estimate individual mass-weighted single-age/metallicity of the bulge and the disk. \citet{FraserMcKelvie2018} studied 279 S0 field galaxies of the integral-field MaNGA survey \citep{Bundy2015}. They separated the MaNGA datacube into a bulge region considering all the spaxels within one bulge effective radius and a disk region outside two bulge effective radii. They used these spectra to measure light-weighted ages and metallicities derived from Lick indices. 

We apply the kinematic corrections to the SAMI spectra as described in Section~\ref{sec:Corrections for velocity and velocity dispersion}. We do not separate the bulge and disk regions according to the bulge effective radius like in \citet{FraserMcKelvie2018}, but we spatially bin the SAMI spectra as described in Section~\ref{binning} and we make use of the estimated bulge-disk contributions in mass. For the bulge spectrum, we select the central-most annular bin, which has the highest fraction of light coming from the bulge. The representative 1D disk spectrum is the one from the outermost spatial bin with the highest stellar mass contribution from the disk. To obtain reliable results and avoid contamination from the other component, we consider only representative bins of the bulge (disk) with S/N$>$20 and with a contribution in galaxy $M_{*}$ from the disk (bulge) of at least 60\%. Applying these criteria to galaxy 9091700038, the \textit{upper panel} of Figure~\ref{9091700038_ppxf} represents the 1D bulge spectrum, while the \textit{second to last} represents the 1D disk spectrum. We use pPXF and the same procedure described in Section~\ref{sec:Galaxy age and metallicity maps} to fit the 1D separated spectra for mass-weighted ages and metallicities with the 16th and 84th percentiles as uncertainties derived from the 1000 Monte Carlo simulations. 

Using this method, the bulge-disk properties of only 54/192 galaxies have been reliably estimated, since for the remaining 138 galaxies the selection criteria of bulge/disk bins with $f^{bulge/disk}_{M*, bin}>0.6$ and S/N$>$20 are not satisfied. This is an empirical method which allows us to study fewer galaxies compared to the other two methods described in Sections~\ref{A new method} and \ref{A method based on light weights}. A major limitation of this method is that the bulge (disk) spectra are contaminated by the disk (bulge) light. Therefore, the stellar population parameters derived from these spectra will be less pure when compared with methods that attempt to decompose the spectra.

\section{Results}
\label{sec:Results}
Our aim is to study the stellar population properties of the bulge and the disk in order to understand the formation of S0 galaxies in the cluster environment. We focus on disentangling the age-metallicity degeneracy inherent in analyses using only colors. In this Section, we compare the results from three different methods and we analyze the differences between the bulge and disk stellar population properties.

\subsection{$g-i$ colors}
\label{colors}
Since we are interested in studying the stellar population properties of the bulge and the disk, we explore the separate $g-i$ colors of the two components for the 192 SAMI double-component galaxies. The bulge and disk $g-i$ colors are estimated from the 2D photometric bulge-disk decomposition of Barsanti et al. in preparation, where both the apparent magnitudes have been corrected for the extinction of the Milky Way using the dust maps of \citet{Schlegel1998} and for the K-correction \citep{Blanton2007}.

\begin{figure}
\includegraphics[width=0.45\textwidth]{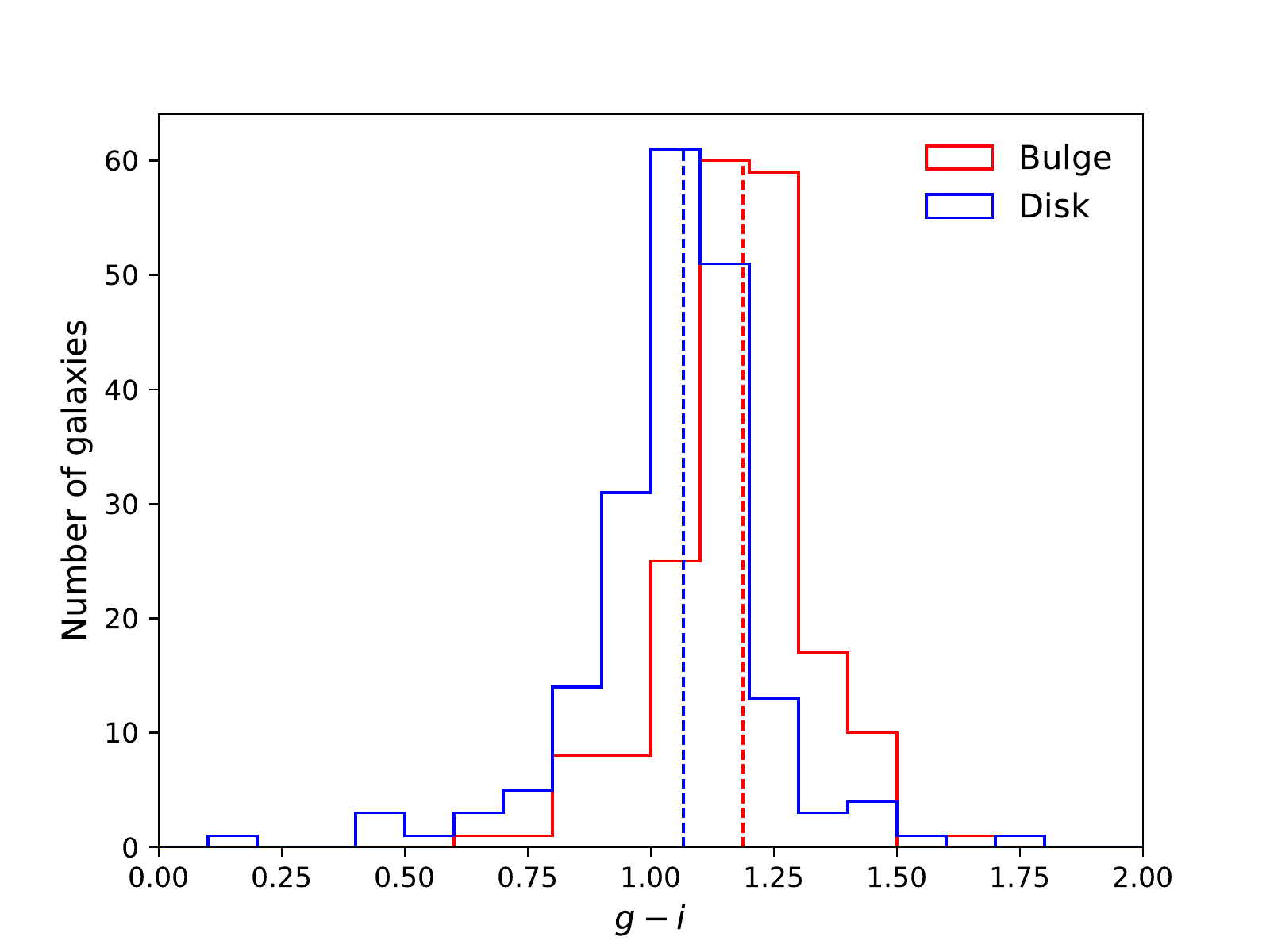}
\caption{Histogram in $g-i$ color for the bulges (red) and the disks (blue) of the 192 SAMI cluster galaxies. The dashed lines represent the median values. Bulges have redder colors than disks.}
\label{colorsDouble}
\end{figure}

We observe that 73$\pm$3\% of the galaxy sample have bulges on average 1.4 times redder compared to their surrounding disks. Figure~\ref{colorsDouble} shows the $g-i$ color distributions for the bulges and the disks. The disk color distribution in blue is shifted towards bluer colors compared to the bulge color distribution. The median $g-i$ offset separating the two distributions is 0.12$\pm$0.02 mag.

\subsection{Comparison between the three methods}
\label{Comparison with other methods}
We report the one-to-one galaxy comparison for the bulge/disk mass-weighted single-age/metallicity results from the three different methods discussed in Sections~\ref{A new method},~\ref{A method based on light weights} and \ref{A method based on radial separation}. 

\begin{figure*}
\centering
\includegraphics[width=0.75\columnwidth]{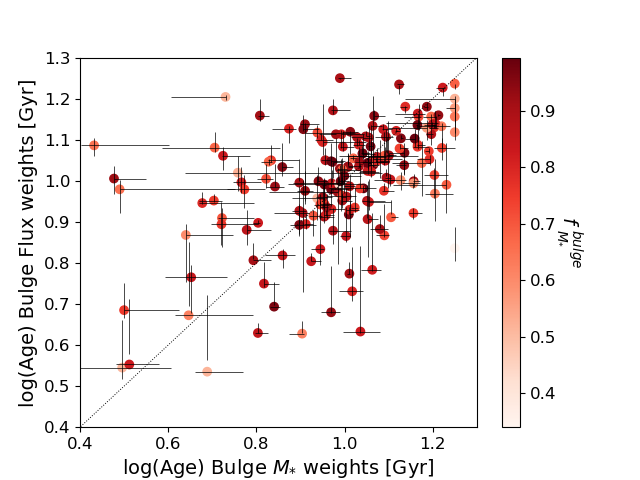}
\includegraphics[width=0.75\columnwidth]{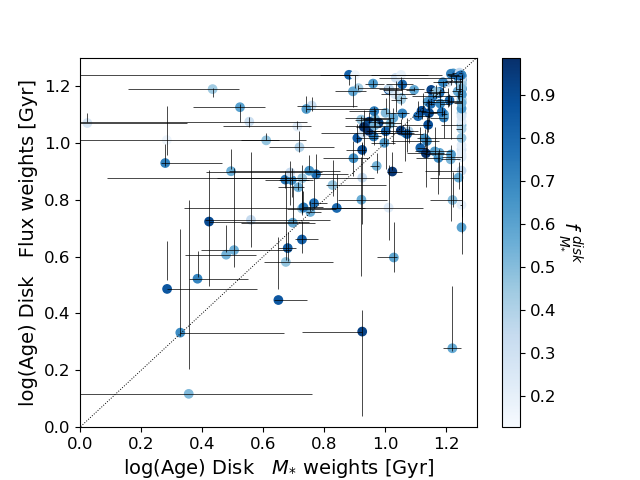}\\
\includegraphics[width=0.75\columnwidth]{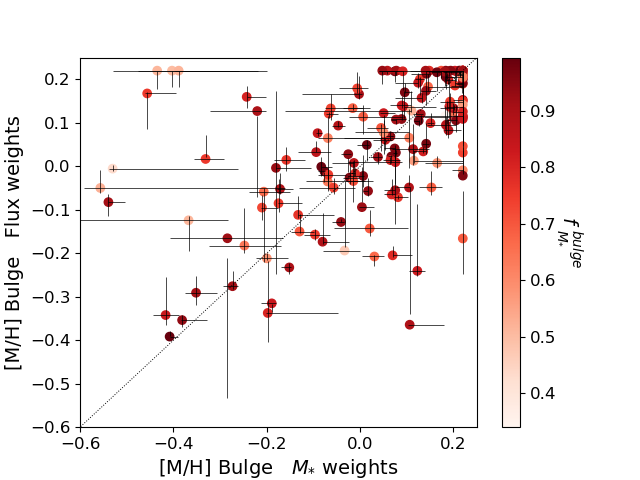}
\includegraphics[width=0.75\columnwidth]{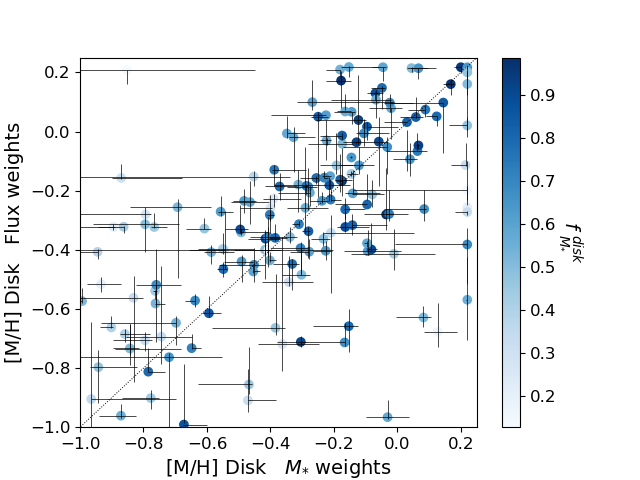}
\caption{Comparison for the age and metallicity of the bulge (\textit{upper and lower left}) and of the disk (\textit{upper and lower right}). The x and y axis represent the results from the mass and flux weights methods, respectively. Points with partially missing errors are due to the 16th and 84th percentiles matching with the result. The dotted black line represents the bisector. Points are color-coded according to the highest contribution per bin to $M_{*}$ from the bulge and the disk. The results clustered along the bisector suggest agreement between the two methods. The largest offsets are for galaxies with $f^{bulge/disk}_{M_{*}, bin}<0.6$.}
\label{Mendez}
\end{figure*}

\begin{figure*}
\centering
\includegraphics[width=0.7\columnwidth]{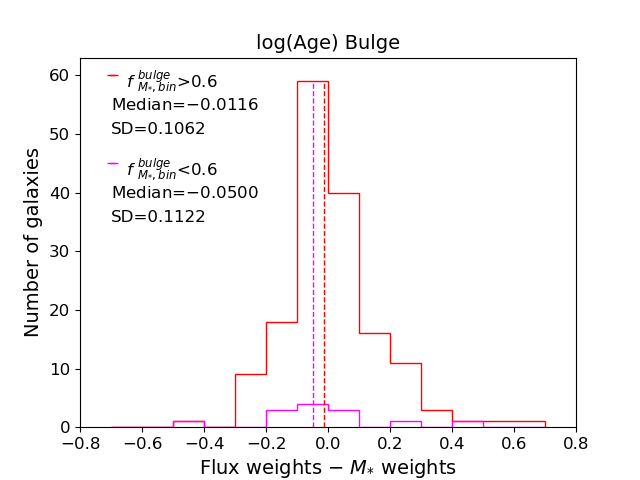}
\includegraphics[width=0.7\columnwidth]{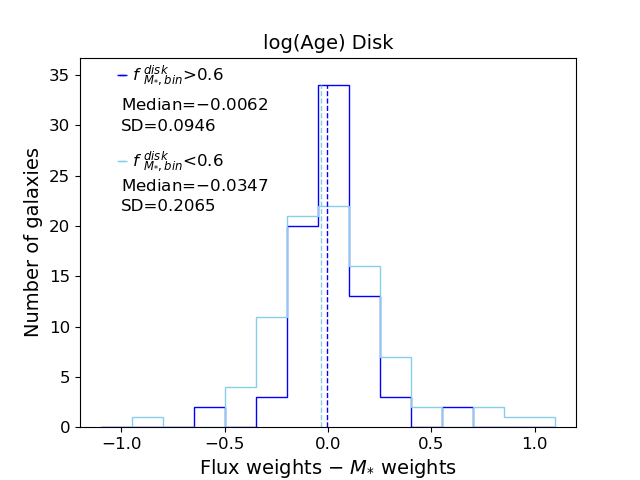}\\
\includegraphics[width=0.7\columnwidth]{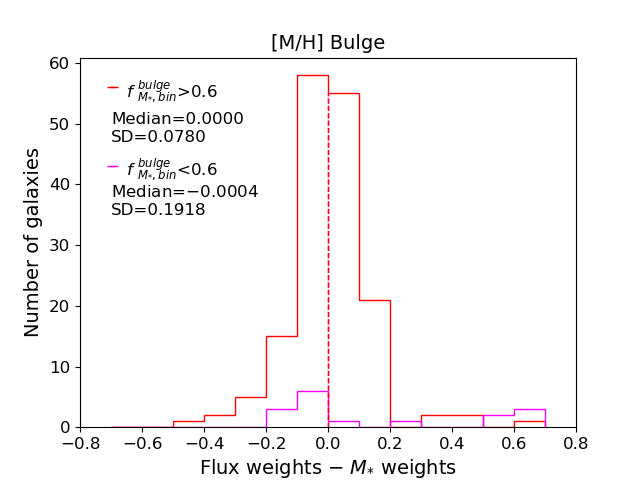}
\includegraphics[width=0.7\columnwidth]{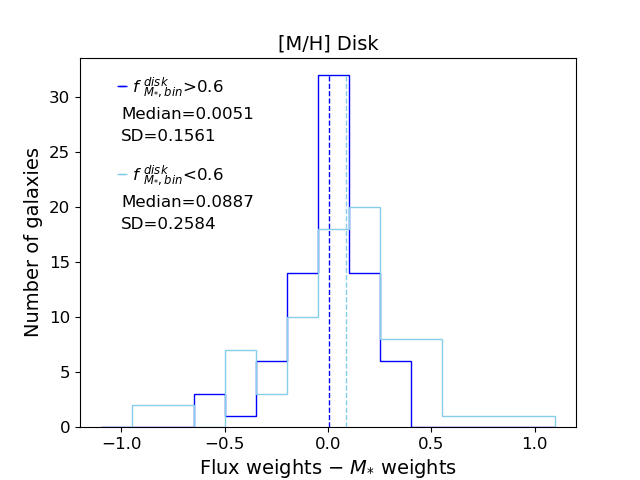}
\caption{Distributions of the differences for the bulge/disk ages and metallicities estimated with the mass and flux weights methods. Galaxies with $f^{bulge/disk}_{M_{*}, bin}>0.6$ (red/blue) and $<$0.6 (magenta/light blue) are shown separately. Largest medians and standard deviations are measured for galaxies with $f^{bulge/disk}_{M_{*}, bin}<0.6$, driving the largest offsets for the whole galaxy population.}
\label{Mendez2}
\end{figure*}

Figure~\ref{Mendez} shows the comparison between the ages and metallicities of the bulges and disks estimated with the methods based on mass and flux weights described in Sections~\ref{A new method} and \ref{A method based on light weights}, respectively. 181/192 SAMI double-component cluster members are considered since 181 galaxies could be decomposed with the method based on flux weights. The values are clustered along the bisector, with those characterized by a lower contribution to the galaxy stellar mass from the components $f_{M_{*}}$ showing the highest deviation since they are harder to constrain. The galaxies with pegged bulge/disk age values have also the smallest $f^{bulge/disk}_{M_{*}}$ contributions, in agreement with the outcomes of Appendix~\ref{Simulations}. This might suggest that for the purpose of stellar population study a more stringent cut on B/T to select double-component galaxies than $0.2<B/T<0.8$ applied by Barsanti et al. in prep is needed. The impact is larger for the disk because of the combination of the lower contribution from the disk and the declining S/N with radius. We show in Figure~\ref{Mendez2} the distributions of the differences between the properties estimated with the two different methods, with plotted medians and robust standard deviations. Galaxies are divided into those with $f^{bulge/disk}_{M_{*}, bin}>0.6$ and those with $f^{bulge/disk}_{M_{*}, bin}<0.6$. For galaxies with $f^{bulge/disk}_{M_{*}, bin}>0.6$ the distributions show a Gaussian shape with medians $\sim$0 dex and standard deviations $\sim0.1$ dex, suggesting no particular shift towards higher or lower values for one method. The distributions for the galaxies with $f^{bulge/disk}_{M_{*}, bin}<0.6$ have larger medians and standard deviations $\sim0.2$ dex, driving the largest offsets for the whole galaxy population.

\begin{figure*}
\centering
\includegraphics[width=0.75\columnwidth]{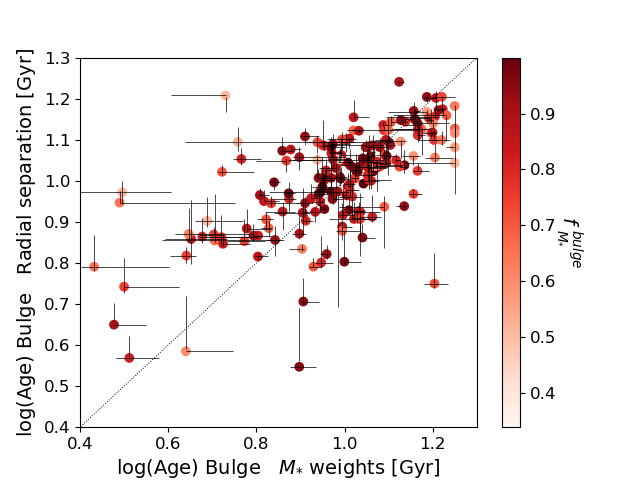}
\includegraphics[width=0.75\columnwidth]{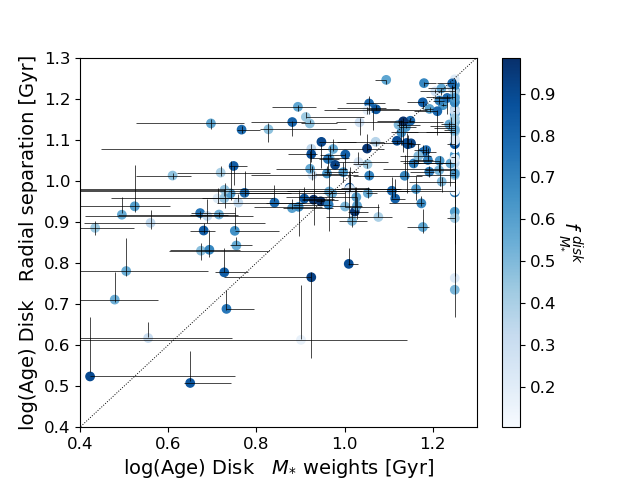}\\
\includegraphics[width=0.75\columnwidth]{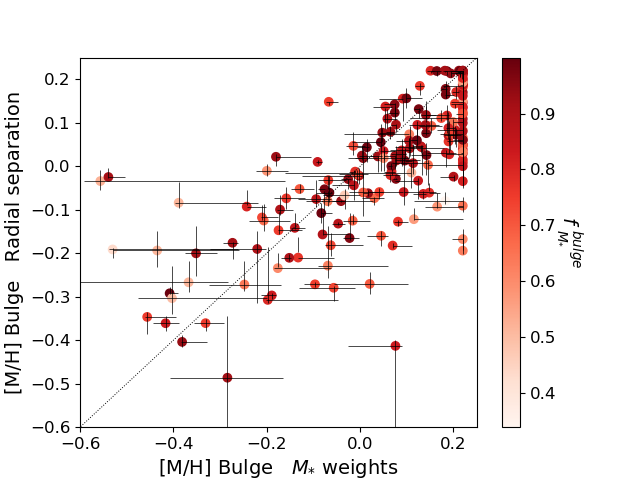}
\includegraphics[width=0.75\columnwidth]{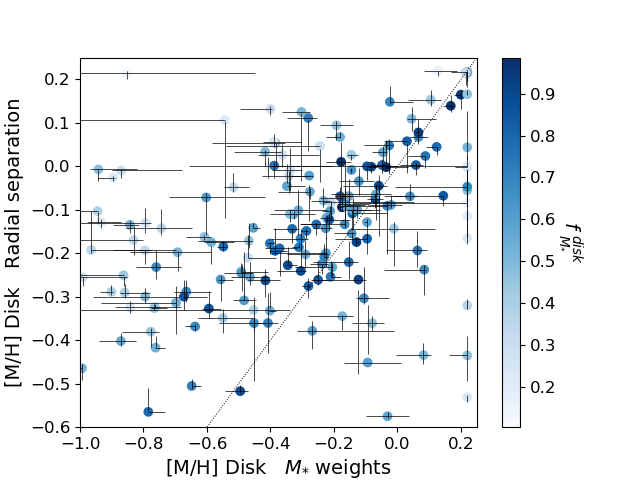}
\caption{Comparison for the age and metallicity of the bulge (\textit{upper and lower left}) and of the disk (\textit{upper and lower right}). The x and y axis represent the results from the mass wights and radial separation methods, respectively. The dotted black line represents the bisector. Points are color-coded according to the contribution in the galaxy stellar mass from the bulge and the disk and are group along the bisector, indicating agreement between the two methods. The largest offsets are for galaxies with $f^{bulge/disk}_{M_{*}, bin}<0.6$.}
\label{McKelvie}
\end{figure*}

\begin{figure*}
\centering
\includegraphics[width=0.7\columnwidth]{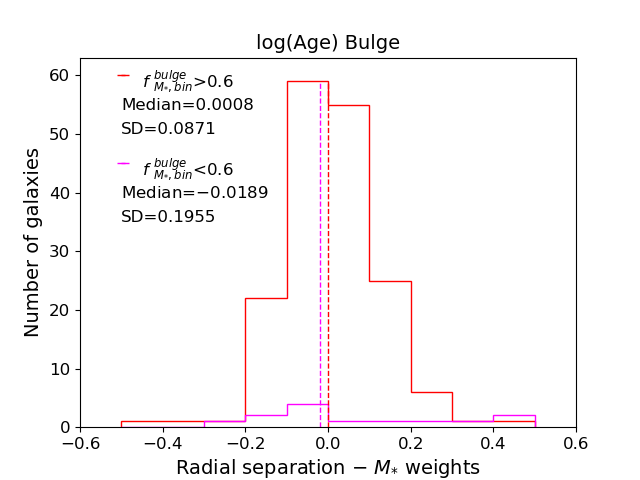}
\includegraphics[width=0.7\columnwidth]{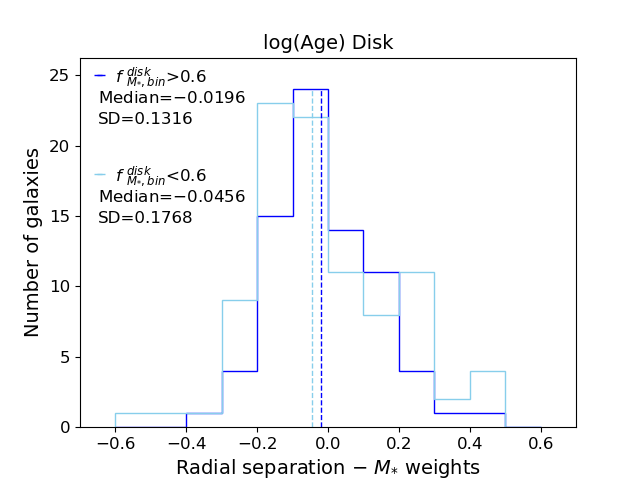}\\
\includegraphics[width=0.7\columnwidth]{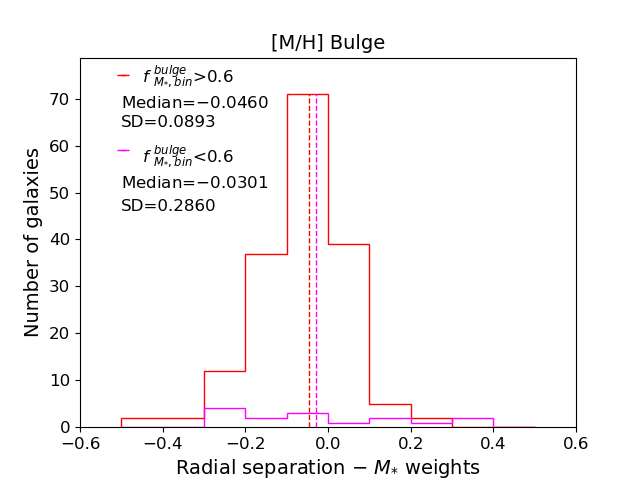}
\includegraphics[width=0.7\columnwidth]{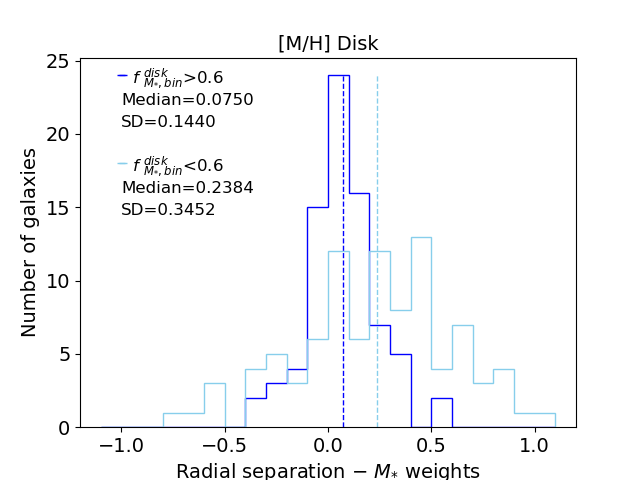}
\caption{Distributions of the differences for the bulge/disk ages and metallicities estimated with the methods based on mass weights and radial separation. Galaxies with $f^{bulge/disk}_{M_{*}, bin}>0.6$ (red/blue) and $<$0.6 (magenta/light blue) are shown separately. Largest medians and standard deviations are measured for galaxies with $f^{bulge/disk}_{M_{*}, bin}<0.6$.}
\label{McKelvie2}
\end{figure*}

Figures~\ref{McKelvie} and \ref{McKelvie2} show the comparison and the difference distributions for the results according to the methods based on mass weights and radial separation (Section~\ref{A method based on radial separation}). We consider the whole sample of 192 SAMI double-component galaxies, considering the most central and outermost bins with S/N$>$20 and without any cut on $f_{M_{*}}$. In Figure~\ref{McKelvie} the values are clustered along the bisector indicating agreement between the results. The largest offsets are mainly due to galaxies with low $f^{bulge/disk}_{M_{*}, bin}$. In Figure~\ref{McKelvie2} galaxies with $f^{bulge/disk}_{M_{*}, bin}>0.6$ show a Gaussian shape with medians$\sim$0 dex, indicating agreement between the methods. Larger medians and standard deviations are measured for galaxies with $f^{bulge/disk}_{M_{*}, bin}<0.6$, indicating poorer agreement. This justifies our choice in Section~\ref{A method based on radial separation} to consider only galaxies with $f^{bulge/disk}_{M_{*}, bin}>0.6$ for the most central and outermost bins in the radial separation method. Moreover, the distributions in Figure~\ref{McKelvie2} show larger medians and standard deviations compared to the distributions in Figure~\ref{Mendez2}. Therefore, the mass wights method shows a better agreement with the flux weight method with respect to the radial separation method.

\begin{figure*}
\centering
\includegraphics[width=0.75\columnwidth]{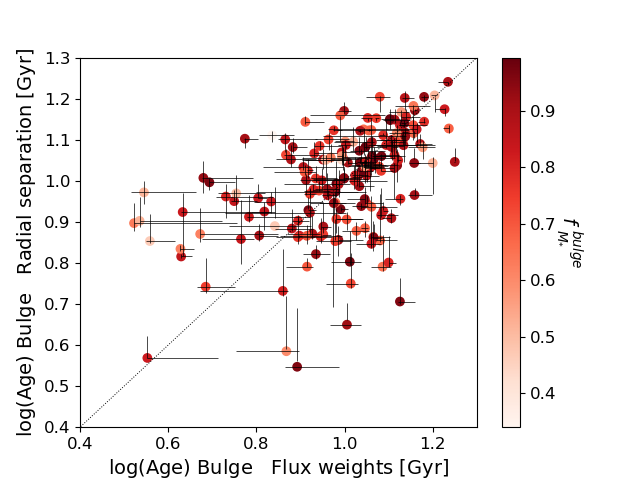}
\includegraphics[width=0.75\columnwidth]{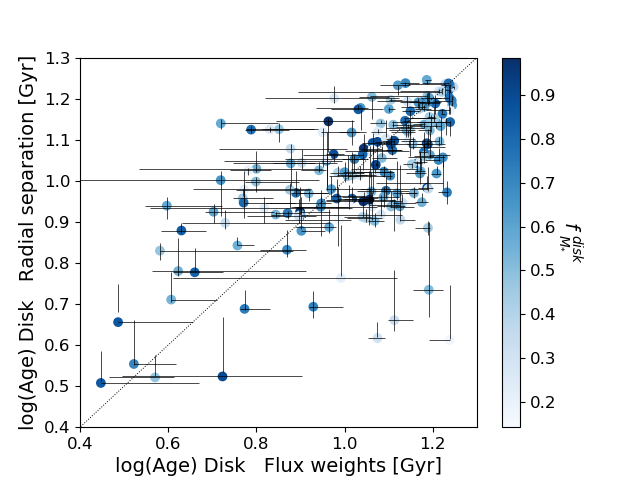}\\
\includegraphics[width=0.75\columnwidth]{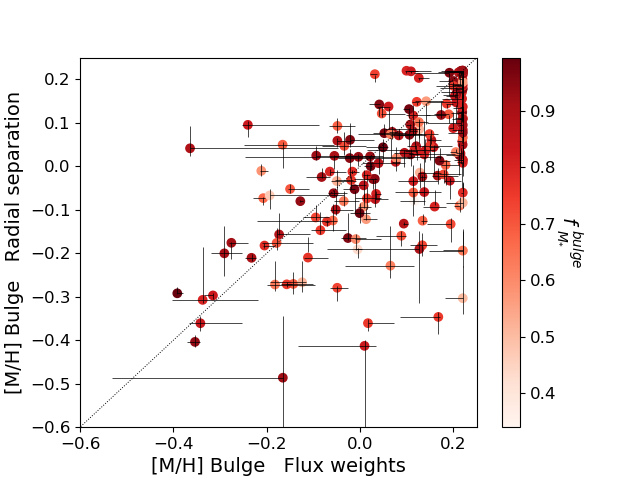}
\includegraphics[width=0.75\columnwidth]{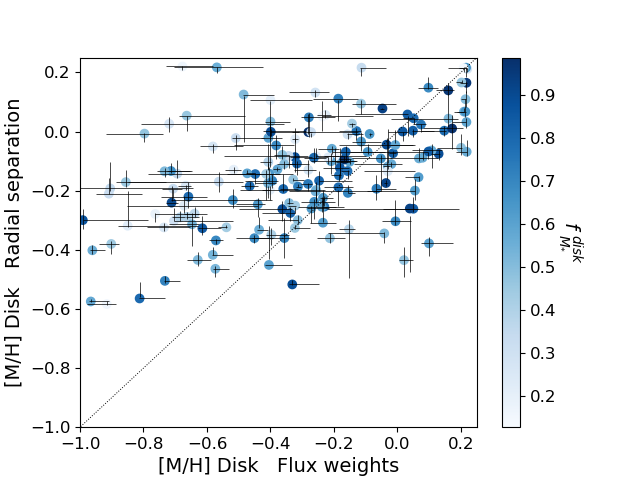}
\caption{Comparison for the age and metallicity of the bulge (\textit{upper and lower left}) and of the disk (\textit{upper and lower right}). The x and y axis represent the results from the flux wights and radial separation methods, respectively. The dotted black line represents the bisector. Points are color-coded according to the contribution in the galaxy stellar mass from the bulge and the disk and are group along the bisector, indicating agreement between the two methods. The largest offsets are for galaxies with $f^{bulge/disk}_{M_{*}, bin}<0.6$.}
\label{McKelvie3}
\end{figure*}

\begin{figure*}
\centering
\includegraphics[width=0.7\columnwidth]{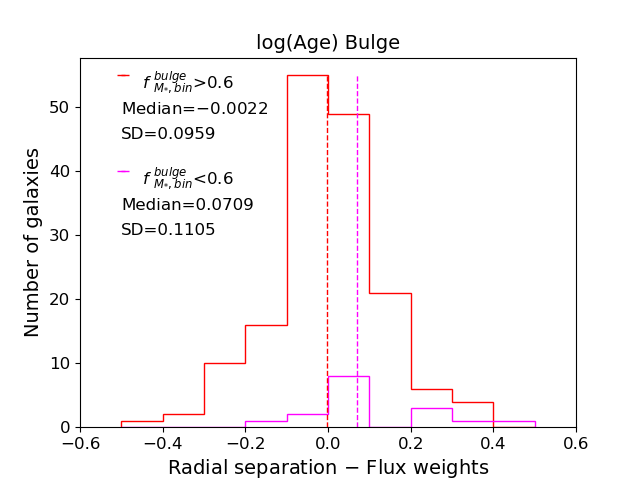}
\includegraphics[width=0.7\columnwidth]{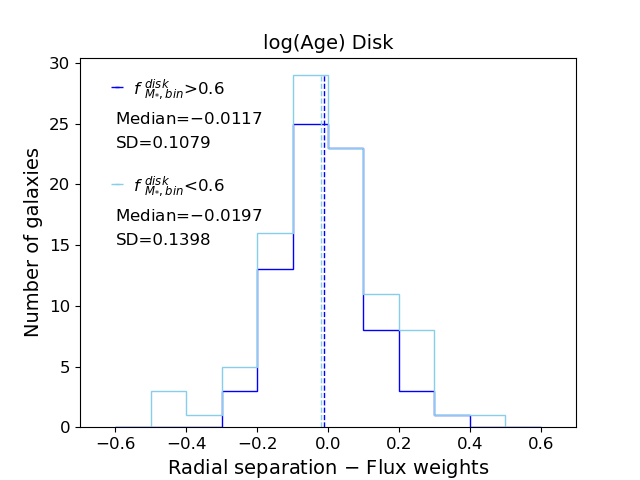}\\
\includegraphics[width=0.7\columnwidth]{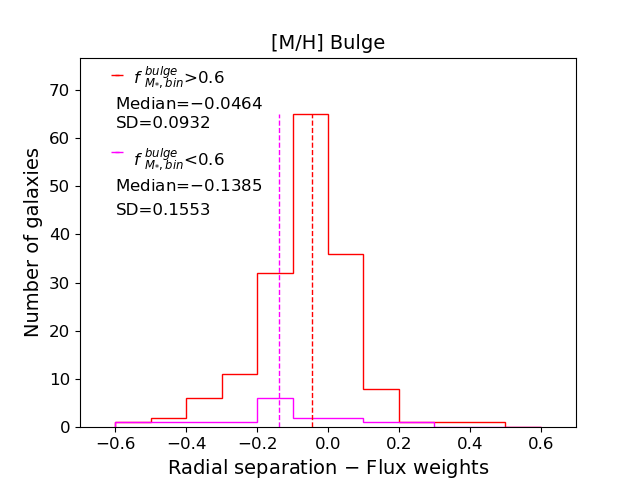}
\includegraphics[width=0.7\columnwidth]{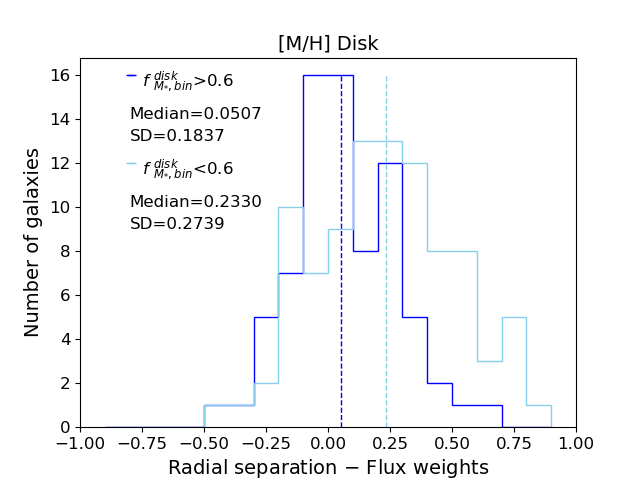}
\caption{Distributions of the differences for the bulge/disk ages and metallicities estimated with the methods based on flux weights and radial separation. Galaxies with $f^{bulge/disk}_{M_{*}, bin}>0.6$ (red/blue) and $<$0.6 (magenta/light blue) are shown separately. Largest medians and standard deviations are measured for galaxies with $f^{bulge/disk}_{M_{*}, bin}<0.6$.}
\label{McKelvie4}
\end{figure*}

Finally, Figures~\ref{McKelvie3} and \ref{McKelvie4} show the comparison and the difference distributions for the results according to the methods based on flux weights and radial separation. We consider 181 galaxies, according to the galaxy sample decomposed with the flux weights method. The values are clustered along the bisector indicating agreement between the results. The distributions are separated for galaxies with $f^{bulge/disk}_{M_{*}, bin}>0.6$ and with $f^{bulge/disk}_{M_{*}, bin}<0.6$. They show similar trends as in Figure~\ref{McKelvie2} for the comparison between the mass weights and radial separation method. Largest medians and standard deviations are measured for galaxies with $f^{bulge/disk}_{M_{*}, bin}<0.6$, which are harder to constrain due to the contamination from the bulge/disk flux. 

Overall, the best agreement is between the mass and flux weights methods, having the lowest medians and standard deviations for the difference distributions of their results. The comparisons between the radial separation method and the other methods is similar, highlighting the empirical limits of the radial separation method where the bulge/disk results are contaminated by the flux of the other component and are spatially limited. For all the comparisons the highest deviations between the results are for galaxies with low $f^{bulge/disk}_{M*, bin}$ values. This is particularly evident for the radial separation method, justifying our choice to select only galaxies with $f^{bulge/disk}_{M*, bin}>0.6$ in Section~\ref{A method based on radial separation}.

\subsection{Comparing the stellar populations of bulges and disks}
\label{Ages and metallicities}
Our aim is to explore the formation of S0 galaxies in clusters. Different scenarios imprint different fossil records in the stellar population properties of the bulges and the disks. We analyze the the best-fitting mass-weighted ages and metallicities of the bulges and the disks for the 192 double-component SAMI galaxies obtained with the three methods described in Sections~\ref{A new method}, \ref{A method based on light weights} and \ref{A method based on radial separation}. Table~\ref{Comparison3methods} reports the method, the number of analyzed galaxies, the percentages of galaxies with older, younger, more metal-rich, more metal-poor, redder and bluer bulges with respect to the disks. The age, metallicity and $(g-i)$ results for the three methods are in agreement. The analysis of the age reveals that $\sim$44\% of galaxies are characterized by bulges older than the disks. The remaining 56\% of galaxies are characterized by bulges younger than the disk ages. We find that most bulges, $\sim$82\%, are more metal-rich than the disks. 

In order to quantify the systematics effect due to the chosen method, we estimate the number of galaxies that are considered with different stellar population properties by the different methods. The comparison between the $M_{*}$ weights and flux weights methods shows that $\sim$30\% of the galaxies have a bulge older or more-metal-rich than the disk according to the flux weights method, but they show an opposite result for the $M_{*}$ weights method. However, as shown in Figure~\ref{Mendez} the measurements are close the one-to-one relationship. The remaining $\sim$70\% shows the same age and metallicity properties according to both methods. For the comparison between the flux weights and the radial separation methods we find the same outcome. The number of galaxies that differ is $\sim$20\% between the $M_{*}$ weights and the radial separation methods. Thus, on average the percentages in Table~\ref{Comparison3methods} are characterized for the 75\% by galaxies that show the same bulge and disk stellar population properties according to the three methods.

We investigate the significance of the age and metallicity differences between the two components with respect to their random uncertainties. For the $M_{*}$ weights method the random errors on the $(\log(Age),[M/H])_{bulge/disk}$ results are estimated as the standard deviations of the $(\log(Age),[M/H])_{bulge/disk}$ distributions using a Monte Carlo approach, as described in Section~\ref{Age and metallicity of bulges and disks}. In case of pegged solution and consequent zero standard deviation, we associate to that result the average standard deviation from all the measurements. For the flux weights and radial separation methods the random errors on $(\log(Age),[M/H])_{bulge/disk}$ are estimated as the standard deviations of 1000 Monte Carlo simulations, as described in Sections~\ref{A method based on light weights} and \ref{A method based on radial separation}, respectively. The mean random errors for the bulge results are 0.03 dex, 0.03 dex and 0.02 dex for the $M_{*}$ weights, flux weights and radial separation methods, respectively. The mean random errors for the disk results are 0.10 dex, 0.06 dex and 0.04 dex for the $M_{*}$ weights, flux weights and radial separation methods, respectively. We assess the statistical significance of the differences in age and metallicity between bulges and disks by comparing the difference to the quadrature sum of the measured bulge and disk random uncertainties. Results are significantly different at the $3\sigma$ level.

Table~\ref{ComparisonRandomErrors} lists for each method the percentages of galaxies with bulges significantly (at the $3\sigma$ level) older/younger/similar in age and significantly more metal-rich/more metal-poor/similar in metallicity with respect to the disks. We observe that on average $\sim$23\% of the galaxies have a bulge significantly older than the disk. For $\sim$34\% of galaxies the bulge is significantly younger than the disk. For the remaining $\sim$43\% of galaxies the differences in age are not significant with respect to their random uncertainties. The analysis of the metallicity shows that most galaxies, $\sim$62\%, have a bulge significantly more metal-rich than the disk. Only $\sim$7\% of the galaxies have a bulge significantly more metal-poor than the disk. The remaining $\sim$31\% of galaxies show a metallicity for the bulge consistent with that of the disk. Overall, we find that most bulges are significantly more metal-rich than the disks, while they can be either significantly older or younger than the disks. According to the methods based on mass and flux weights, the bulges are on average $\sim$3 times more metal-rich than disks, whereas they are $\sim$2 times more metal-rich according to the radial separation method.

Finally, we test for consistency between the $(g-i)$ colors estimated using the decomposed photometry, and those estimated using the closest SSP templates to the age/metallicity determined for the bulge and disk. In this paper the $(g-i)$ colors are derived from the 2D bulge-disk decomposition. However, we observe the same percentages of redder bulges for the three methods also when the $(g-i)$ colors of the bulges and the disks are estimated using their age and metallicity. For each galaxy component we associate its estimated age and metallicity from Section~\ref{Stellar population properties} with the MILES SSP template characterized by the closest values. Each MILES SSP template has a predicted $(g-i)$ color \citep{Vazdekis1996}. We conclude that for the three methods bulges tend to be redder and more metal-rich compared to the disks, but we do not see a clear trend with age. 

\begin{table*}
 \centering
  \caption{Differences in mass-weighted ages, metallicities and $(g-i)$ colors between bulges and disks from the three methods. Column 1 lists the method, column 2 the number of analyzed galaxies, columns 3/4/5/6/7/8 the percentages of galaxies with bulges older/younger/more metal-rich/more metal-poor/redder/bluer than the disks. The last line lists the average values for the three methods.}
  \label{Comparison3methods}
  \begin{tabular}{@{}lccccccc@{}}
  \hline
Method   &  N$_{g}$ &  \%(Age) B$>$D &   \%(Age) B$<$D  & \%[M/H] B$>$D  & \%[M/H] B$<$D& \%($g-i$) B$>$D & \%($g-i$) B$<$D\\
\hline
$M_{*}$ weights & 192 &45$\pm$4& 55$\pm$4 & 80$\pm3$ & 20$\pm$3&73$\pm$3&27$\pm$3\\
Flux weights & 181 & 45$\pm$4& 55$\pm$4 & 86$\pm3$ & 14$\pm$3&73$\pm$3&27$\pm$3  \\
Radial separation &54& 43$\pm$7& 57$\pm$7 & 81$\pm5$ & 19$\pm$5&67$\pm$6&33$\pm$6  \\
\hline
Average & & 44$\pm$5& 56$\pm$5 & 82$\pm$4 & 18$\pm$4 &71$\pm$4&29$\pm$4\\
\hline
\end{tabular}
\end{table*}

\begin{table*}
 \centering
  \caption{Significant differences in bulge/disk mass-weighted ages and metallicities for the three methods. Column 1 lists lists the method, column 2 the number of analyzed galaxies, columns 3/4/5 the percentages of galaxies with bulges significantly older/younger/equal than the disks, and columns 6/7/8 the percentages of galaxies with bulges significantly more metal-rich/metal-poor/equal than the disks. The last line lists the average values for the three methods.}
  \label{ComparisonRandomErrors}
  \begin{tabular}{@{}lccccccc@{}}
  \hline
Method   &  N$_{g}$ &  \%(Age) B$\gg$D &   \%(Age) B$\ll$D  & \%(Age) B$\sim$D  & \%[M/H] B$\gg$D& \%[M/H] B$\ll$D & \%[M/H] B$\sim$D\\
\hline
$M_{*}$ weights & 192 &26$\pm$3& 36$\pm$3 & 38$\pm3$ & 66$\pm$3&12$\pm$2&22$\pm$3 \\
Flux weights & 181 & 18$\pm$3& 37$\pm$4 & 45$\pm4$ & 70$\pm$3&6$\pm$2&24$\pm$3 \\
Radial separation &54& 26$\pm$6& 30$\pm$6 & 44$\pm7$ & 50$\pm$7&2$\pm$2&48$\pm$7 \\
\hline
Average & & 23$\pm$4& 34$\pm$4 & 43$\pm$5 & 62$\pm$4&7$\pm$2&31$\pm$4\\
\hline
\end{tabular}
\end{table*}

\subsection{Comparing the stellar population ages of bulges and disks}
\label{Exploring bulges older or younger than the disks}
\begin{figure*}
\centering
\includegraphics[width=\columnwidth]{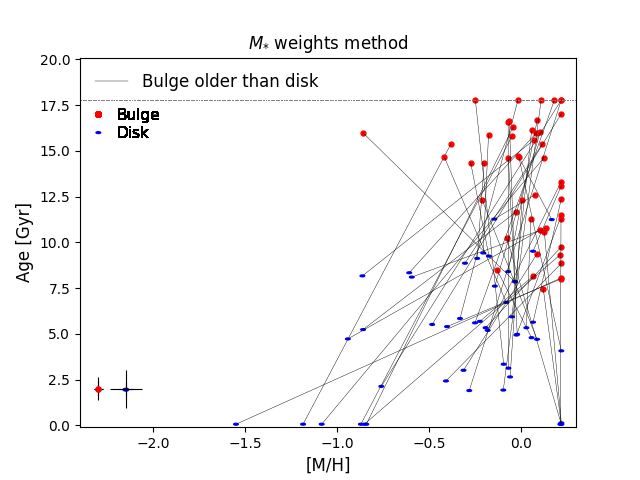}
\includegraphics[width=\columnwidth]{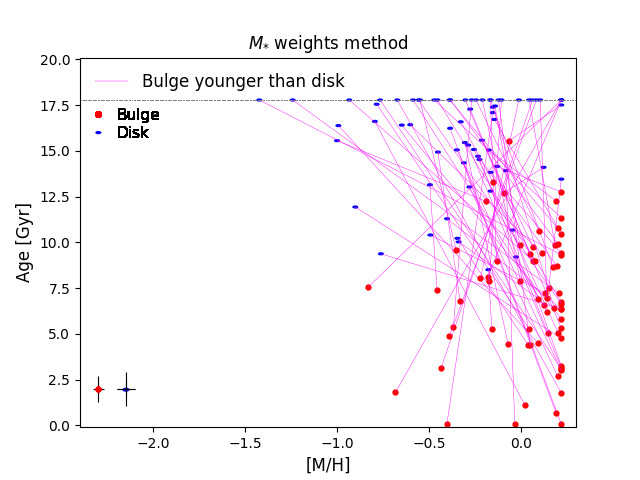}
\includegraphics[width=\columnwidth]{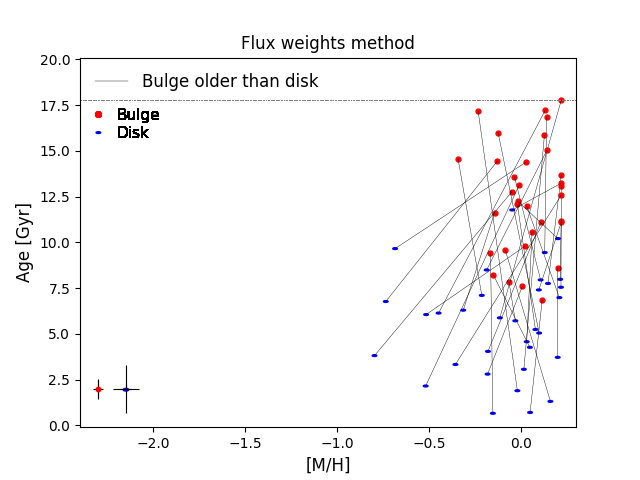}
\includegraphics[width=\columnwidth]{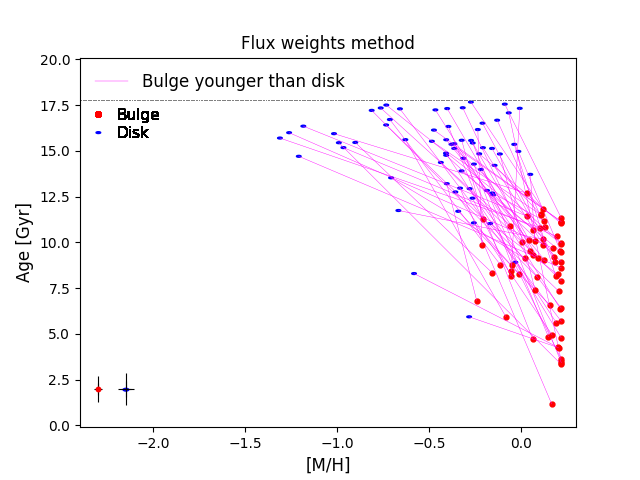}
\includegraphics[width=\columnwidth]{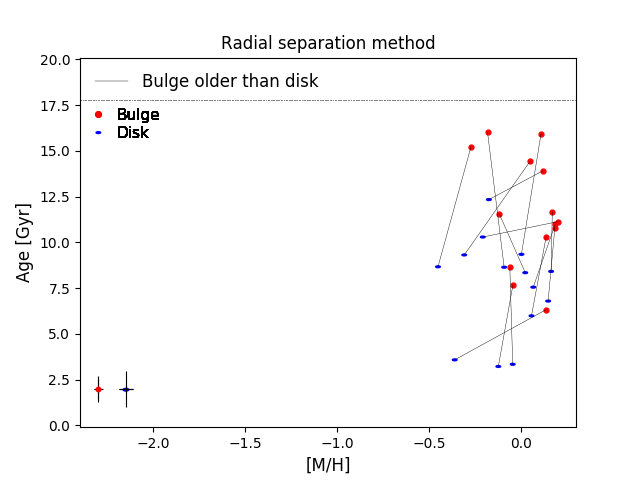}
\includegraphics[width=\columnwidth]{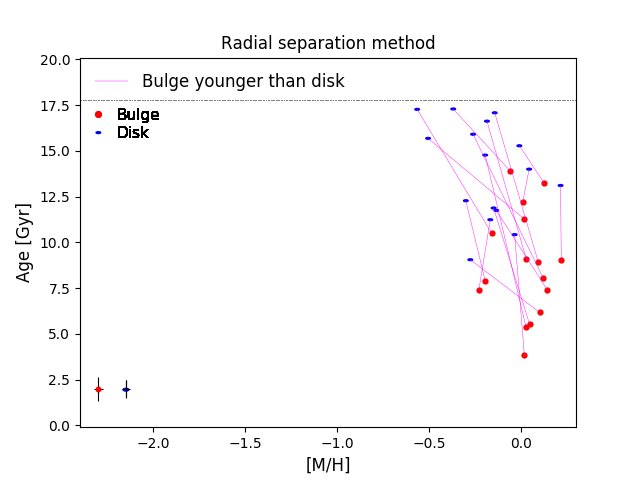}

\caption{Age-metallicity plots for the three methods (\textit{top to bottom panels}) where galaxies have the bulge significantly older (\textit{left panels}) and younger (\textit{right panels}) than the disk. The mean errors on the bulge and disk properties are plotted on the bottom left. In both galaxy populations, bulges are mainly more metal-rich compared to the disks, in agreement for the three methods.}
\label{AgeMetalPlot}
\end{figure*}

In order to disentangle galaxies where the bulge age differs significantly from the disk age, we analyze the age-metallicity plots for the three methods in Figure~\ref{AgeMetalPlot}. Bulges are represented by red dots, the disks by blue ellipses and the grey/magenta lines connect bulge and disk of the same galaxy. The \textit{left panels} show the galaxies where the bulge is significantly older than the disk, while the \textit{right panels} show the galaxies where the bulge is significantly younger than the disk counterpart. The pegged solutions at the upper/lower fit limits might be caused by the assumption of single-age/metallicity instead of radial gradients as reported in Section~\ref{Age and metallicity of bulges and disks}. Moreover, as shown in Appendix~\ref{Simulations}, pegged solutions depend on the bulge-to-total flux ratio of the galaxy and they are generated for low $M_{*}$ weights. For both the galaxy populations with significantly younger/older bulge than the disk, bulges are mainly more metal-rich with respect to the disks.

Table~\ref{ComparisonOldYoungbulges} lists the method, the number of galaxies with bulges significantly older/younger than the disks and the percentages of bulges significantly more metal-rich, more metal-poor, similar in metallicity, redder and bluer when compared to the disks. Combining the results of the three methods, we find that 46\% of the galaxies with bulges that are significantly older than disks also have bulges that are significantly more metal-rich than their disks. In these cases, the bulge is on average 2.3 times more metal-rich than the disk. For the 43\% of these galaxies the bulge and the disk have similar metallicity. For 70\% of the galaxies the bulge is also redder than the disk by a factor of $\sim 1.4$ on average. For the galaxy population with bulges that are significantly younger than the disks, the percentage of bulges that are also significantly more metal-rich increases to $\sim$75\%. These bulges are $\sim$2.8 times more metal-rich than the disks. In 73\% of these cases, the bulge is also redder than the disk by a factor of $\sim 1.2$. Thus, our results indicate that, regardless of bulge and disk age, in a majority of cases bulges are both $\sim$1.2-1.4 redder and $\sim$2-3 more metal-rich than disks. These results are consistent across all three methods used, with $\sim$70\% of the galaxies showing the same stellar population properties for the bulges and the disks for all the methods.

We investigate whether galaxies with significantly older or younger bulges than the disks trace different formation scenarios for S0 galaxies. To this end, we explore their dependence on the properties of the galaxy, the galaxy components and the cluster environment. We study the galaxy stellar mass and the B/T distributions of the two populations in age. These two parameters trace possible different in situ formation scenarios. We study the bulge and disk properties, such as $g$-band magnitude, S\'ersic index, effective radius and stellar mass. These parameters help us to understand if bulges significantly older or younger than the disks have different features. Finally, we study the projected distance from the cluster centre normalized by $R_{200}$ and the galaxy density. Using these environment metrics, we can assess whether the two galaxy populations in age depend on their environment. The galaxy density is measured as the fifth nearest neighbour surface density $\Sigma_{5}$ by \citet{Brough2013}. To check for significant differences between the two galaxy populations we apply the Anderson$-$Darling test \citep{Stephens1974}. This tests the null hypothesis that the two galaxy populations are drawn from the same parent distribution, and it probes the differences in the distribution tails. We consider the results only for the mass and flux weights methods, since the radial separation method is reliable for only a low number of galaxies.

\begin{figure*}
\centering
\includegraphics[width=\columnwidth]{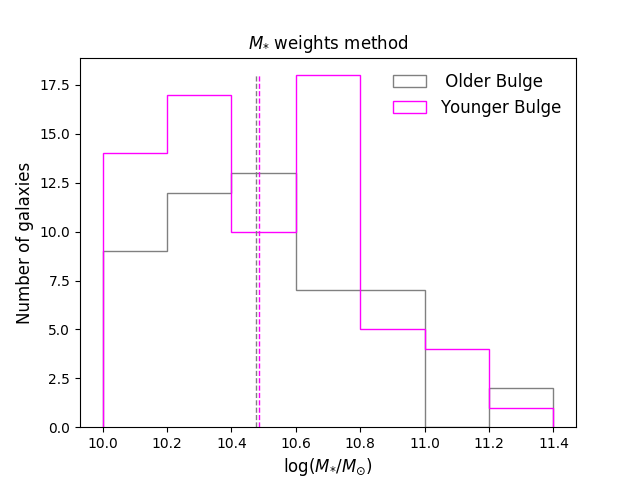}
\includegraphics[width=\columnwidth]{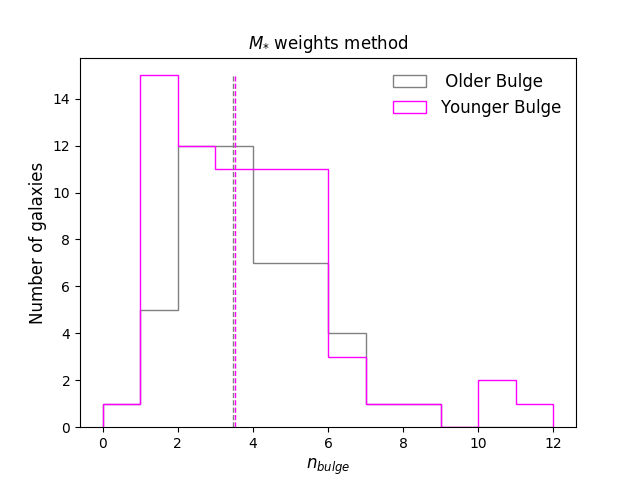}
\includegraphics[width=\columnwidth]{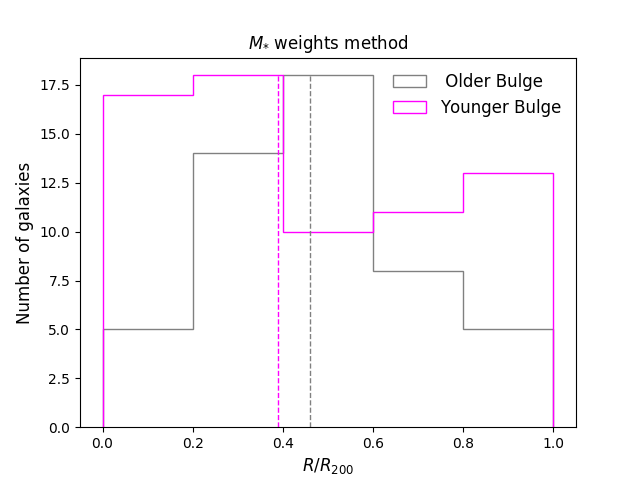}
\includegraphics[width=\columnwidth]{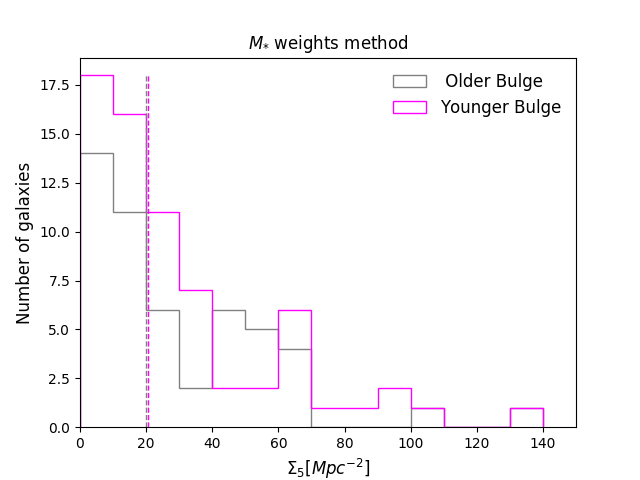}
\caption{Histograms for the galaxies with a bulge significantly older (grey color) or younger (magenta color) than the disk as a function of the galaxy stellar mass (\textit{top left panel}), bulge S\'ersic index (\textit{top right panel}), projected cluster-centric distance (\textit{bottom left panel}) and galaxy density (\textit{bottom right panel}). The dashed lines represent the median values. According to the Anderson$-$Darling test, we measure no significant differences.}
\label{BulgeProp}
\end{figure*}

For both the methods, we observe that there is not a significant difference between the galaxy populations with bulges older or younger than the disks as a function of the galaxy, bulge/disk and environment properties. Figure~\ref{BulgeProp} shows for the galaxy populations with bulges significantly older or younger than the disks the distributions in $M_{*}$, $n_{bulge}$, $R/R_{200}$ and galaxy density. We plot the results only from the mass weights method, since they are in agreement with those from the flux weights method. If we exclude those galaxies where the metallicity difference between the bulge and the disks is not significant, we find consistent results.

\begin{table*}[t!]
 \centering
  \caption{Metallicities and $(g-i)$ colors for bulges significantly older or younger than the disks from the three methods. Column 1 lists the method, column 2 the population with bulges significantly older/younger than the disks, column 3 the number of galaxies, columns 4/5/6 the percentages of bulges significantly more metal-rich/metal-poor/equal than the disks, and columns 7/8 the percentages of bulges redder/bluer than the disks.}
  \label{ComparisonOldYoungbulges}
  \begin{tabular}{@{}lccccccc@{}}
  \hline
Method   & \%(Age) & N$_{g}$ &  \%[M/H] B$\gg$D  & \%[M/H] B$\ll$D&  \%[M/H] B$\sim$D  & \%($g-i$) B$>$D & \%($g-i$) B$<$D\\
\hline
$M_{*}$ weights & B$\gg$D & 50 &56$\pm$7& 14$\pm$5 & 30$\pm6$ & 76$\pm$6& 24$\pm$6\\
$M_{*}$ weights & B$\ll$D & 69 &64$\pm$6& 11$\pm$4 & 25$\pm$5 & 75$\pm$5& 25$\pm$5\\
Flux weights & B$\gg$D & 33 & 39$\pm8$ & 12$\pm$6 & 49$\pm$8& 76$\pm$7& 24$\pm$7 \\
Flux weights & B$\ll$D & 66 & 92$\pm3$ & 0$\pm$1& 8$\pm$3& 74$\pm$5& 26$\pm$5 \\
Radial separation &B$\gg$D & 14& 43$\pm$12& 7$\pm$13 & 50$\pm13$ & 57$\pm$12& 43$\pm$12\\
Radial separation &B$\ll$D & 16& 69$\pm$11& 0$\pm$1 & 31$\pm11$ & 69$\pm$11& 31$\pm$11\\
\hline
\end{tabular}
\end{table*}


\begin{table*}
 \centering
  \caption{Mass-weighted ages and metallicities for redder bulges from the three methods. Column 1 lists the method, column 2 the number of galaxies with redder bulges than the disks, columns 3/4/5/6 the percentages of galaxies with bulges significantly older/younger/more metal-rich/more metal-poor than the disks.} 
  \label{RedderBulges}
  \begin{tabular}{@{}lccccccc@{}}
   \hline
Method   &  N$_{g}$ \%($g-i$) B$>$D &  \%(Age) B$\gg$D & \%(Age) B$\ll$D & \%[M/H] B$\gg$D & \%[M/H] B$\ll$D\\
\hline
$M_{*}$ weights & 141/192 &27$\pm$4 & 36$\pm$4  & 70$\pm$4  & 11$\pm$3  \\
Flux weights & 132/181  & 19$\pm$3 &37$\pm$4  &71$\pm$4  & 6$\pm$2 \\
Radial separation &36/54 &22$\pm$7 & 31$\pm$8 & 56$\pm$8 & 0$\pm$3 \\
\hline
\end{tabular}
\end{table*}

\subsection{Breaking the age-metallicity degeneracy}
As shown in Figure~\ref{colorsDouble}, the majority ($\sim$70\%) of the galaxies in our sample have bulges that are redder than their disks. We now seek to understand whether this is due to the bulges being more metal-rich and/or older than their surrounding disks. For galaxies with redder bulges Table~\ref{RedderBulges} lists for each method the percentages of galaxies with bulges significantly older, younger, more metal-rich and more metal-poor than the disks. For the galaxies with bulges that are redder than disks, $\sim$66\% have bulges that are also significantly more metal-rich when compared with the disks. On the other hand, only $\sim$23\% of redder bulges are also significantly older than the disks. We conclude that the redder color in bulges is mainly driven by their metallicity. Consistent results are obtained if the $(g-i)$ colors are those predicted for the closest MILES SSP templates associated with the bulge/disk age and metallicity, instead of the values from the 2D bulge-disk decomposition. 

\begin{figure*}
\centering
\includegraphics[width=\columnwidth]{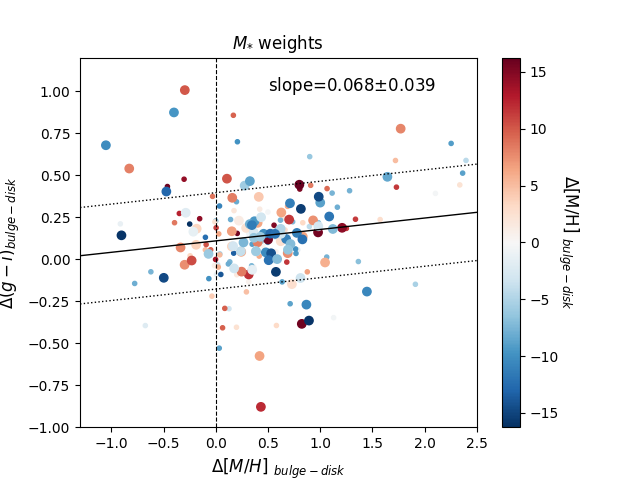}
\includegraphics[width=\columnwidth]{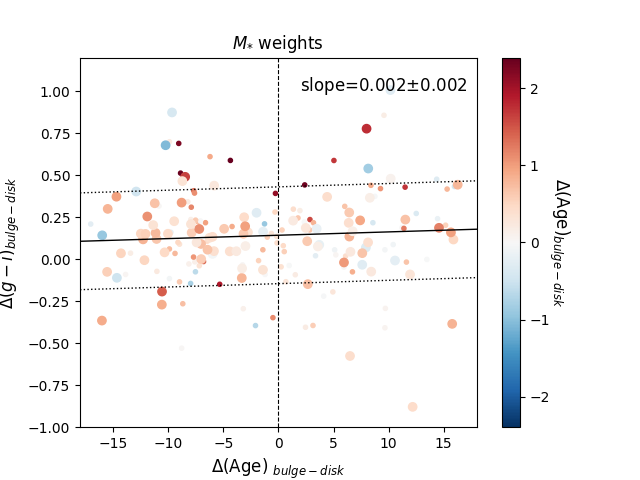}
\includegraphics[width=\columnwidth]{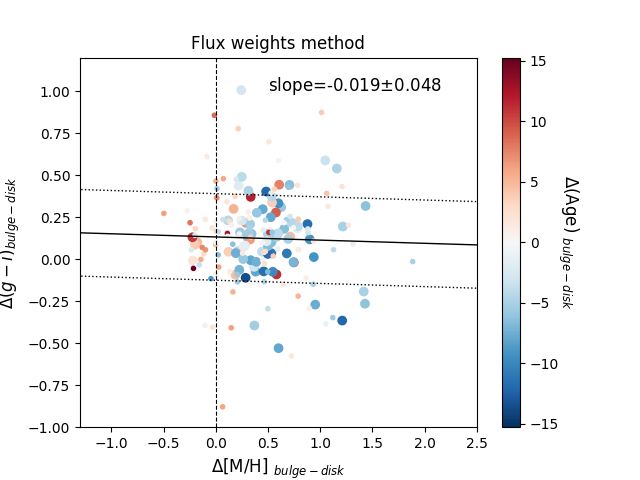}
\includegraphics[width=\columnwidth]{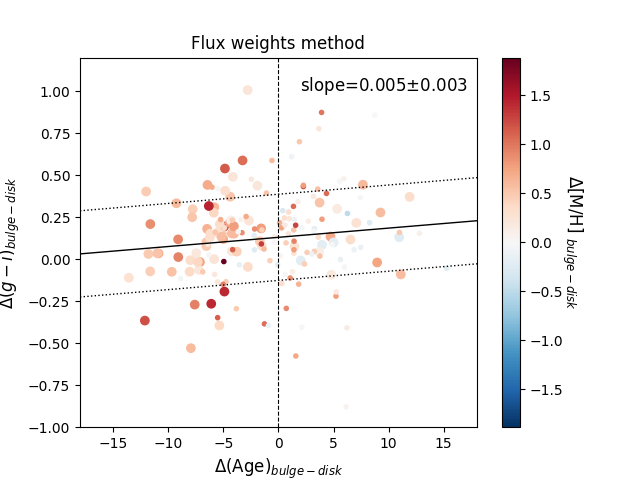}
\includegraphics[width=\columnwidth]{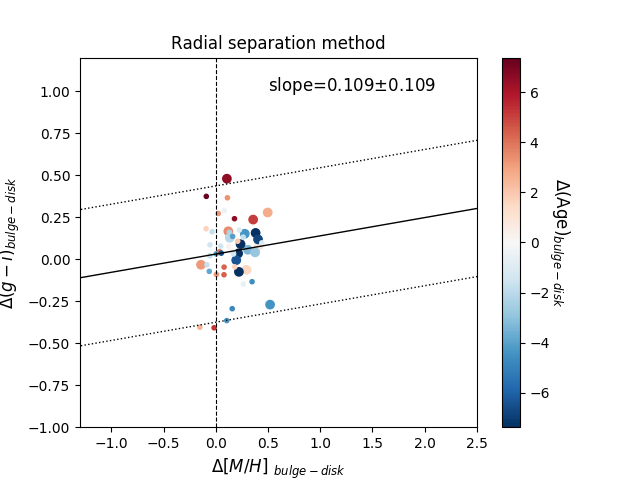}
\includegraphics[width=\columnwidth]{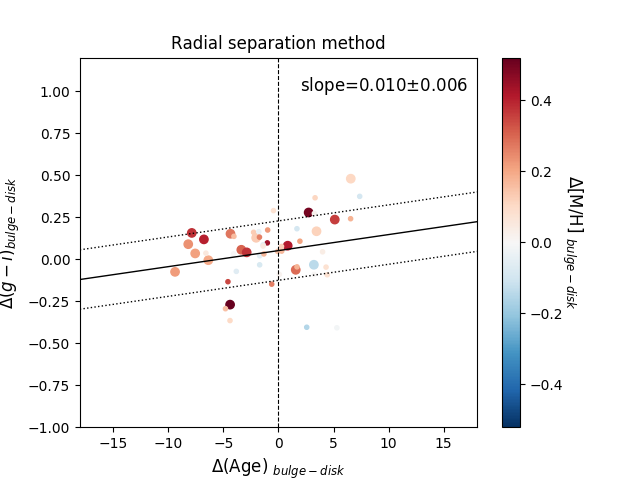}
\caption{$\Delta(g-i)=(g-i)_{bulge}-(g-i)_{disk}$ versus $\Delta[M/H]=[M/H]_{bulge}-[M/H]_{disk}$ (\textit{left panel}) and $\Delta(Age)=Age_{bulge}-Age_{disk}$ (\textit{right panel}) for the three methods. Points are color-coded according to the age and metallicity difference, respectively. Larger points correspond to significant differences in age or metallicity. The dashed black line marks $\Delta[M/H]=0$ and $\Delta(Age)=0$, respectively. Redder bulges are mainly more metal-rich than the disk, while they cover all the age values. The  best-fitting line is filled black and estimated with the Hyper-Fit software \citep{Robotham2015}. The dotted black lines mark the 1$\sigma$ interval.   Flat trends are observed for $\Delta(g-i)$ versus $\Delta[M/H]$ and versus $\Delta(Age)$.}
\label{RedderBulgesMetalAgeplots}
\end{figure*}

To explore the correlation between colors, age and metallicity Figure~\ref{RedderBulgesMetalAgeplots} shows for the three methods the difference in the $g-i$ color $\Delta(g-i)=(g-i)_{bulge}-(g-i)_{disk}$ as a function of the metallicity difference $\Delta[M/H]=[M/H]_{bulge}-[M/H]_{disk}$ (\textit{left panel}) and age difference $\Delta(Age)=Age_{bulge}-Age_{disk}$ (\textit{right panel}). The points are color-coded according to the age and metallicity difference, respectively. Redder points correspond to bulges that are older or more metal-rich than the disks. Larger points correspond to significant differences in age or metallicity. In agreement for the three methods, galaxies with bulges redder than the disks are mainly more metal-rich since most of them have $\Delta[M/H]>0$. For the $\Delta(g-i)$ versus $\Delta(Age)$ plots we do not observe any particular trend as a function of $\Delta(Age)$, with galaxies assuming either $\Delta(Age)>0$ or $\Delta(Age)<0$.

If the age is the primary driver of the bulges being redder than the disks, we expect a general trend towards larger age differences (in the sense that bulges are older than disks) for bulges that are redder than disks. This would show as a positive slope in the $\Delta(g-i)$ versus $\Delta(Age)$ plots. If the metallicity is the main driver for bulges being redder than the disks, we expect a positive slope in the $\Delta(g-i)$ versus $\Delta[M/H]$ plots. Considering all the bulges, the slopes of the best-fitting lines gave us hints on what we expect in terms of trends. Especially for the $M_{*}$ weights method, we observe a stronger positive trend for the $\Delta(g-i)$ versus $\Delta[M/H]$ compared to the slope for the $\Delta(g-i)$ versus $\Delta(Age)$. However, considering the uncertainties all the best-fitting lines are not statistically significant and they are consistent with flat trends.

\section{Discussion}
\label{sec:Discussion}
In this Section we discuss the limitations and strengths of the three methods based on $M_{*}$ weights, flux weights and radial separation implemented in Section~\ref{A new method}, \ref{A method based on light weights} and \ref{A method based on radial separation}, respectively. We compare our results with the previous findings from the literature. We discuss their interpretations in light of simulations and physical processes. Finally, we address the caveats of this study.

\subsection{Suggestions for the usage of the methods}
We explore three methods to estimate mass-weighted single-age/metallicity of the bulges and the disks. The method based on $M_{*}$ weights has several limitations and requires high mass weights from both the galaxy components to constrain their solutions. The mass weights depend on the galaxy B/T but also on how well the data are able to spatially recover the two components. This method allows for more complicated models to be tested when compared with the linear model explored in this paper, adding radial and/or time gradients to the age/metallicity of the bulge and the disk. The method based on flux weights uses fewer assumptions, but it requires high signal-to-noise ratio at each wavelength range to produce spectra not contaminated by systematic noise. The separation between the bulge and disk contributions is most contaminated for the method based on radial separation, reducing the analysis to be valid for a small galaxy sample. However, this method is the most simple method to be implemented and it allows to obtain similar outcomes. 

In conclusion, the use of the $M_{*}$ and flux weights methods is preferable, especially foreseeing large IFU surveys observed by spectrographs such as Hector \citep{Bryant2018}. These surveys will be characterized by high spectral resolution in the blue arm and high S/N at each wavelength, allowing the methods to overcome some of the previous data-dependent limitations.

\subsection{Comparison with previous works}
The results for the bulge/disk $g-i$ colors in Section~\ref{colors} are in agreement with previous studies on the cluster environment, finding most bulges being redder than the disks \citep{Hudson2010,Head2014}. 
\citet{Head2014} studied the colors of the bulges and the disks of S0 galaxies in the Coma cluster. They observed bulges with redder colors and a median $g-i$ offset of 0.097$\pm$0.004 mag in agreement with our median offset. The difference in color between the two galaxy components underlies different star formation histories, implying that redder bulges are older and/or more metal-rich. However, the study of the colors alone is not able to break the age-metallicity degeneracy. Our results confirm that the colour difference is driven by the higher metallicity of the bulge relative to the disk, and not by differences in the stellar population ages. 

Previous works on bulge/disk ages and metallicities study the differences between the age and metallicity results of the two components (e.g., \citealp{Johnston2014,FraserMcKelvie2018}). We exclude differences that are not statistically significant with respect to the random errors on the bulge/disk stellar population properties. Our finding in Section~\ref{Ages and metallicities} of galaxies with a bulge significantly younger and more metal-rich than the disk is in agreement with previous works on S0 galaxies in clusters for age and metallicity gradients \citep{Bedregal2011} and for bulge and disk stellar populations \citep{Silchenko2012,Johnston2014}. \citet{Johnston2014} studied only 13 S0 galaxies of the Virgo cluster, finding that all of them have younger, more metal-rich bulges when compared to the disks. However, we observe that only a fraction ($\sim$30\% considering significant differences) of the whole galaxy sample shows such stellar population properties. For galaxies with a bulge younger than the disk there is the possibility that the component identified as the bulge using the photometric bulge-disk decomposition, either in 1D like in \citet{Johnston2014} or in 2D like in \citet{Silchenko2012} and in this study, is contaminated by the light of the central part of the disk characterized by recent star formation. To explore this possibility a more sophisticated combined photometric and kinematics bulge-disk decomposition would reveal more information. For the CALIFA survey, \citet{Mendez2019b} found that star formation occurs only in the disk, not in the bulge, and that it is not limited to the outer disk regions, but it also occurs in the central regions dominated by the bulge light. For the ATLAS$^{\rm 3D}$ survey, early-type galaxies with younger and more metal-rich stellar populations in the core regions with respect to the outskirt have also been observed \citep{Krajnovic2013,McDermid2015,Krajnovic2020}, in agreement with our outcome. 

We do not observe a trend with galaxy stellar mass in Section~\ref{Exploring bulges older or younger than the disks}, unlike \cite{FraserMcKelvie2018} who observed two different population for 279 low- and high-mass S0 field galaxies in the MaNGA survey. \citet{FraserMcKelvie2018} measured light-weighted averages of Lick indices and translated them into age/metallicity by bi-linearly interpolating the model lines of the stellar population templates. They found that galaxies with $M_{*}/M_{\odot}>10^{10}$ are characterized by older bulge and younger disk, while galaxies with $M_{*}/M_{\odot}<10^{10}$ mainly have younger bulge and older disk. In this work we consider only galaxies with $M_{*}/M_{\odot}>10^{10}$ due to signal-to-noise ratio limitations on SAMI data \citep{Owers2019}. The percentage of high-mass galaxies with older bulge is 85\% in the field from \citet{FraserMcKelvie2018}. This percentage drops to 23\% in the cluster environment from this study, considering statistically significant results. This comparison suggests that there may be differences in the formation mechanisms of S0 galaxies in the field with respect to the cluster environment. This conclusion is supported by the work of \citet{Coccato2019}, who compared the kinematics of cluster and field S0s finding differences between the two populations and suggesting different formation processes in different environments. 

\subsection{Physical interpretations}
To explain bulges being redder than disks, \citet{Head2014} speculated that bulges need to be either $\sim$2-3 times older or $\sim$2 times more metal-rich. Our results support the latter scenario where redder bulges are $\sim$2-3 times more metal-rich than disks, but they can be either older or younger. 

S0 galaxies with a bulge that is older than the disk can be formed from faded spiral galaxies through an inside-out quenching process in situ to the galaxy where the bulge quenches first \citep{Mendel2013,Tacchella2015,Spindler2018}. Once these galaxies enter the cluster environment their star formation is mainly quenched via interactions with the intra-cluster medium, where environmental processes such as ram pressure stripping or strangulation, remove the disk gas \citep{Gunn1972,Larson1980}. Evidence of outside-in star formation quenching due to ram pressure stripping processes acting on SAMI cluster galaxies have been observed by \citep{Owers2019}. 
 
The observation of the galaxy population with a bulge that is younger and more metal-rich than the disk implies that the bulge hosts a more recent star formation fueled by enriched material than the disk where the activity has stopped \citep{Bedregal2011}. In this context, \citet{Johnston2014} proposed a scenario where in the process of quenching the disk part of the galaxy, the gas collapses towards the galaxy centre. This leads to a final episode of star formation in the bulge, leading to its younger age compared to the disk. S0 galaxies with a younger bulge than the disk can be formed from spiral galaxies through multiple tidal interactions in dense environments according the simulations of \citet{Bekki2011}. These encounters can trigger repetitive starbursts within the bulges of spirals, leading the bulges to grow with younger stellar populations than the disks. 
\\
\subsection{Caveats}
\label{Caveats}
We address the three caveats that mainly characterized this study: (i) the definitions of ``bulge'' and ``disk'', (ii) the estimate of single-ages/metallicities and (iii) the interpretation of the age/metallicity measurements.

We assume that galaxies are characterized by only two components: a central bulge and a surrounding disk. We use a purely photometric approach to separate the components. However, we need to keep in mind that these definitions are limited to an extended ``disk'' and a centrally-concentrated ``bulge'', and they might not respect the entire complexity of the galaxy components \citep{Head2015,Fisher2019}. In particular, the fit of the exponential component is dominated by the outer regions of the galaxy and its extrapolation at the centre can only capture a limited amount of the complex central structure that might be associated with that disk \citep{Mendez2019b}. The recent work of \citet{Breda2020} shows insights about the invalidity of the standard exponential disk profile within the galaxy central region, observing a down-bending trend beneath the bulge radius for a large fraction of late-type galaxies. Assigning all the left-over light to a ``bulge'' is how we often proceed, but we may need to balance the interpretation of the two photometric components in terms of what we classically call ``bulge'' and ``disk''. A combined analysis of their kinematic and photometric properties would shed more light on their definition \citep{Tabor2019,Poci2019,Oh2020}. 

We consider single values for the mass-weighted age and metallicity of the bulge and the disk, without considering radial gradients which might be needed for a better modelling of these properties. In particular, radial gradients might limit the effects of the photometric extrapolation of the disk to the centre on the stellar population measurements associated with the bulge. Figure~\ref{DataModelResidual} shows that using Equations (1) and (2) we are able to reproduce approximately radial gradients in age and metallicity for the 9091700038 galaxy. However, for a few spatial bins the residuals indicate differences between the data and model at the 5$\sigma$ level. This large discrepancy for some bins is consistent with the expectation that radial gradients exist within the galaxy, but the proposed linear model is only able to reproduce the data using single $Age_{bulge/disk}$ and $[M/H]_{bulge/disk}$ values. For a better agreement, more complicated versions of the Equations (1) and (2) that take into account radial gradients for the age/metallicity of the bulge/disk should be fitted.

In regards to the interpretation of the age/metallicity measurements, we make use of the MILES SSP templates in Section~\ref{sec:Galaxy age and metallicity maps}. These SSP models make predictions up to $\sim$17.78 Gyr, which is higher than the upper age limit measured for the Universe 13.798$\pm$0.037 Gyr (Planck Collaboration, \citealp{Planck2014}). \citet{McDermid2015} explored galaxies with an estimated age $>14$ Gyr, finding that they are consistent with the age of the Universe once observational uncertainties in the data and systematic uncertainties in the models are considered (see their Appendix D). Moreover, as stated by \citet{Johnston2014}, different SSP models lead to different absolute measurements of age/metallicity because of the uncertainties that characterized the stellar astrophysics into these models. In this context, the age/metallicity measurements of the bulge and the disk should be interpreted in a relative sense and considered as constraints on the different stellar populations rather than the absolute results. In our analysis we exclude age/metallicity differences between the galaxy components that are not statistically significant compared to their random errors. These differences highlight possible limitations in the used data or methods, since we are not able to discern between the age/metallicity of the bulge and the disk. The use of photometric and spectroscopic data with higher S/N and spatial resolution as well as a larger galaxy sample might help us to obtain more statistically significant results.

\section{Summary and conclusions}
\label{Summary and conclusions}
We explore the stellar population properties of the bulge and disk components in cluster galaxies. Specifically, we investigate their colors, mass-weighted ages and metallicities separately. We study the SAMI cluster sample characterized by eight low-redshift clusters. This sample allows us to combine photometric with spectroscopic information and to conduct a statistical study on the bulge and disk stellar populations. 

The characterization of the double-component galaxy sample is based on the 2D photometric bulge-disk decomposition performed by Barsanti et al. in preparation for the SAMI clusters. They use a Bayesian approach and the packages {\sc ProFound} for source detection and {\sc ProFit} for galaxy light modelling. From the double-component sample of 469 galaxies within $2.5\,R_{200}$ identified by Barsanti et al. in preparation, we select only primary SAMI targets with $R<R_{200}$, $M_{*}/M_{\odot}>10^{10}$ and $\sigma<400$ km $\rm{s^{-1}}$. We find 192 double-component galaxies, where the bulge + disk components are preferentially modelled by the S\'ersic + exponential profile. These double-component galaxies are mainly characterized by fast rotators, passive spectra and S0 morphology.

To measure mass-weighted single-age and single-metallicity for the bulges and the disks we investigate three methods:
\begin{enumerate}
\item We develop a new method based on stellar mass weights. The use of the 2D bulge-disk decomposition performed in the $r$, $g$ and $i$-bands allows to shift from light to mass. We estimate the spatially-resolved contributions separately from the bulge and the disk to the galaxy stellar mass. We spatially bin and kinematically correct the SAMI galaxy spectra. For each spatial bin the bulge-disk mass fractions are used as weights on the mass-weighted age and metallicity of the galaxy, derived using full spectral fitting. Finally, mass-weighted single-age and single-metallicity values are obtained for the bulge and the disk of each galaxy.
\item We test a method similar to that of \citet{Mendez2019} based on flux weights for the two components. Separate 1D SAMI aperture spectra are obtained for the bulge and the disk according to their spatially-resolved bulge-to-total and disk-to-total flux ratios. 
\item  We test a method similar to that of \citet{FraserMcKelvie2018} based on radial separation. Separate 1D bulge and disk spectra are obtained considering the galaxy spectra from the most central and outermost bins, respectively. To avoid the contamination of the flux from the other components, representative bins of the bulge (disk) are selected having a contribution in galaxy $M_{*}$ from the bulge (disk) of at least 60\%.
\end{enumerate}

Comparing the three methods, we observe that the $M_{*}$ and flux weights methods show the best agreement. The highest deviations between the results are for galaxies with low $f^{bulge/disk}_{M*, bin}$ values. This is particularly evident for the radial separation method. The implementation of the $M_{*}$ and flux weights methods is preferable, especially foreseeing large IFU surveys characterized by higher S/N and higher spectral resolution in the blue arm than the SAMI data (e.g., Hector, \citealp{Bryant2018}).

Our results can be summarised as follows.
\begin{enumerate}
    \item Bulges are mainly redder than the disks. The median $g-i$ offset separating the color distributions of the bulges and the disks is 0.12$\pm$0.02 mag.
    \item According to all three methods used to measure mass-weighted single-age and single-metallicity of the bulge and the disk, we observe that $\sim$23\% of galaxies have bulges significantly older than the disk counterpart and $\sim$34\% have significantly younger bulges. For the remaining $\sim$43\% of galaxies the differences in age are not significant with respect to their random uncertainties. We find $\sim$62\% of bulges on average $\sim$2-3 times significantly more metal-rich than the disks. Only $\sim$7\% of the galaxies have a bulge significantly more metal-poor than the disk.
    \item According to all three methods, both galaxy populations with older or younger bulges compared to the disks show redder and more metal-rich bulges. The two populations do not show significant differences when comparing galaxy, bulge/disk and environment properties. Limitations of this analysis are: the exclusion of galaxies with $M_{*}<10^{10}\,M_{\odot}$ and a sample not large enough to perform a reliable study as a function of environment metrics.
    \item According to all three methods, redder bulges are more metal-rich than the disks, whereas they can be either younger or older than the disks. Our analysis suggests that the redder color in bulges is mainly driven by their metallicity.
\end{enumerate}

Further analysis is required to understand the role played by the galaxy stellar mass and the environment in affecting the stellar population properties of the bulge and the disk. A complete study of high- and low-mass S0 galaxies in high- and low-galaxy density environments is necessary to investigate possible differences. A study of the projected-phase space would be ideal to shed light on the possible environmental processes acting in quenching star formation activity within the bulge and the disk, but a larger sample of double-component galaxies is required to extrapolate reliable conclusions.

Overall, the three different methods tested to estimate age and metallicity of the bulge and the disk suggest the same results. Bulges are mainly redder and $\sim$2-3 times more metal-rich than their surrounding disk, while we do not find any particular trend with age. We conclude that the redder color in bulges is mainly due to this higher metal content of the bulge stellar populations. We observe galaxies with bulges either older or younger than the disks, suggesting that they might experience different formation scenarios. Analysis of the star formation histories of both components may help us understanding possible different galaxy evolution paths.

\section*{Acknowledgements}
We thank the referee for the useful comments. SB acknowledges the International Macquarie University Research Training Program Scholarship (iMQRTP 2017537) for the support. M.S.O. acknowledges the funding support from the Australian Research Council through a Future Fellowship (FT140100255). LC is the recipient of an Australian Research Council Future Fellowship (FT180100066) funded by the Australian Government. JBH is supported by an ARC Laureate Fellowship that funds Jesse van de Sande and an ARC Federation Fellowship that funded the SAMI prototype. JJB acknowledges support of an Australian Research Council Future Fellowship (FT180100231). JvdS is funded under Bland-Hawthorn's ARC Laureate Fellowship (FL140100278). NS acknowledges support of an Australian Research Council Discovery Early Career Research Award (project number DE190100375) funded by the Australian Government and a University of Sydney Postdoctoral Research Fellowship. The SAMI Galaxy Survey is based on observations made at the Anglo-Australian Telescope. The Sydney-AAO Multi-object Integral field spectrograph (SAMI) was developed jointly by the University of Sydney and the Australian Astronomical Observatory, and funded by ARC grants FF0776384 (Bland-Hawthorn) and LE130100198. The SAMI input catalogue is based on data taken from the Sloan Digital Sky Survey, the GAMA Survey and the VST/ATLAS Survey. The SAMI Galaxy Survey is supported by the Australian Research Council Centre of Excellence for All Sky Astrophysics in 3 Dimensions (ASTRO 3D), through project number CE170100013, the Australian Research Council Centre of Excellence for All-sky Astrophysics (CAASTRO), through project number CE110001020, and other participating institutions. The SAMI Galaxy Survey website is http://sami-survey.org/. This study uses data provided by AAO Data Central (http://datacentral.org.au/).

%



\software{2d{\sc fdr} package \citep{2015ascl.soft05015A}, astrolibR \citep{AstroLib}, {\sc astropy} \citep{Astropy2013, Astropy2018}, Hyper-Fit \citep{Robotham2015}, K-corrections \citep{Blanton2007}, pPXF \citep{Cappellari2004,Cappellari2017}, {\sc ProFit} \citep{Robotham2017}, {\sc ProFound} \citep{Robotham2018}, SAMI {\sc python} package \citep{Allen2014}.}


\newpage
\appendix
\section{Required Signal-to-Noise ratio for spatial bin}
\label{sec:Required Signal-to-Noise}
For IFU surveys, the outer spaxels where the galaxies are fainter have lower continuum signal-to-noise ratio (S/N) when compared with the brighter central regions. Improving the S/N of the SAMI spectra requires spatial binning so that reliable stellar population parameters can be estimated. However, there is also a trade-off in the amount of binning required and the spatial resolution necessary for disentangling bulge and disk properties. Therefore, we wish to determine the minimum S/N required to produce reliable SSP information without the unnecessary loss of spatial information. As defined in Section~\ref{A new method}, S/N is measured taking the median value of the flux divided by the noise in the rest-frame wavelength range 4600-4800~\AA\hspace{0.5mm}.

We investigate the accuracy of stellar ages estimated by pPXF in order to find the minimum S/N required for each spatial bin. We make use of the MILES single stellar population (SSP) templates \citep{Vazdekis2010}. We consider 6 metallicities [M/H]=0.22, 0.00, -0.40, -0.71, -1.31, -1.71 and for each metal value we select 10 ages from 1 to 16 Gyr. We change these templates in order to reproduce simulated spectra representing the features of the observed SAMI spectra. Each simulated spectrum is convolved with the typical velocity dispersion of a galaxy $\sigma_{gal}=200$ km $\rm{s^{-1}}$. We use the formula of \citet{Owers2019}, which considers the convolution with the instrument resolution for the blue arm of the spectrograph $\sigma_{inst}=1.13$~\AA\hspace{1mm}\citep{vandeSande2017} and the resolution of the MILES templates $\sigma_{MILES}=1.06$~\AA\hspace{1mm}\citep{Falcon2011}. The final wavelength-dependent velocity dispersion is:
\begin{equation}
\sigma^2=\Big (\frac{\lambda \sigma_{gal}}{c} \Big)^2 + \sigma_{inst}^2-\sigma_{MILES}^2
\end{equation}
where $\lambda$ is the wavelength in \AA\hspace{0.5mm} and $c$ is the speed of light in km $\rm{s^{-1}}$. Then, to make each simulated spectrum similar to the SAMI spectrum, the simulated spectrum is interpolated onto a grid with the same wavelength spacing as in the blue SAMI spectrum and in the same rest-frame wavelength range 3650-7000~\AA\hspace{0.5mm}. 

The initial noise associated with each simulated spectrum is extracted from the central region of the same galaxy as the square root of the variance, combining the red and blue SAMI data cubes. For each simulated spectrum we consider the different S/N values=10, 20, 30, 40, 50 scaling the noise every time. For each S/N value we run 100 iterations adding noise to the simulated spectrum using a random Gaussian array. We use pPXF to fit the simulated spectrum. First, we fit for the velocity and velocity dispersion using 0 and 200 km $\rm{s^{-1}}$ as inputs, considering the MILES library, excluding those pixels with noise=NaN and using a 12th degree additive polynomial to minimize template mismatch. Then, we fix the kinematic components and we fit for mass-weighted ages without considering any gas emission line and using no multiplicative polynomial and no regularization. 

\begin{figure}
\centering
\includegraphics[width=0.49\columnwidth]{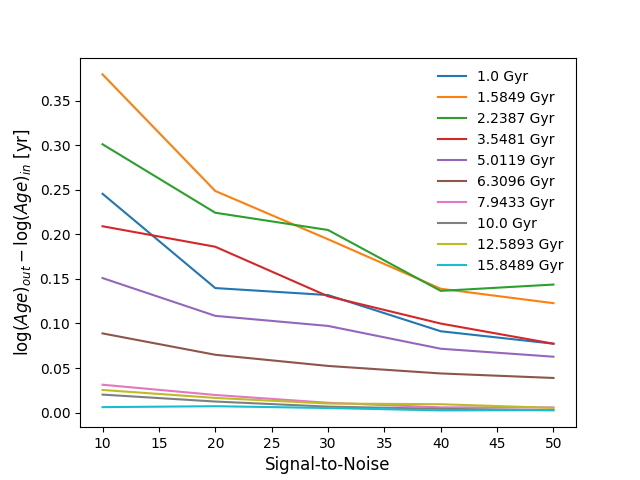}
\includegraphics[width=0.49\columnwidth]{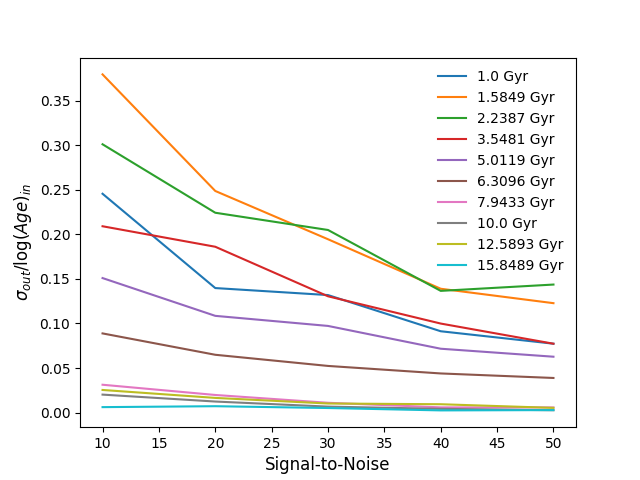}
\caption{For the fixed metallicity [M/H]=0.00, the \textit{left panel} shows the offset log(Age)$_{out}-$ log(Age)$_{in}$ and the \textit{right panel} shows the scatter around log(Age)$_{out}$ normalized by log(Age)$_{in}$. Both panels are as a function of S/N values. The S/N=20 is chosen as the minimum required S/N value for spatially binned spaxels.}
\label{SNoffset_scatter}
\end{figure}

For each S/N value, we estimate the mean log(Age)$_{out}$ and the scatter $\sigma_{out}$ as standard deviation of the output log age distribution and compare these quantities against their input values. Figure~\ref{SNoffset_scatter} shows for the fixed metallicity [M/H]=0.00 the offset log(Age)$_{out}-$ log(Age)$_{in}$ and the scatter $\sigma_{out}$ normalized by log(Age)$_{in}$ as a function of S/N in the \textit{left} and \textit{right panels}, respectively. As expected, offset and scatter decrease with increasing S/N. Considering the  \textit{left panel}, there is a sharp decline in the offset from S/N=10 to S/N=20. Beyond S/N=20, the decline is less steep. Considering the \textit{right panel}, $\sigma_{out}/\log(\rm{Age})_{in}\sim$ 0.002$-$0.020 at S/N=20 implies that we are able to recover the old ages within an uncertainty of $\sim$5\% and for young ages of $\sim$50\%. Similar values are found also for the other metallicities. Thus, we consider S/N=20 as the minimum required S/N for each spatial bin.

\begin{figure}
\centering
\includegraphics[width=0.5\columnwidth]{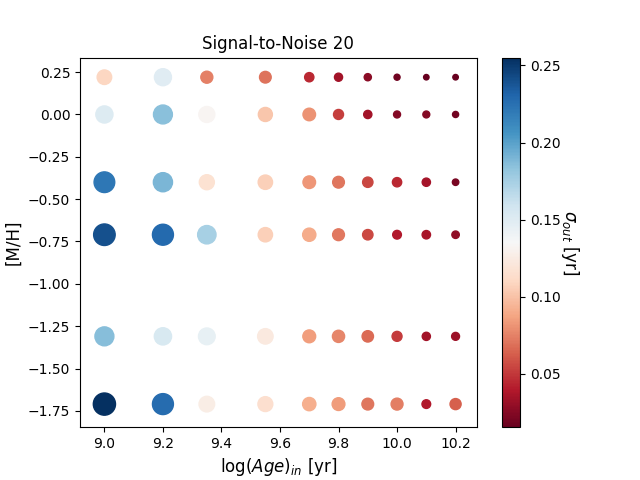}
\caption{Scatter for all the considered ages and metallicities for fixed S/N=20. Bluer and bigger the circle, larger is the scatter. The circle sizes range according $\sigma_{out}$ from 0.015 to 0.255 yr. Younger ages are characterized by a larger scatter compared to older ages.}
\label{SN20}
\end{figure}
Our results are in agreement with \citet{Ge2018}; younger ages are characterized by a larger offset and scatter compared to older ages. Young stellar populations have strong Balmer lines through which they can easily identified, however they show a larger stellar mass-light ratio compared to old stars, especially at low S/N, which makes the fit harder to constrain. Figure~\ref{SN20} shows the scatter for all the ages and metallicities for fixed S/N=20, where the highest offset is $\sim$0.25 for young ages. The youngest age (1 Gyr) does not show the largest discrepancy for [M/H]=0.25 and 0.00 (see also Figure~\ref{SNoffset_scatter}). This could be due to the strong peak of the age-sensitive Balmer absorption features that better constraint the fits. \citet{Ge2018} suggested S/N=30 as preferable value for spatially binned spectral fitting of spaxels. In our case S/N=20-30 have similar median scatters $\sigma_{out}$=0.08-0.06 yr over the ages, thus we prefer S/N=20 to preserve the spatial resolution. We find larger scatters and offsets compared to \citet{Ge2018} due to the fact that we consider mass-weighted ages, which are less directly linked to the light and therefore harder to constrain than luminosity-weighted ages (see Figure 11 of \citealp{Ge2018}). We do not test for the shape of the error spectrum and the level of dust extinction, since \citet{Ge2018} found that pPXF results are nearly independent of these quantities.

The same approach to investigate the accuracy of stellar ages estimated by pPXF is also used by \citet{Woo2019}, who built synthetic spectra from different star-formation histories. \citet{Woo2019} analyzed ages from 0.56 to 17.7 Gyr and S/N from 0 to 60, observing an increase in scatter for both younger ages and lower S/N, although the increase is not as significant as that seen in Figures~\ref{SNoffset_scatter} and~\ref{SN20}. This is likely because \citet{Woo2019} excluded catastrophic failures from their analysis.

\section{Test for kinematic corrections}
\label{Test for kinematics corrections}
We use the method based on flux weights described in Section~\ref{A method based on light weights} to test the accuracy of the kinematic corrections applied to the SAMI spectra in Section~\ref{sec:Corrections for velocity and velocity dispersion}. 181/192 double-component galaxies have been reliably decomposed with this method into separate 1D bulge/disk spectra. We apply pPXF to fit for the kinematic components of the 1D bulge/disk spectra separately and we run only 100 Monte Carlo simulations to estimate the 16th and 84th percentiles as uncertainties, since the kinematic fits are time consuming to compute due to their non-linear determination. We compare the results for the two components and implement the same test of \citet{Johnston2014}. If the kinematic corrections are applied correctly, then the 1D spectra of the bulge and the disk are expected to have matched zero velocity and maximum velocity dispersion measured within the galaxy.

\begin{figure*}[h!]
\centering
\includegraphics[scale=0.55]{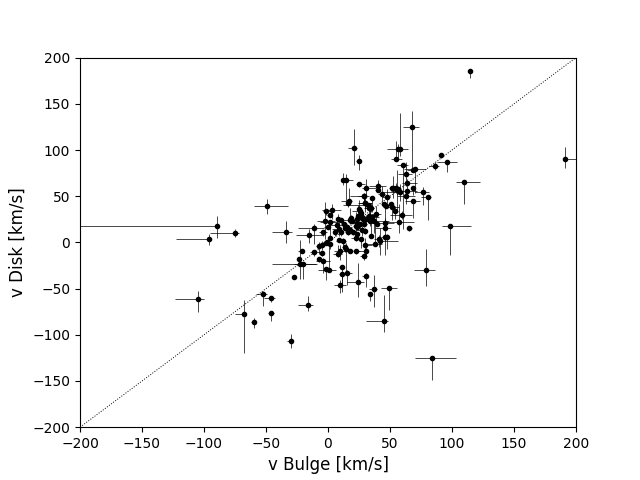}
\includegraphics[scale=0.55]{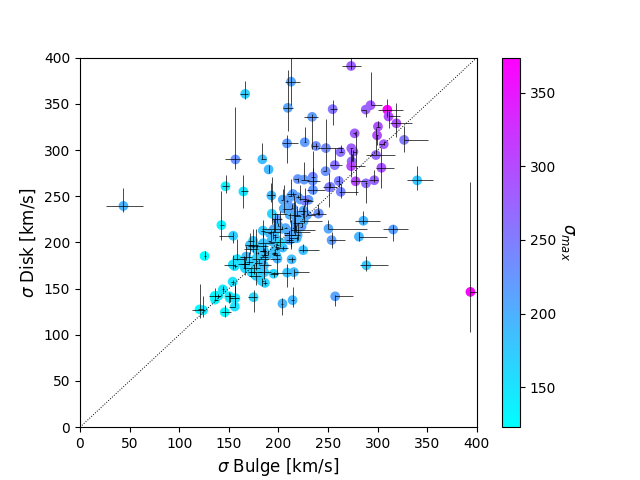}
\caption{Comparison between the velocity of the bulge and that of the disk (\textit{left panel}) and for the velocity dispersion (\textit{right panel}). The dotted black line represents the bisector. In the \textit{right panel} points are color-coded according to the maximum velocity dispersion within the galaxy. For both plots, the points are grouped along the bisectors with few outliers, confirming the accuracy of the applied kinematic corrections.}
\label{velsigma}
\end{figure*}

Figure~\ref{velsigma} displays the comparison between the velocities of the bulge and the disk (\textit{left panel}), and for the velocity dispersions (\textit{right panel}). For most of the galaxies, the bulge and disk velocities are grouped around the bisector with an offset of $\pm$200 km $\rm s^{-1}$. The bulge and disk velocity dispersions match along the bisector and correspond to the maximum value. The offset and the outliers are due to the low signal-to-noise ratio of the bulge/disk spectra. These plots highlight the reliability of the applied kinematic corrections, but the deviations show the reason we fit for these components in the step (2) of Section~\ref{sec:Galaxy age and metallicity maps}, before to estimate the age/metallicity values of the galaxy components.

\section{Testing the method based on $M_{*}$ weights}
\label{Simulations}
The method based on $M_{*}$ weights described in Section~\ref{A new method} estimates single-age/metallicity values of the bulge and the disk, according to Equations (1) and (2). We implement simulated SAMI galaxy spectra to test this method, comparing input age/metallicity values for the bulge and the disk with the respective outputs. Specifically, we explore how the results might be affected by low $M_{*}$ weights from the bulge and the disk, i.e. low $f^{bulge/disk}_{M_{*}, bin}$ values. We consider three galaxies having bulge-to-total flux ratio B/T=0.3, 0.5, 0.7 and for each galaxy we consider their projected bulge and disk stellar mass factions maps representing $f^{bulge/disk}_{M_{*}, bin}$ (similar to the \textit{right panels} of Figure~\ref{DataModelResidual}). For each B/T we use two MILES SSP templates representing the bulge and disk spectra, respectively, and their inputs for the age/metallicity. For each spatial bin in the simulated galaxies, we generate fake spectra by multiplying the bulge/disk input spectra with the associated $f^{bulge/disk}_{M_{*}, bin}$ and summing the flux. We make the simulated galaxy spectra similar to the SAMI spectra and characterized by S/N=20, as described in Appendix~\ref{sec:Required Signal-to-Noise}. For the noise associated to each simulated spectrum we make use of the blue and red SAMI data cubes of the considered galaxy. We extract the noise from a spaxel in the respective bin as the square root of the variance. 

We apply the pPXF steps to find $Age_{bin}$ and $[M/H]_{bin}$ as described in Section~\ref{sec:Galaxy age and metallicity maps}. We run 1000 Monte Carlo simulations adding noise to each simulated galaxy spectrum per bin using a random Gaussian array. For each simulation we fit Equations (1) and (2) as in Section~\ref{Age and metallicity of bulges and disks} to find $(Age,[M/H])_{bulge}$ and $(Age,[M/H])_{disk}$. The final outputs for the age/metallicity of bulge/disk are represented by the mean values of the $Age_{bulge/disk}$ and $[M/H]_{bulge/disk}$ distributions obtained from the simulations. The scatter associated with the mean values are obtained as the 16th and 84th percentiles of these distributions. For each B/T value, we consider the same 24 input combinations for the age/metallicity of the bulge/disk.

\begin{figure*}
\centering
\includegraphics[width=0.49\columnwidth]{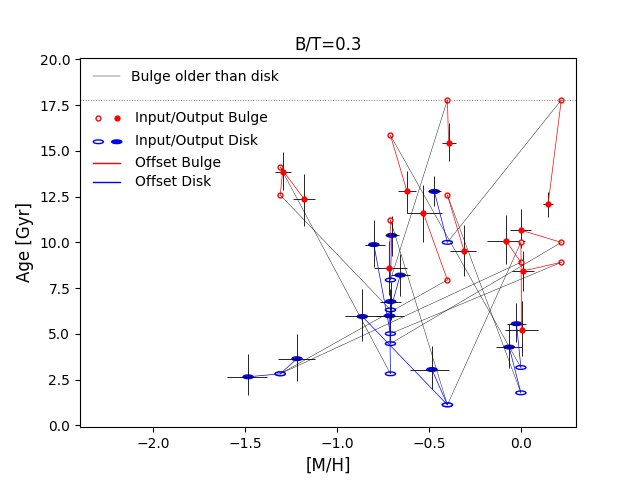}
\includegraphics[width=0.49\columnwidth]{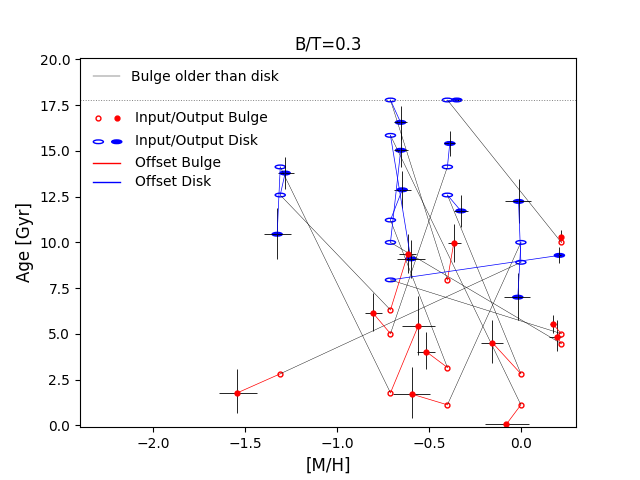}
\includegraphics[width=0.49\columnwidth]{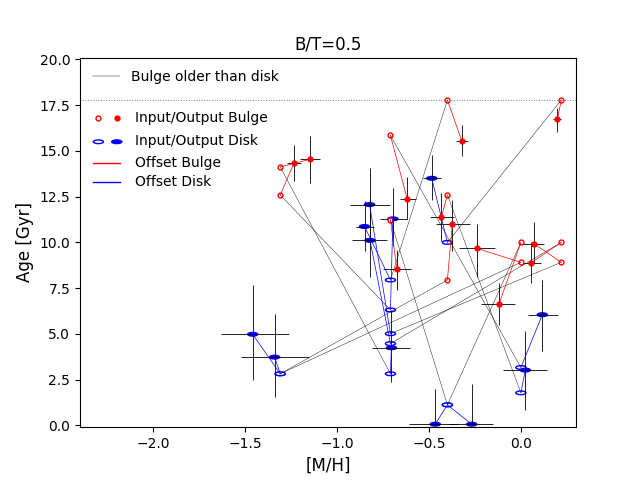}
\includegraphics[width=0.49\columnwidth]{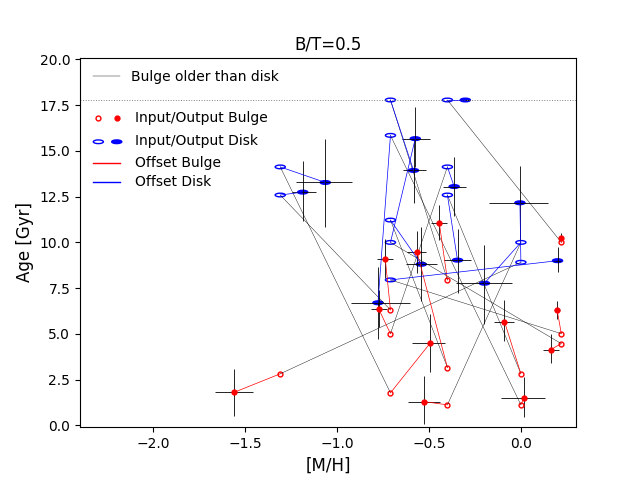}
\includegraphics[width=0.49\columnwidth]{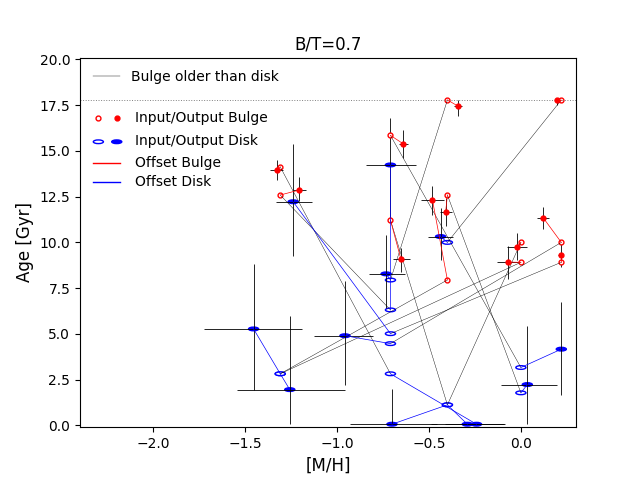}
\includegraphics[width=0.49\columnwidth]{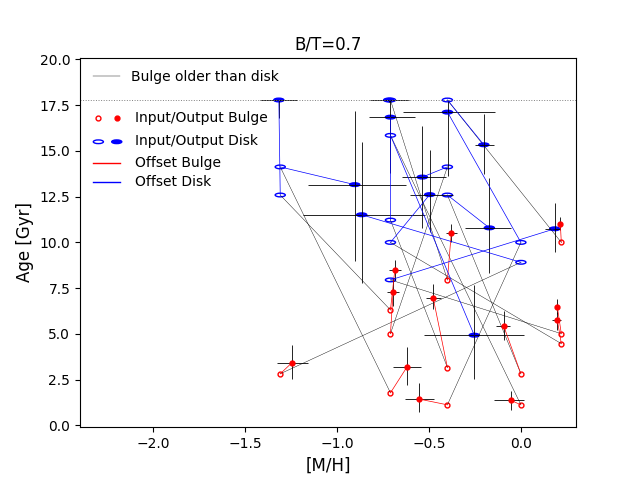}

\caption{Age-metallicity plots for B/T=0.3, 0.5, 0.7 (\textit{top to bottom panels}) separating the inputs with ages of the bulge older or younger compared to the disk (\textit{left and right panels}, respectively). The bulge/disk inputs are represented by circles/ellipses with red/blue contours and they are connected by the grey line. The bulge/disk outputs are represented by filled red/blue circles/ellipses with the associated 16th and 84th percentiles. Bulge/disk inputs and outputs are connected by red/blue lines representing the offsets, respectively. The bulge inputs are more closely recovered for high B/T, while for the disk the results are better constrained for low B/T.}
\label{AgeMetalPlotSimulations}
\end{figure*}

Figure~\ref{AgeMetalPlotSimulations} shows the age-metallicity plots for B/T=0.3, 0.5, 0.7 (\textit{top to bottom panels}) separating the inputs with ages of the bulge older or younger compared to the disk (\textit{left and right panels}, respectively). For increasing B/T values, the measured age/metallicity values more closely match the input values: both the offsets (red lines) and the scatter associated with the outputs (black lines) decrease. On the other hand, considering the age/metallicity disk results, for increasing B/T values both the offsets (blue lines) and the scatter associated with the outputs (black lines) increase and some outputs are pegged to the upper or lower fit limits. 

Pegged solutions for the bulge/disk are obtained for low/high B/T values, i.e. when the bulge/disk results are harder to constrain, respectively. Table~\ref{OffsetSimulations} lists for each B/T value the associated maximum values for $f^{bulge/disk}_{M_{*}, bin}$ and the mean offsets for the bulges and the disks from the 24 input combinations and the respective mean scatter. The bulge/disk offsets are estimated as the difference between $([M/H],Age)_{input}$ and $([M/H],Age)_{output}$ for the bulge/disk results, respectively. The bulge/disk mean scatter is estimated as the mean of the scatter associated with $([M/H],Age)_{output}$ for the bulge/disk results, respectively. For increasing B/T values and increasing maximum $f^{bulge}_{M_{*}, bin}$, the mean offset for the bulges decreases with the outputs being closer to their inputs, while the mean offset for the disks increases. The same trend is observed for the mean scatter. For increasing B/T, the mean scatter for the bulges decreases with the outputs being more precise, while the mean scatter for the disks increases meaning more spread output solutions. Low $f^{bulge/disk}_{M_{*}, bin}$ implies that the results are harder to constrain and might lead to pegged results at the lower or upper fit limits. We conclude that this method depends on the contributions from the bulge and the disk to the galaxy stellar mass, which might depend on the galaxy B/T but also on how well the data are able to spatially recover the two components.

\begin{table}[h!]
 \centering
  \caption{Testing the method based on $M_{*}$ weights. Column 1 lists B/T, columns 2/3 list the maximum values for $f^{bulge/disk}_{M_{*}, bin}$, columns 4/5 and columns 6/7 list the mean offsets and scatters for the bulges and the disks from the 24 input combinations.} 
  \label{OffsetSimulations}
  \begin{tabular}{@{}lcccccc@{}}
  \hline
 B/T & MAX($f^{bulge}_{M_{*}, bin}$) & MAX($f^{disk}_{M_{*}, bin}$) & $<$offset$>_{bulge}$ & $<$offset$>_{disk}$ & $<$scatter$>_{bulge}$ & $<$scatter$>_{disk}$\\
 & & & Gyr & Gyr & Gyr & Gyr\\
\hline
0.30 & 0.58 & 0.90 &1.96&2.50&0.09 &0.07\\
0.50 & 0.80 & 0.65 &1.87&2.86&0.09&0.15\\
0.70 & 0.91 & 0.51 &1.28&2.88&0.05&0.21\\
\hline
\end{tabular}
\end{table}

\bibliography{biblioBDssp}{}
\bibliographystyle{aasjournal}



\end{document}